\documentclass[useAMS,usenatbib]{mn2e}

\usepackage{graphicx}
\usepackage{times}
\usepackage{caption}
\usepackage{color}
\usepackage{float}
\usepackage{verbatim}
\usepackage{hyperref}
\usepackage{fixltx2e}
\usepackage{comment}
\usepackage{amsmath}

\title[Measuring environments in photometric surveys]{Measuring galaxy environments in large scale photometric surveys}

\author[J. Etherington and D. Thomas]{\parbox[t]{\textwidth}{James~Etherington\textsuperscript{\thanks{E-mail: james.etherington@port.ac.uk}}~and~Daniel~Thomas}\\
Institute of Cosmology and Gravitation, University of Portsmouth, Portsmouth, PO1 3FX, UK}

\begin{document}

\maketitle

\begin{abstract}

The properties of galaxies in the local universe have been shown to depend upon their environment. Future large scale photometric surveys such as DES and Euclid will be vital to gain insight into the evolution of galaxy properties and the role of environment. Large samples come at the cost of redshift precision and this affects the measurement of environment. We study this by measuring environments using SDSS spectroscopic and photometric redshifts and also simulated photometric redshifts with a range of uncertainties. We consider the $N$\textsuperscript{th} nearest neighbour and fixed aperture methods and evaluate the impact of the aperture parameters and the redshift uncertainty. We find that photometric environments have a smaller dynamic range than spectroscopic measurements because uncertain redshifts scatter galaxies from dense environments into less dense environments. At the expected redshift uncertainty of DES, $0.1$, there is Spearman rank correlation coefficient of $0.4$ between the measurements using the optimal parameters. We examine the galaxy red fraction as a function of mass and environment using photometric redshifts and find that the bivariate dependence is still present in the SDSS photometric measurements. We show that photometric samples with a redshift uncertainty of $0.1$ must be approximately $6-16$ times larger than spectroscopic samples to detect environment correlations with equivalent fractional errors.

\end{abstract}

\section{Introduction} \label{sec:introduction}

Fluctuations in the early universe were the seeds of the structure that is observed today. After a period of rapid inflation \citep{Guth1981}, in some regions, under the influence of gravity, dark matter accumulated and subsequently collapsed into haloes \citep{White1978}. This in turn, deepened the gravitation potential in these regions. In the hierarchical model of structure growth; haloes grew by accreting dark matter and also through merging with satellite haloes. Baryons fell into the potential wells of the haloes and eventually galaxies formed and evolved \citep{Press1974}. It seems plausible that there should be a measurable connection between the properties of galaxies and their environments.

It is for this reason that galaxy environment has been an active research field for several decades. The pioneering work in this field was the discovery of the galaxy morphology-density relationship \citep{Oemler1974, Dressler1980}. This work showed that elliptical galaxies are preferentially found in dense environments, whereas spiral galaxies are preferentially found in less dense environments. Large volume galaxy surveys such as the Sloan Digital Sky Survey (SDSS) \citep{York2000}, utilizing modern CCDs have revolutionised the study of galaxy environment. Nearly complete galaxy catalogues, including the angular and redshift information for large fractions of the sky, for the local universe, have made it possible to study the environmental dependence of simple galaxy properties, such as colour \citep{Hogg2004, Blanton2005, Chuter2011}, morphology, \citep{Park2007, Kovac2010, Bamford2009, Skibba2009} star formation rates \citep{Kauffmann2004, Peng2010} and stellar populations \citep{Thomas2010, Rogers2010}. Statistically significant samples can be obtained from today`s datasets, even after the data has been binned on several axes. A galaxy bimodality has been firmly established \citep{Strateva2001, Baldry2004, Mignoli2009, Pozzetti2010}. There is a ``red sequence'' population of galaxies which tend to be redder, more massive, elliptical and live in more dense environments. The stellar populations in these galaxies formed on shorter timescales \citep{Thomas2005} and the stars evolved rapidly until the gas and dust was depleted. After this their evolution was passive. Conversely the ``blue cloud'' population of galaxies tend to be bluer in colour, less massive, have spiral morphologies and live in less dense environments. They have evolved more slowly and still have active star formation. 

Both mass and galaxy environment are considered as drivers of galaxy evolution. However their precise roles and relative importance have been strongly debated. The definition of the mass and environment is vital. Galaxy environment is considered to be a second order driver but is potentially important for low mass galaxies \citep{Thomas2010, Rogers2010, Vulcani2012}. The fraction of red galaxies, in samples from the SDSS, has been shown to be a bivariate function of stellar mass and projected local density \citep{Baldry2006, Peng2010}. Hydrodynamical and semi analytical simulations, however, suggest that halo mass is the fundamental parameter. Observationally it is difficult to disentangle the effects of halo mass from galaxy environment, although \citet{Haas2012} have supplied a possible approach. Evidence that galaxy environment plays a significant role in driving the evolution of satellite galaxies is mounting \citep{VanDenBosch2008, Faltenbacher2010, Peng2012, Carollo2013}.  

Finding correlations between the properties of galaxies and their environments does not, however, imply causation. It is difficult to isolate and quantify, in a statistically robust manner, the relative importance of physical mechanisms such as tidal stripping \citep{Read2006}, harassment \citep{Farouki1981} and strangulation \citep{Larson1980} that transform galaxy properties, driven by a galaxy`s environment. However there is substantial evidence for ram-pressure stripping \citep{Abramson2011, Boselli2014, Fumagalli2014} in local clusters obtained from multiwavelength observations. Dwarf galaxies in falling into clusters can be completely stripped of their gas in {\raise.17ex\hbox{$\scriptstyle\sim$}}$1\;$Gyr \citep{Kenney2014}.  

To study the redshift evolution of galaxy properties as a function of environment statistically significant samples of galaxies are required for a large redshift range. The current state of the art surveys are pencil beam surveys such as DEEP2 \citep{Newman2013}, UKIDSS Ultra deep survey (UDS) \citep{Chuter2011} and the COSMOS survey \citep{Koekemoer2007}. These surveys are complete to relatively high redshifts so facilitate studies of the evolution of galaxy properties \citep[e.g.][]{Cooper2007}. However they target galaxies within relatively narrow areas (a few squared degrees). Larger areas (thousands of square degrees) are required to establish galaxy samples that can be investigated further (e.g. by binning in mass, galaxy type, environment etc.). This is feasible only with photometric surveys. Currently, spectroscopic surveys with these requirements would be too costly and slow.   

The next generation of large scale photometric surveys includes the Dark Energy Survey (DES) and ESA`s space mission Euclid. The DES aims to survey $5000$ square degrees of the sky by imaging $300$ million galaxies up to a redshift of {\raise.17ex\hbox{$\scriptstyle\sim$}}$1.5$ whereas Euclid will be a full sky survey to a redshift of {\raise.17ex\hbox{$\scriptstyle\sim$}}$2.0$. These surveys will provide the opportunity to study the evolution of galaxy properties as a function of environment. Photometric surveys make it possible to survey an enormous number of galaxies as the average measurement time per galaxy is considerably less than it is for spectroscopic measurements. This comes at the cost of redshift precision. In the DES the distribution of the difference between the photometric redshifts and the true redshifts is expected to approximately follow a Gaussian distribution with a standard deviation {\raise.17ex\hbox{$\scriptstyle\sim$}}$0.1$ \citep{Banerji2008}. 

\begin{figure*}
\begin{minipage}[t]{0.49\linewidth}
   \vspace{8pt} 
   \centering 
   \includegraphics[width=.80\linewidth]{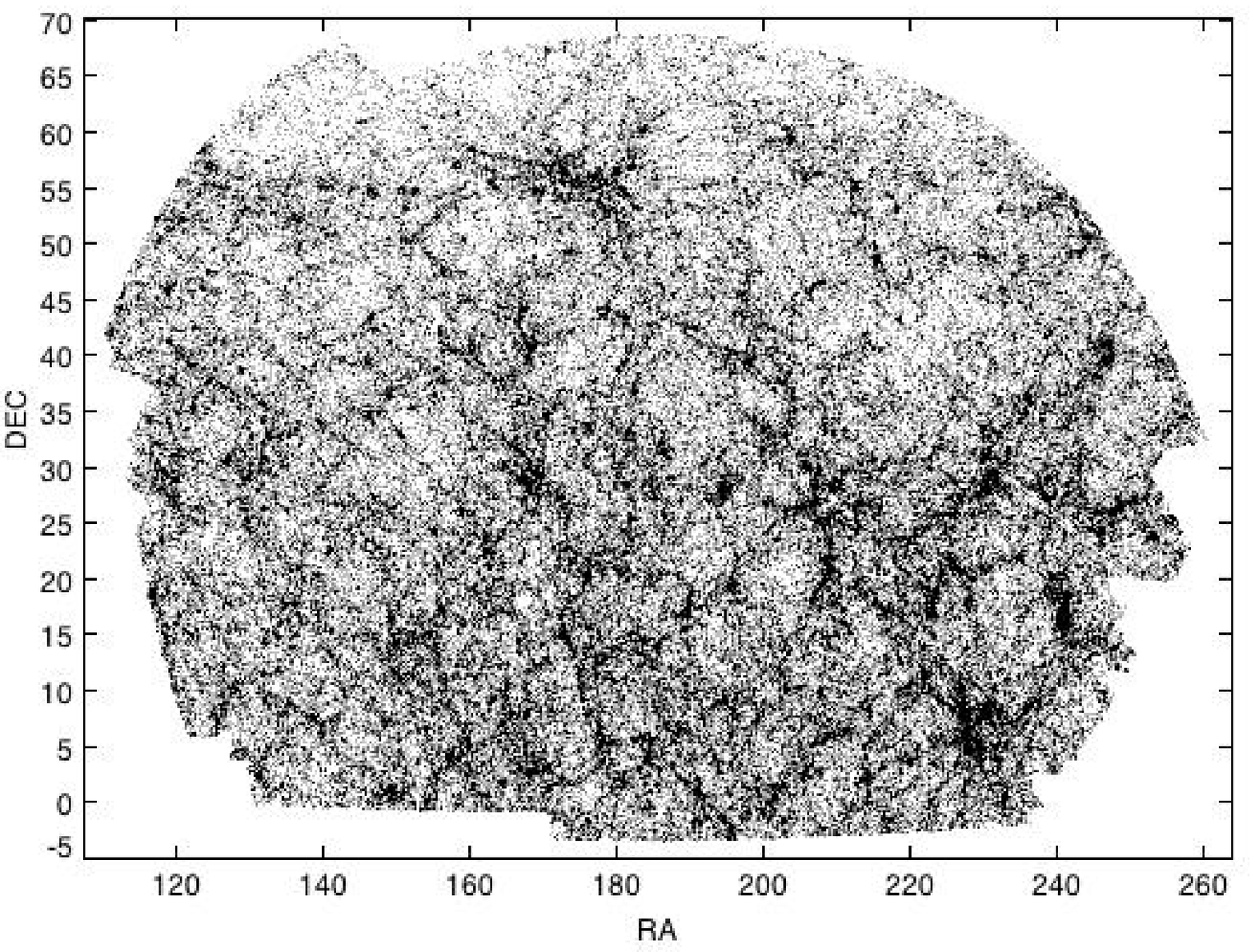}
\end{minipage}
\begin{minipage}[t]{0.49\linewidth}
   \vspace{0pt}  
   \centering  
  \includegraphics[width=.99\linewidth]{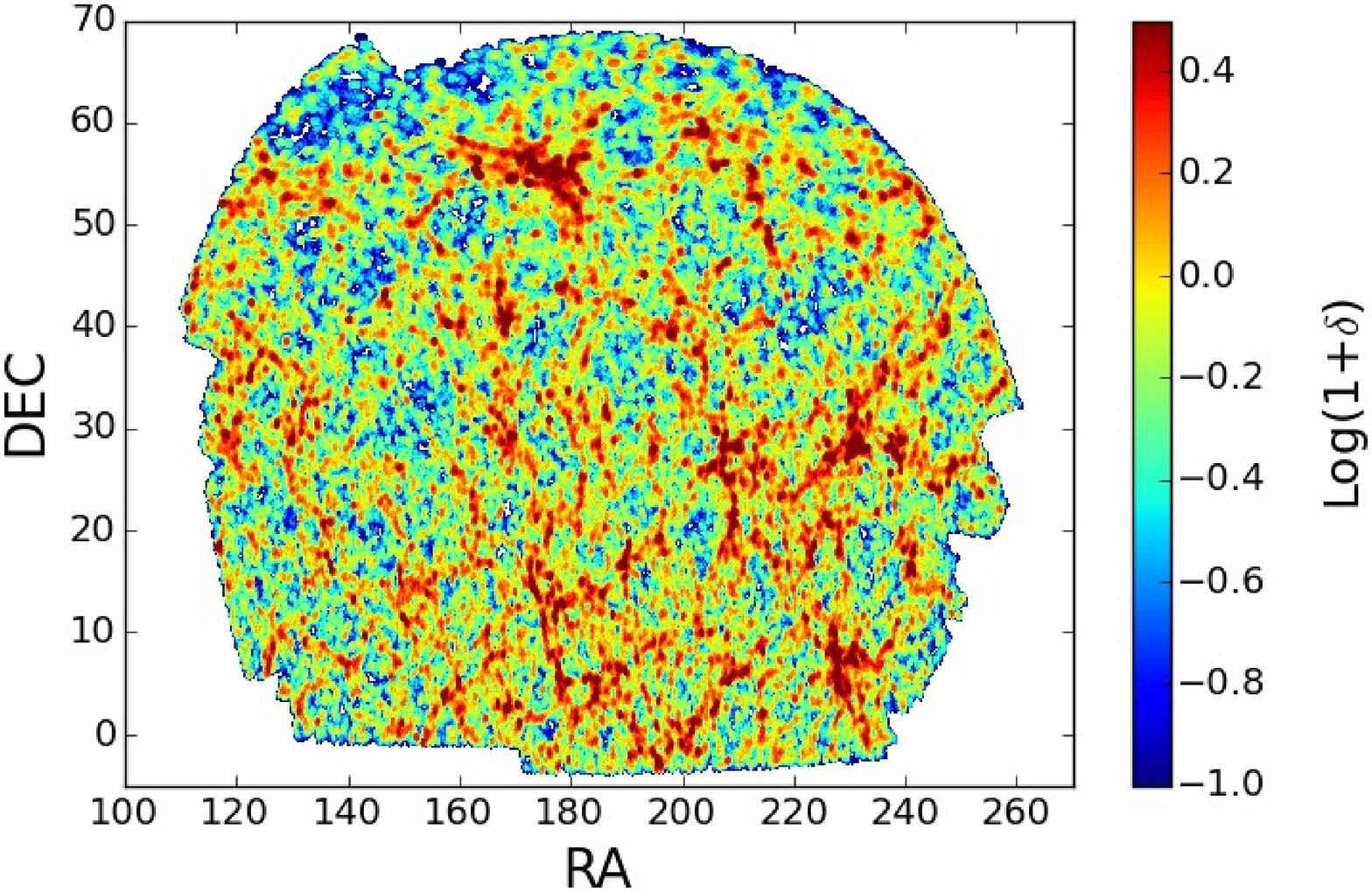}
\end{minipage}
\caption{Galaxy distribution for the SDSS main footprint (left) and the projected galaxy environment (right) for the redshift range: $0.02<z<0.085$. Dense environments are shown in red and sparse environments are in blue. }  
\label{fig:visualization_environment}
\end{figure*}

Errors in the redshifts of the galaxies result in errors in the environment measurements. A large error on a single redshift could result in a galaxy cluster member being measured in a void; a completely different environment. The situation becomes worse when multiple galaxies in the same spatial vicinity are assigned redshift errors. The environment of a particular target can vary dramatically. Redshift information derived from photometric techniques will be available in the next generation of photometric galaxy surveys but spectroscopic measurements will be limited. The purpose of this paper is to investigate two commonly used methods to measure galaxy environment and explore the impact of using photometric redshifts. Previous work examining this theme with considerations of the DEEP2 survey includes \cite{Cooper2005}. Here we reexamine the impact of photometric redshifts on galaxy environment measurements using SDSS data and consider the next generation of large scale photometric surveys. 

The paper consist of six sections. In the next section a brief review of the methods to measure galaxy environment is presented. Section \ref{sec:data_selection} describes the data selection and galaxy properties used in this work. The results are described in Section \ref{sec:results} and further discussion is presented in Section \ref{sec:discussion}. We present the conclusions in Section \ref{sec:conclusions}. 

In this work we have assumed a cosmology with $\Omega_{m}=0.3$, $\Omega_{\Lambda}=0.7$ and $H_{0} = 70\;$kms$^{-1}$Mpc$^{-1}$. We adopt comoving coordinates to calculate the distances between galaxies as this system is suitable for this low redshift work ($z<0.1$).

\section{Review of environment methods} \label{sec:review_environment_methods}

In essence, a galaxy`s environment is the density field in which a galaxy resides. We illustrate this in Figure \ref{fig:visualization_environment} using SDSS data. The left hand plot shows the SDSS footprint and the right hand plot shows the average projected galaxy environment for the redshift range: $0.02-0.085$. Several research fields, although often considered as separate, probe this density field. For example, there is significant effort in the weak lensing community to produce density maps \citep[e.g.][]{Szepietowski2014} and the 2-point correlation function of the galaxy distribution is routinely used in studies of large-scale structure \citep[e.g.][]{Tojeiro2014}. However, the environmental dependence of galaxy properties has been typically studied with two broad families of methods: the $N$\textsuperscript{th} nearest neighbour methods and the fixed aperture methods. Within these two broad families of methods there are a large number of variants \citep[e.g.][]{Muldrew2012} and this makes it challenging for researchers to compare results. The environment measurements computed via different methods should be correlated. However, they often probe different scales. For example, $N$\textsuperscript{th} nearest neighbour methods intrinsically adapt the scale probed for each target whereas the scale probed by aperture methods is fixed. 

\subsection{Fixed aperture}

Perhaps the simplest method to measure galaxy environment is to count the number of galaxies (excluding the target) within a fixed volume centred on the target.  A number density can be computed by dividing the number of galaxies found by the comoving volume. This type of approach is known as a fixed aperture method. A variety of apertures have been used including: spheres \citep{Croton2005}, cylinders \citep{Gallazzi2009} and annuli \citep{Wilman2010}. These fixed aperture methods are discrete measures of environment as the number count is either a positive integer or zero and the volume is fixed. The discrete nature of these environment measures makes it difficult to distinguish between low density environments. Field galaxies can exist in isolation with no other galaxies present within an aperture, centred on the galaxy.

The size of the aperture used in a study, for example, the radius of the sphere or the radius and height of the cylinder, must be chosen. Choosing a small aperture will result in a large number of the target volumes being devoid of galaxies. However the measure would be sensitive to the most dense environments. Conversely, a large aperture will reduce the number of target volumes that are devoid of galaxies, enabling measurements of a range of low density environments. Probing increasingly large scales, however, tends to homogenise the environments. A large range of scales from $<1-10\;$Mpc \citep{Muldrew2012} have been studied in previous works probing individual haloes through to structures in the cosmic web. Studies show that environmental effects on galaxy properties (e.g. colour) are strongest on scales smaller than {\raise.17ex\hbox{$\scriptstyle\sim$}}$1\;$Mpc \citep{Blanton2007, Wilman2010}. Ideally the choice of scale should match the scale of the physical processes that are believed to drive the evolution of the galaxy properties. However considerations of the sample size, signal-to-noise and the dynamic range of environments often leads to a compromise, particularly when volume limited samples are constructed \citep{Kauffmann2004}. 

\subsection{$N$\textsuperscript{th} Nearest neighbour}

The other main family of methods is the $N$\textsuperscript{th} nearest neighbour methods. For each target galaxy in the dataset the $N$\textsuperscript{th} nearest neighbour galaxy is identified. Spherical volumes can be used with simulations \citep{Haas2012} as the distances between galaxies can be determined accurately. More often in observational studies cylindrical volumes are used to constrain the nearest neighbours, to mitigate the effects of redshift distortions or redshift errors. In either case the comoving distance to the $N$\textsuperscript{th} nearest neighbour is computed and this is used as the radius to compute a comoving volume centred on the target galaxy. Dividing $N$ by this volume gives a number density. A dense environment is obtained when the $N$\textsuperscript{th} nearest neighbour is close to the target.  

A consequence of the $N$\textsuperscript{th} nearest neighbour method is that if there are $N-1$ neighbours in a group, near the target, the $N$\textsuperscript{th} neighbour will be located in the next, closest group. Hence the volume targeted will be based on the inter and not the intra group distance. A careful choice of the value of $N$ enables studies to focus on a particular range of scales.

\section{Data selection} \label{sec:data_selection}

To examine the local universe we followed the approach adopted by \cite{Baldry2006} and also by \cite{Peng2010}. Spectroscopic and photometric datasets were extracted from the seventh data release (DR7) \citep{Abazajian2009} of the SDSS using the CasJobs website. Specifically this data includes the SDSS spectroscopic and photometric redshift measurements. 

First the photometric data was obtained from the ``Galaxy View'' in the SDSS database. Only objects with a Petrosian, $r$, magnitude in the range: $10.0<r<18.0$ were selected and to ensure stars were excluded only those objects where $r$\textsubscript{PSF} - $r$\textsubscript{Model}$>0.25$ were included \citep[e.g.][]{2002Scranton}. We refer to this sample as the photometric parent sample as all of the objects in the subsequent samples in this work can be found in this sample. Spectroscopic data was extracted for all of the objects in the photometric parent sample where measurements were available. The photometric parent sample contained $1,607,820$ objects of which $803,939$ objects had spectroscopic measurements. These selections were reduced to obtain two volume limited samples (see Section \ref{subsec:volume_limited_samples}). One sample was based on the spectroscopic redshifts and one was based on the photometric redshifts. During this reduction the redshift range was restricted to: $0.02<z<0.085$ and galaxies where the stellar mass and absolute magnitude were unavailable were excluded. This ensures that the photometric and spectroscopic samples were reasonably well matched.   

To investigate a range of redshift uncertainties, in addition to the spectroscopic and photometric data extracted from the SDSS, simulated photometric redshift catalogues were created (see Section \ref{subsec:simulated_photoz_catalogues}). Volume limited samples for the simulated redshift catalogues were created using the same methodology (see Section \ref{subsec:volume_limited_samples}).

\begin{figure}
   \centering 
   \includegraphics[width=.99\linewidth]{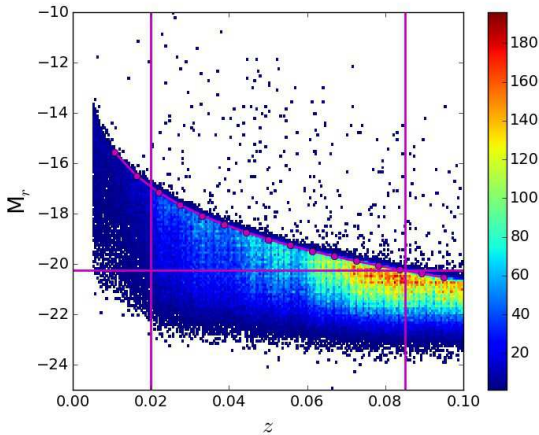}
\caption{Sample selection in the $M\textsubscript{r}$-redshift plane for the spectroscopic catalogue. The vertical magenta lines mark the redshift bounds of the selection. The connecting magenta curve shows the absolute magnitude detection limit. The horizontal magenta line binds the density defining population. The number of galaxies in each 2-dimensional bin is represented with a colour as indicated in the colour bar.}  
\label{fig:sample_selection}
\end{figure}

\subsection{Stellar masses, absolute magnitudes and k-corrections}

Stellar masses are adopted from \citep{Maraston2013}. These have been derived through a standard $\chi^{2}$ SED fitting procedure from the SDSS observed u,g,r,i and z band magnitudes using the fitting code HYPERZ \citep{Bolzonella2000} based on the SDSS spectroscopic redshifts. In this work we used the output based on the the passive
templates from \cite{Maraston2005} consisting of a metal-poor and a metal-rich old population as described in \cite{Maraston2009}. These are called ``passive'' because they exclude very recent and current star formation, but still allow for a wide range of stellar population ages of $3-15\;$Gyr. It is shown in \citet{Maraston2013} that these are adequate to properly model the stellar populations also of (star forming) galaxies in the blue cloud as they minimise the effect of underestimating the age (and hence the galaxy mass) owing to recent star formation \citep{Maraston2013}. Hence these provide the most accurate mass estimates also for star forming galaxies. We have verified that our masses are consistent with the publicly available MPA-JHU\footnote{http://www.mpa-garching.mpg.de/SDSS/DR7/} SDSS DR7 mass estimates \citep{Kauffmann2003, Salim2007} for both red sequence and blue cloud galaxies (scatter {\raise.17ex\hbox{$\scriptstyle\sim$}}$0.2$ dex).

In addition to the stellar masses the absolute magnitudes and K-corrections were derived from the output model stellar populations of this fitting procedure. K-corrections for each galaxy were obtained by interpolating the model u,g,r,i,z magnitudes. For this work we require the $k\textsubscript{ug} = k\textsubscript{u} - k\textsubscript{g}$ correction to obtain the u-g rest frame colour and also $k\textsubscript{r}$ to construct volume limited samples. We note that K-corrections are very small for the relatively low redshift sample studied here ($0.02<z<0.085$) and the analysis presented in this paper does not critically depend on the accuracy in K-correction calculations.

\subsection{Photometric redshifts}

The SDSS database provides two different types of photometric redshift measurements \citep{McCarthy2007}. One of the measurements is based on a template fitting approach \citep{Csabai2003} and the other on an artificial neural network approach \citep{Collister2004}. The fitting approach uses a set of galaxy spectral templates with a range of redshifts, types and luminosities. Expected colours are derived from the templates and fit to the observed galaxy colours. The templates with the best fits are found, from which photometric redshifts are obtained.

The artificial neural network consists of a number of layers of nodes with connections between nodes in adjacent layers. The network accepts the galaxy photometry and outputs photometric redshifts. A representative training set of galaxies is required to tune the connections between the nodes. Once the connections have been tuned the network can quickly compute photometric redshifts from galaxy photometry.  

In this work we show results based on photometric redshifts obtained from the template fitting method. Although not presented in this paper we have also checked some results using photometric redshifts obtained from an artificial neural network method. Quantitatively similar results were obtained and hence the choice of method does not affect the conclusions. 

For the SDSS photometric redshift sample the derived r-band absolute magnitudes based on the spectroscopic redshifts were adjusted to account for the difference in distance modulus between the spectroscopic and photometric redshift measurements. 

We calculated the redshift uncertainty of the photometric redshifts by taking half the difference between the $16^{th}$ and $84^{th}$ percentiles of the distribution of the difference between the spectroscopic and photometric redshifts. We find a redshift uncertainty of $0.0185$ for the SDSS photometric redshift sample.

\subsection{Simulated photometric redshifts} \label{subsec:simulated_photoz_catalogues}

In addition to the photometric redshifts from the SDSS photometry simulated photometric catalogues were created by displacing galaxies from their spectroscopic positions. Displacements were obtained for all of the galaxies that had spectroscopic redshift measurements. The redshift displacements were generated by drawing them from a Gaussian distribution with a particular standard deviation. The standard deviation ($\sigma_z$) is the redshift uncertainty of the catalogue. Simulated photometric redshifts were created by adding the redshift displacements to the spectroscopic redshifts. For the simulated photometric redshift samples the derived r-band absolute magnitudes based on spectroscopic redshifts were adjusted to account of the difference in distance modulus between the spectroscopic and simulated photometric redshift measurements. In total $14$ simulated catalogues were created with the following redshift uncertainties: $0.0025$, $0.005$, $0.0075$, $0.01$, $0.0125$, $0.0150$, $0.0175$, $0.02$, $0.03$, $0.04$, $0.05$, $0.06$, $0.08$ and $0.1$.

\subsection{Volume limited samples} \label{subsec:volume_limited_samples}

Intrinsically faint galaxies become undetectable as redshift increases. An apparent magnitude limited sample would, therefore, suffer from a volume bias. To mitigate this effect volume limited samples were constructed. This was achieved by plotting the K-corrected r-band absolute magnitude, $M_{r}$, as a function of the spectroscopic redshift. Figure \ref{fig:sample_selection} shows the selection of the spectroscopic redshift sample. The vertical magenta lines mark the lower ($z=0.02$) and upper ($z=0.085$) bounds of the sample. We compute the r-band absolute magnitude detection limit using the apparent magnitude limit of the survey ($17.77$), the distance modulus and the $95$\textsuperscript{th} percentile of the r-band K-corrections. The magenta connecting curve in Figure \ref{fig:sample_selection} shows this detection limit. The limiting magnitude, $M_{r}\textsubscript{(limit)}$, at the upper redshift boundary ($z=0.085$) was found to be $-20.26$. Selecting only those galaxies brighter than this value gives the volume limited sample. The same limiting magnitude obtained for the spectroscopic catalogue was then applied to the photometric and simulated catalogues. The galaxies in the volume limited samples are used to trace the density field and are hence also the density defining population.

\begin{table}
\caption{Environment methods and parameters} \label{table:env_methods}
\begin{center}
    \begin{tabular}{@{}lll}
    \hline
    &  Method & Parameters \\ 
    \hline
   (i) & $N$\textsuperscript{th} nearest neighbour & $N$=1-10; $dv$=1000-20000 km/s \\
   (ii) & Fixed aperture & $r$=0.1-3.0 Mpc; $dv$=1000-20000 km/s \\ 
    \hline
    \end{tabular}
\end{center}
\end{table}

\subsection{Target sampling rate} \label{subsec:target_sampling_rate}

The observational setup used in the SDSS consists of optical fibers mounted on aluminium plates. The finite size of the fibers makes it difficult to target more than one galaxy within an angular radius of 55 arcsecs using a single plate. Targeting two galaxies separated by less than this would result in a fiber collision. This targeting results in systematic under sampling that depends on environment. A target sampling correction is required to account for this. 

Following previous work \citep{Baldry2006, Peng2010}, the target sampling rate was calculated for each target in the spectroscopic redshift, photometric redshift and simulated photometric redshift volume limited samples. The target sampling rate is obtained by dividing the number of galaxies within the volume limited sample and within 55 arcsec of the target by the number of galaxies in the photometric sample, including galaxies with no redshift but would have bright enough absolute magnitudes to enter the volume limited sample within our redshift range, and within 55 arcsec of the target. A target sampling rate correction is required for {\raise.17ex\hbox{$\scriptstyle\sim$}}$10$ percent of the objects. The target sampling rate correction was used in two places: (i) to weight (x 1/(target sampling rate)) the contribution of each galaxy in the computation of galaxy environment. (ii) to weight (x 1/(target sampling rate)) each galaxy when representing the overall galaxy population. 

\subsection{Environment methods}

In this work we calculate a number density ($\rho_{i}$) for each galaxy in the photometric, spectroscopic and simulated samples using two methods: (i) the $N$\textsuperscript{th} nearest neighbour method and (ii) the fixed aperture method. Both methods have two aperture parameters: one to constrain the projected size of the aperture on the sky and the other to constrain the depth of the aperture (see Table \ref{table:env_methods}). 

In the $N$\textsuperscript{th} nearest neighbour method velocity cuts ($\pm dv\;$km/s) centered on the target galaxy are used to constrain possible neighbours. The Hubble flow velocity is assumed to be the speed of light multiplied by the redshift. The angle between the target and these galaxies are then computed. The galaxy with the $N$\textsuperscript{th} smallest angle is identified. This is the $N$\textsuperscript{th} nearest neighbour. This angle is used to define the aperture volume which is approximately a conical frustum centred on the target. The number density is then simply the sum of the $N$ target sampling corrections divided by the volume of the aperture. In this study we adopt $N=1-10$ and $dv=1000-20,000\;$km/s. For galaxies close to the redshift boundaries we adjust the aperture volumes as described in Section \ref{subsec:survey_edge}.

In the fixed aperture method again velocity cuts ($\pm dv\;$km/s) centered on the target galaxy are used to constrain the depth of the aperture. The projected size is fixed by choosing a value for the radius ($r\;$Mpc) of the aperture at the redshift of the target galaxy. Again the aperture is approximately a conical frustum. We use $r=0.1-3.0\;$Mpc and $dv=1000-20,000\;$km/s. For galaxies close to the redshift boundaries we adjust the aperture volumes as described in Section \ref{subsec:survey_edge}. Fixed apertures that are devoid of density defining galaxies are assigned a nominal minimum density of $0.5$ galaxies per aperture.

After we have computed a density for each galaxy we then evaluated an overdensity, $\delta$, using the equation below:

\begin{equation} \label{eq:overdensity}
\delta=(\rho_{i}-\rho_{m})/\rho_{m}
\end{equation}

The number density, $\rho_{i}$, is the density described above calculated using either the $N$\textsuperscript{th} nearest neighbour or fixed aperture methods. $\rho_{m}$ is the mean density of galaxies (weighted by the target sampling rate corrections) within a redshift window centered on the target galaxy utilizing all of the available area. Typically the size of the redshift window was set to be the same scale as the velocity cuts used by the method. We finally compute the galaxy environment using the expression:

\begin{equation} \label{eq:environment}
Log10(1+\delta)
\end{equation}

\subsection{Survey Edge} \label{subsec:survey_edge}

The SDSS footprint is not a simple geometry in right ascension and declination. It consists of a series of parabolic strips. The edge of the footprint is irregular (see Figure \ref{fig:visualization_environment}). To manage the edges of the survey often a survey mask populated with a set of random galaxies is employed. This is especially common in scenarios where the two-point correlation function is required \citep[e.g.][]{Wang2013}. For simplicity, however, we constructed a high order polygon ($>1000$) within the footprint to define the edge. This polygon has an area of $6926.64$ square degrees. Galaxies outside of this perimeter are discarded. In total {\raise.17ex\hbox{$\scriptstyle\sim$}}$10$ percent of the galaxies were discarded but many of these galaxies were in stripes outside of the main footprint. The angular separations between galaxies inside the region and the closest points on the perimeter of the polygon were computed.

For a galaxy close to the perimeter a fraction of the aperture volume used to calculate the environment would reside outside of the survey where there is no data. Environments computed for galaxies at the edges of the survey footprint would, therefore, tend to be underestimated. To manage this, the fraction, $f$, defined as the distance to the edge of the polygon divided by the projected radius of the aperture was calculated. We adopt a conservative approach and discard targets with $f<1.0$. Other authors use similar approaches. For example \cite{Peng2010} consider only those galaxies where $f>0.9$ and applies a correction to the aperture volumes where $f>0.9$ but $f<1.0$. Another approach \citep{Cooper2005} is to discard galaxies that are within a certain distance from the perimeter (e.g. $1\;$Mpc). 

In addition to the angular boundaries we manage the redshift boundaries. In cases where the target galaxies are close to a redshift boundary (i.e. the aperture would cross over the boundary) we reduce the depth of the half of the aperture infringing the boundary. We fix the half depth of this half of the aperture to be the comoving distance from the target galaxy to the boundary. This ensures the aperture fits inside the redshift range. The depth of the other half of the aperture that resides within the redshift range is not altered.

\begin{figure}
   \centering 
   \includegraphics[width=.99\linewidth]{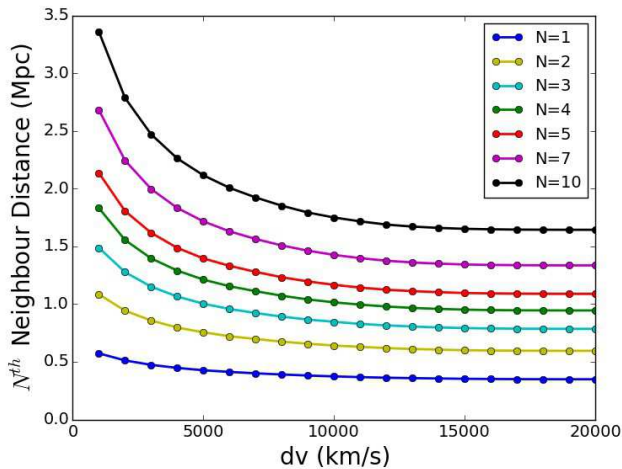}
\caption{Shows the median comoving distance to the $N$\textsuperscript{th} nearest neighbour for a range of values of $N$ as a function of the velocity cut ($dv$). }  
\label{fig:nth_dist_scales}
\end{figure}

\begin{figure}
\includegraphics[width=0.95\linewidth]{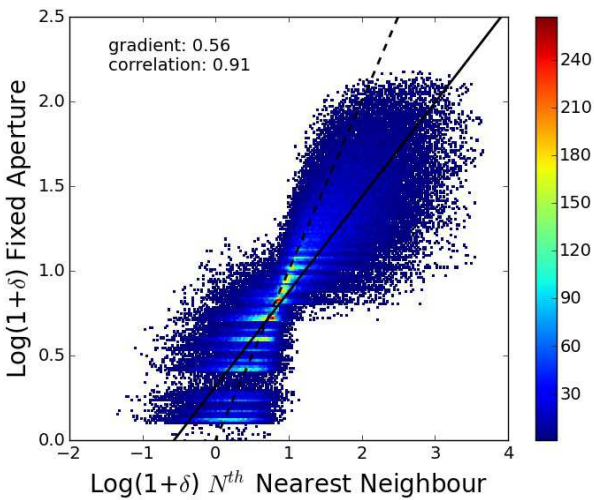} 
\caption{Shows the fixed aperture ($r=1.8\;$Mpc and $dv=1000 \;$km/s) spectroscopic environment measurements versus the nearest neighbour ($N=4$ and $dv=1000\;$km/s) spectroscopic environment measurements. The number of galaxies in each 2-dimensional bin is represented with a colour as indicated in the colour bar. The linear best fit is shown with the solid black line and the 1-to-1 is shown with the dashed line. The correlation quoted is Spearman rank correlation coefficient.}
\label{fig:spec_methods_comparison}
\end{figure}

\begin{figure*}
\begin{minipage}[t]{0.49\linewidth}
   \centering 
   \includegraphics[width=.99\linewidth]{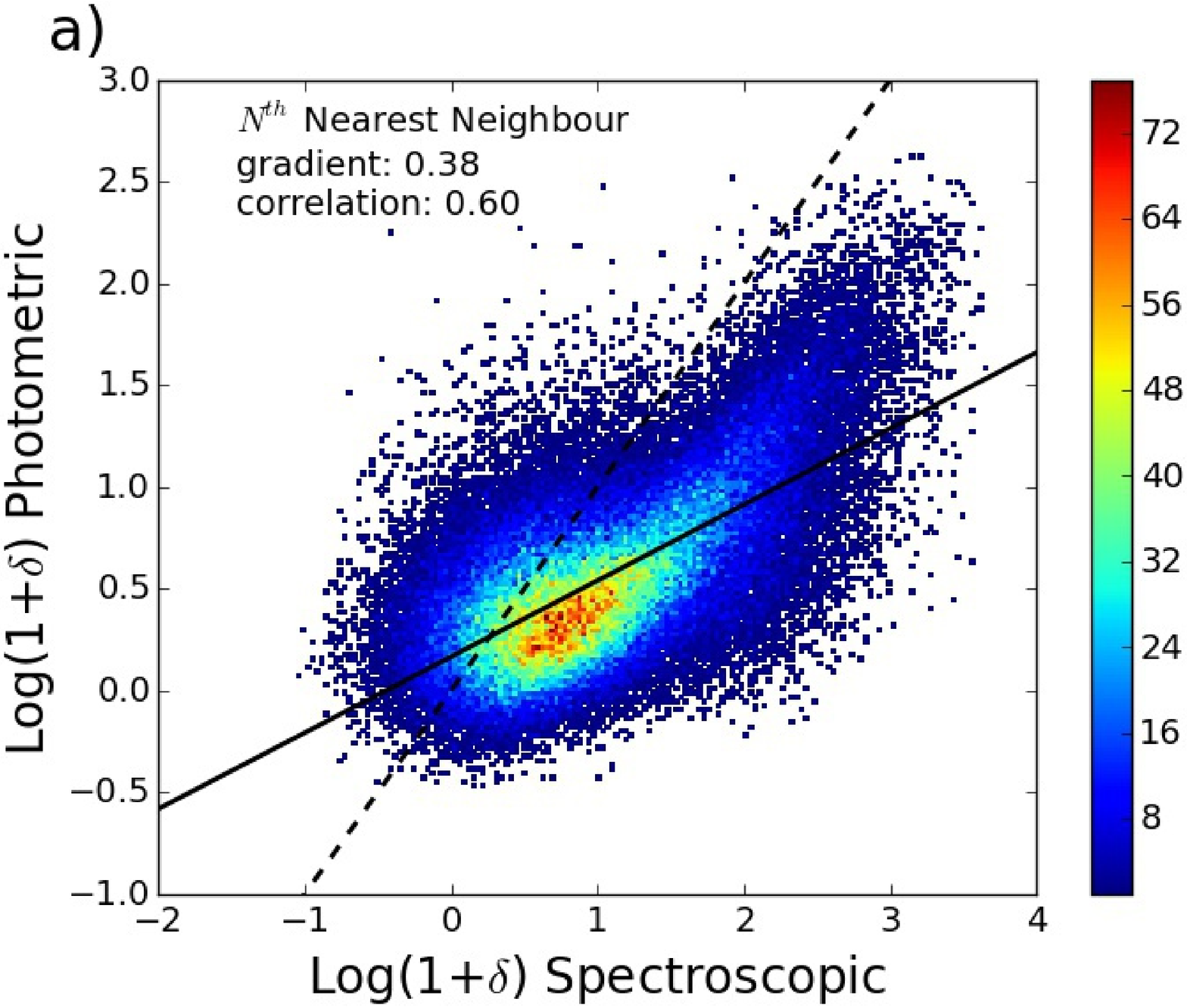}
\end{minipage}
\begin{minipage}[t]{0.49\linewidth}
  \centering  
  \includegraphics[width=.99\linewidth]{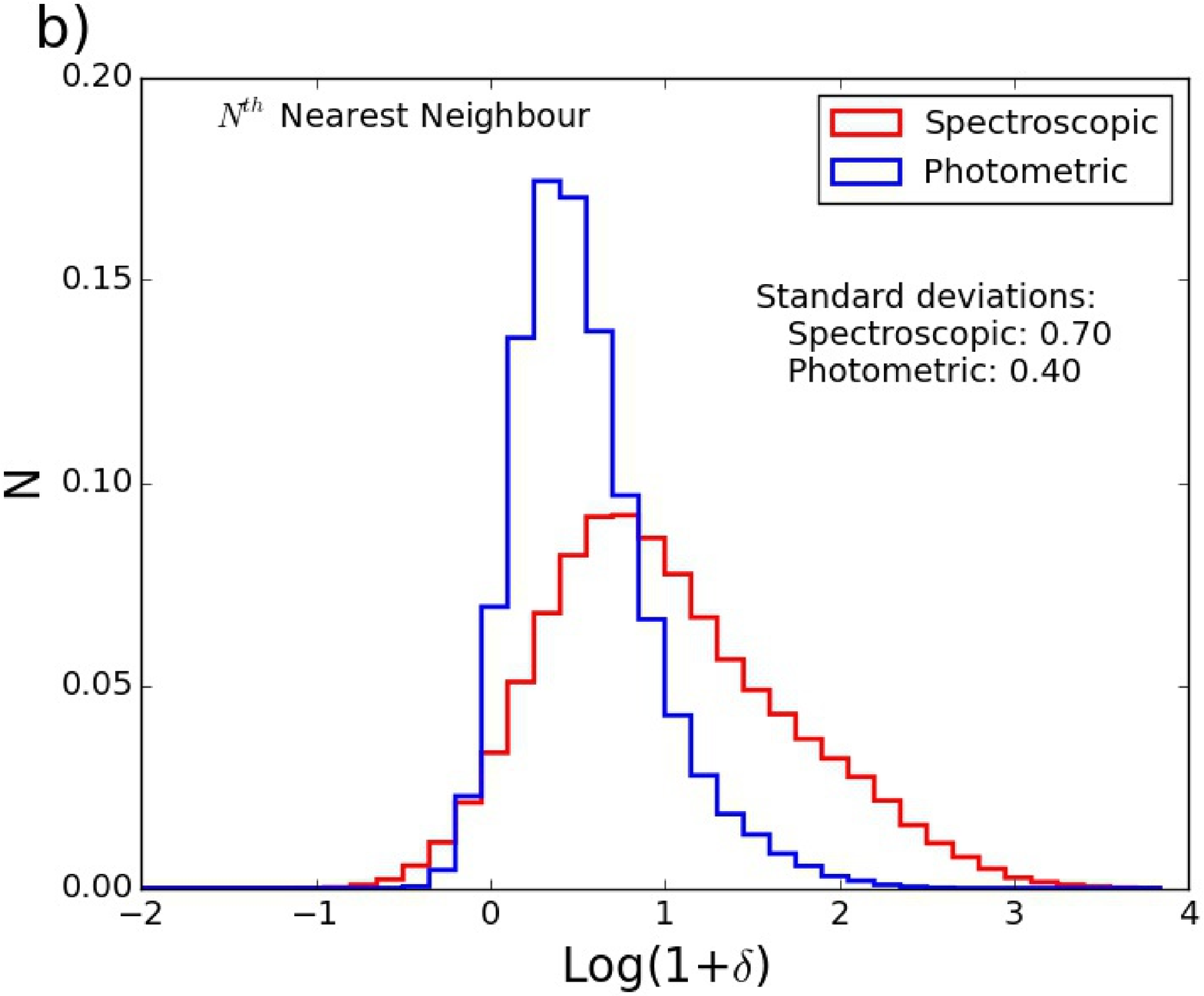}
\end{minipage}
\begin{minipage}[t]{0.49\linewidth}
  \centering  
  \includegraphics[width=.99\linewidth]{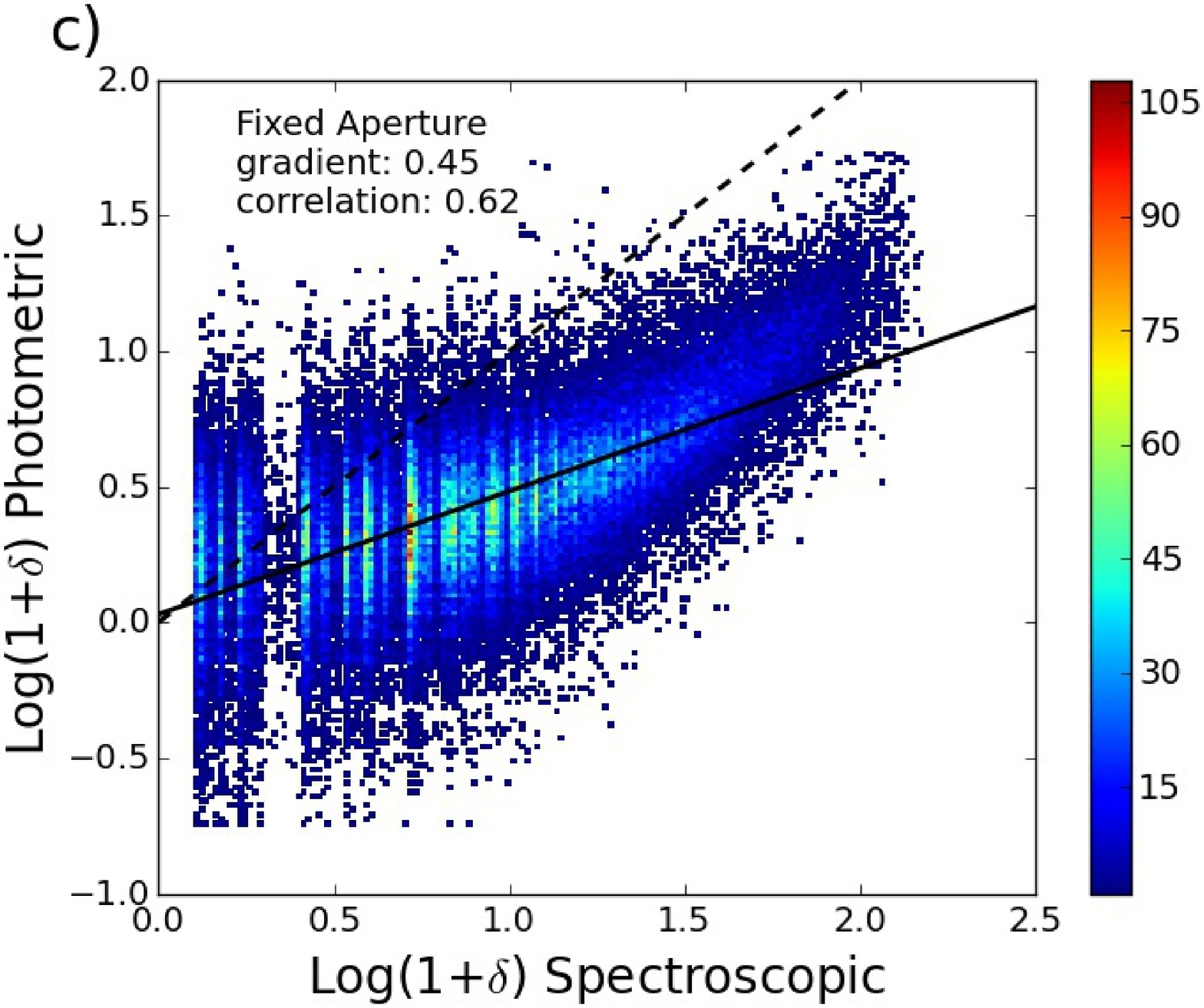}
\end{minipage}
\begin{minipage}[t]{0.49\linewidth}
  \centering  
  \includegraphics[width=.99\linewidth]{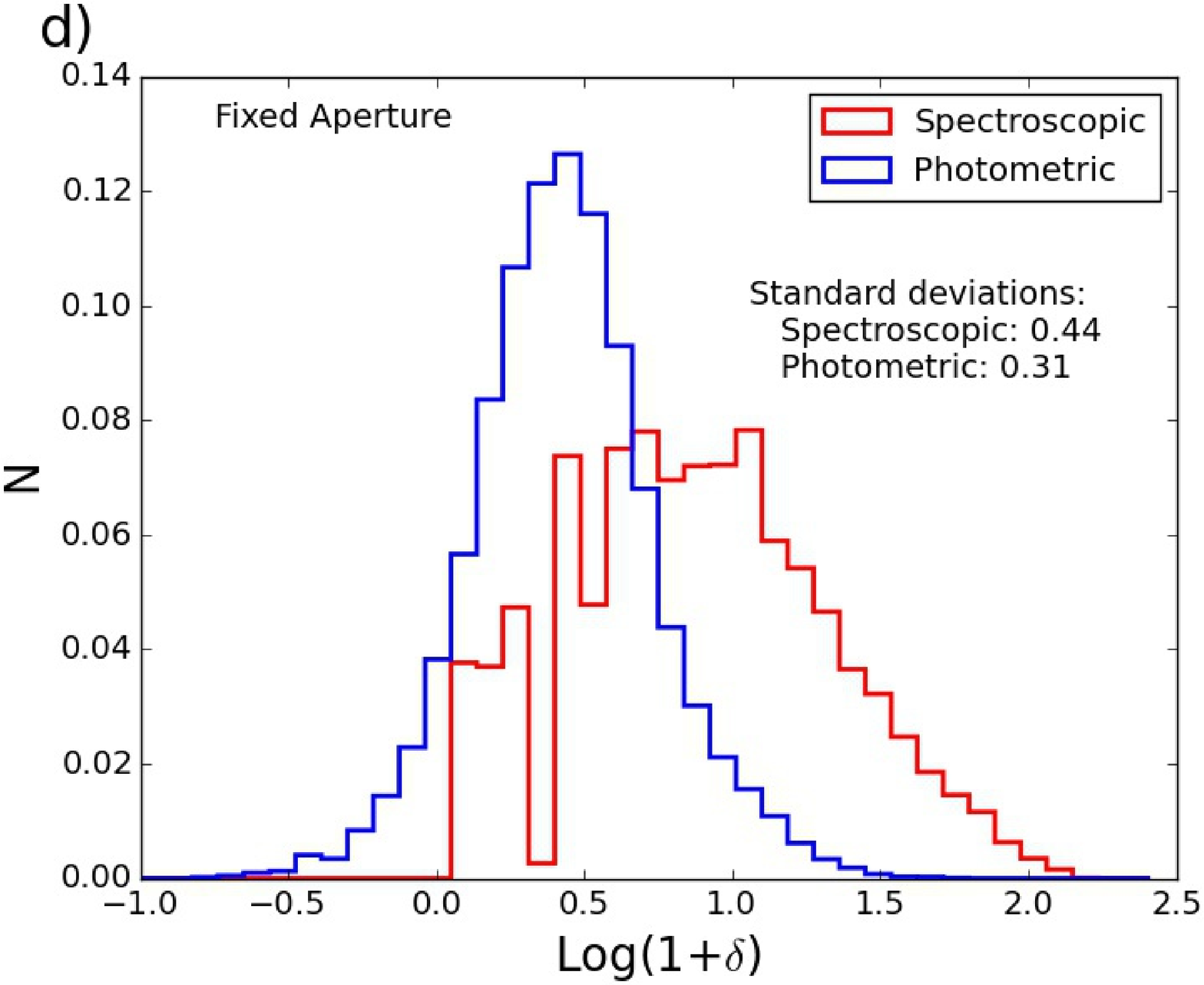}
\end{minipage}
\caption{Density plots of the spectroscopic environment measurements versus the photometric environment measurements for the $N$\textsuperscript{th} nearest neighbour method (a) and the fixed aperture method (c). The corresponding environment distributions for the spectroscopic (red) and photometric (blue) measurements are shown in plots (b) and (d) for the $N$\textsuperscript{th} nearest neighbour method and the fixed aperture method respectively. The aperture parameters used for the spectroscopic measurements are $N=4$ and $dv=1000\;$km/s for the $N$\textsuperscript{th} nearest neighbour method and $r=1.8\;$Mpc and $dv=1000\;$km/s for the fixed aperture method. The aperture parameters for the photometric measurements are $N=7$ and $dv=7000\;$km/s for the $N$\textsuperscript{th} nearest neighbour method and $r=1.9\;$Mpc and $dv=6000\;$km/s for the fixed aperture method. The linear best fit lines for the measurements are shown in black and 1-to-1 lines are dashed. The correlations quoted are Spearman rank correlation coefficients. }  
\label{fig:spec_vs_photo_envs}
\end{figure*}

\begin{figure*}
\centering
\begin{minipage}[t]{0.80\linewidth}
\includegraphics[width=\linewidth]{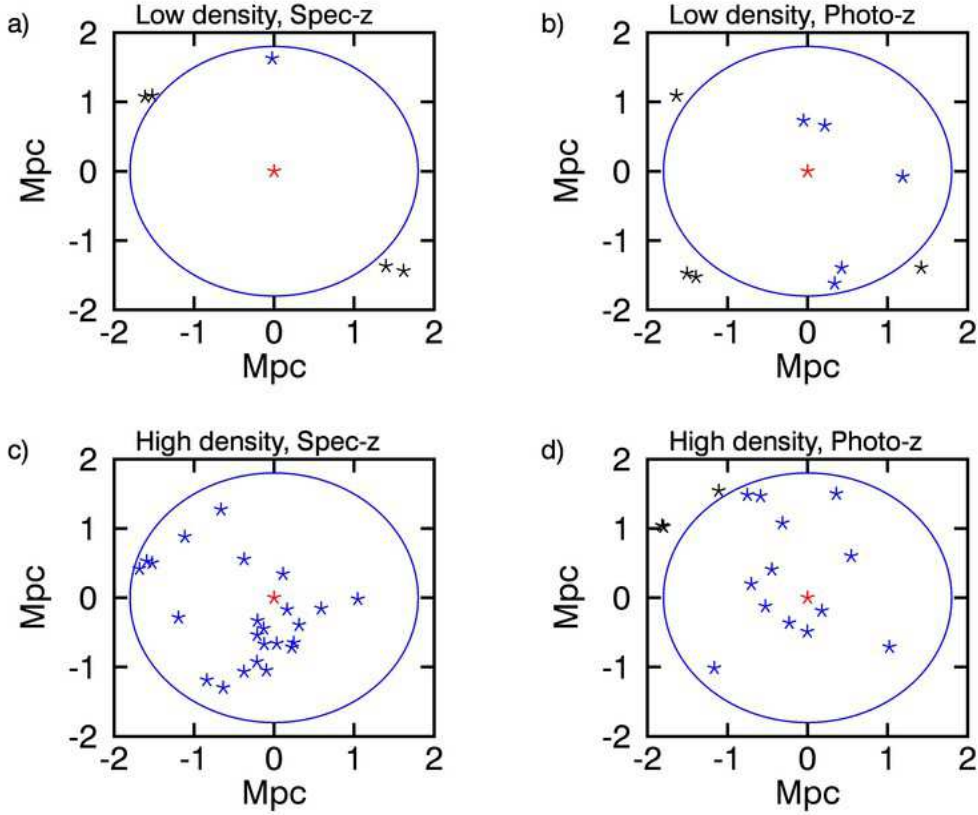}
\end{minipage}
\caption{Environment of two target galaxies: one in a low density environment (top row) and one in a high density environment (bottom row) measured with spectroscopy (left column) and photometry (right column). The targets are marked in red. The blue circles mark a comoving distance of $1.8\;$Mpc from the target. The blue symbols mark galaxies that are inside the circle. The black symbols mark galaxies that are outside the circle. Only those galaxies that are in the density defining population and are also constrained by a velocity cut of $1000\;$km/s are show. The number of galaxies found within $1.8\;$Mpc of the target for plots a) to d) are: $1$, $5$, $24$ and $13$.}  
\label{fig:density_example}
\end{figure*}

\begin{figure*}
\begin{minipage}[t]{0.99\linewidth}
\includegraphics[width=0.49\linewidth]{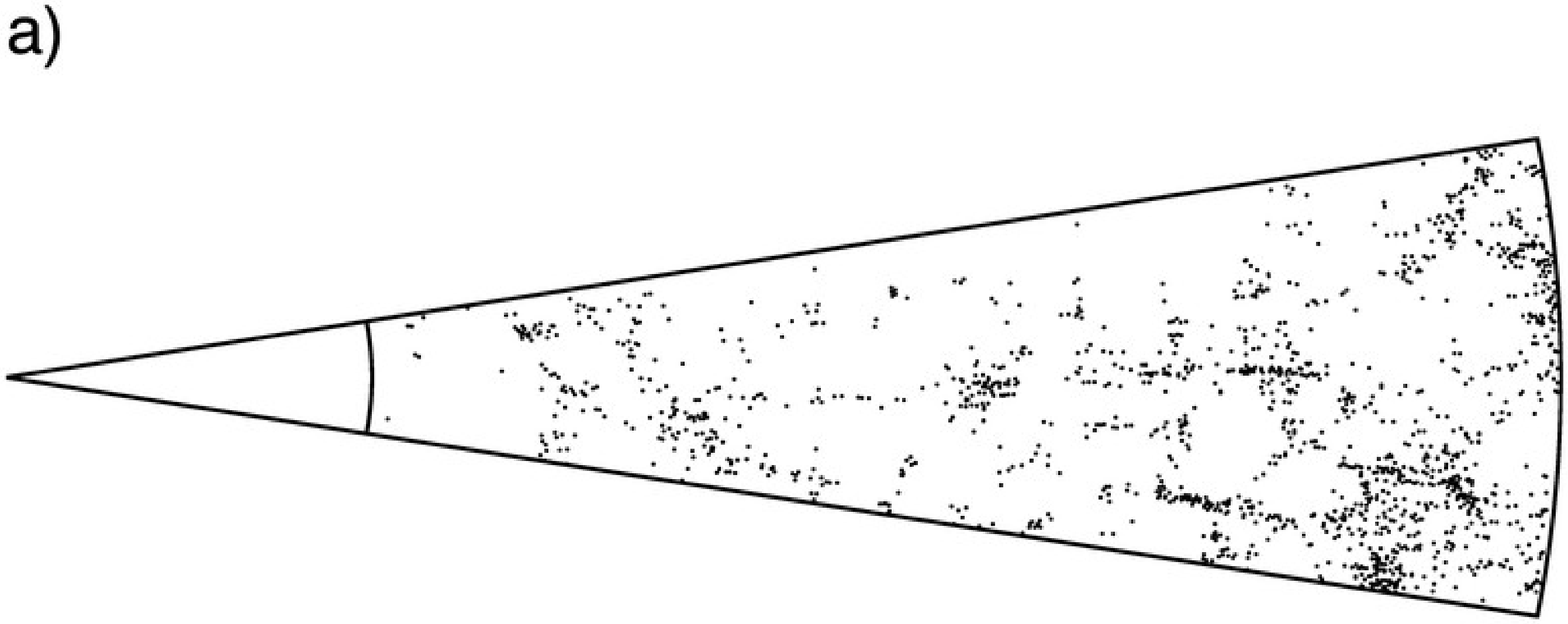}
\includegraphics[width=0.49\linewidth]{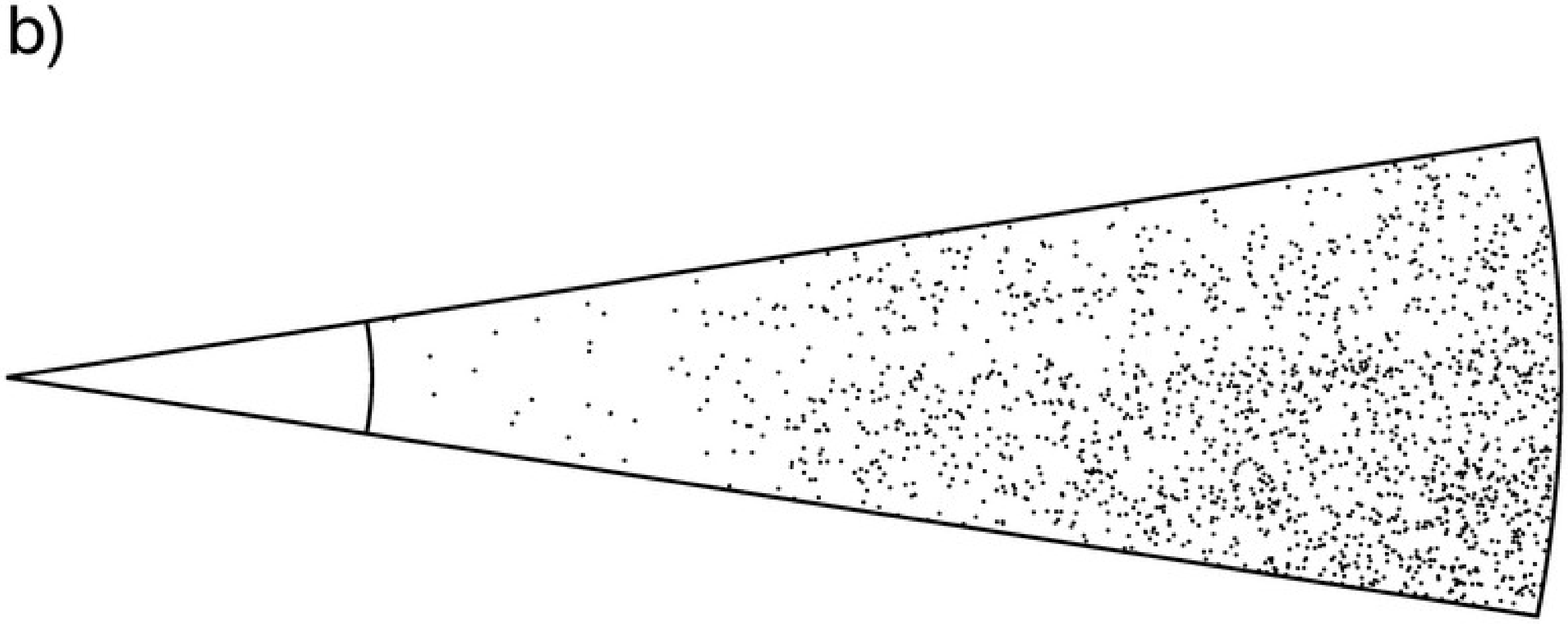}
\end{minipage}
\caption{Shows the redshift and declination of galaxies within a wedge of volume using spectroscopic measurements (a) and photometric measurements (b). The redshift range is: $0.02-0.085$ and extends along the radial direction of the cones. The declination range is $20-40$ degrees. The right ascension is restricted to the range $215-220$ degrees.}  
\label{fig:wedge}
\end{figure*}

\begin{figure*}
\begin{minipage}[t]{0.33\linewidth}
   \centering 
   \includegraphics[width=.99\linewidth]{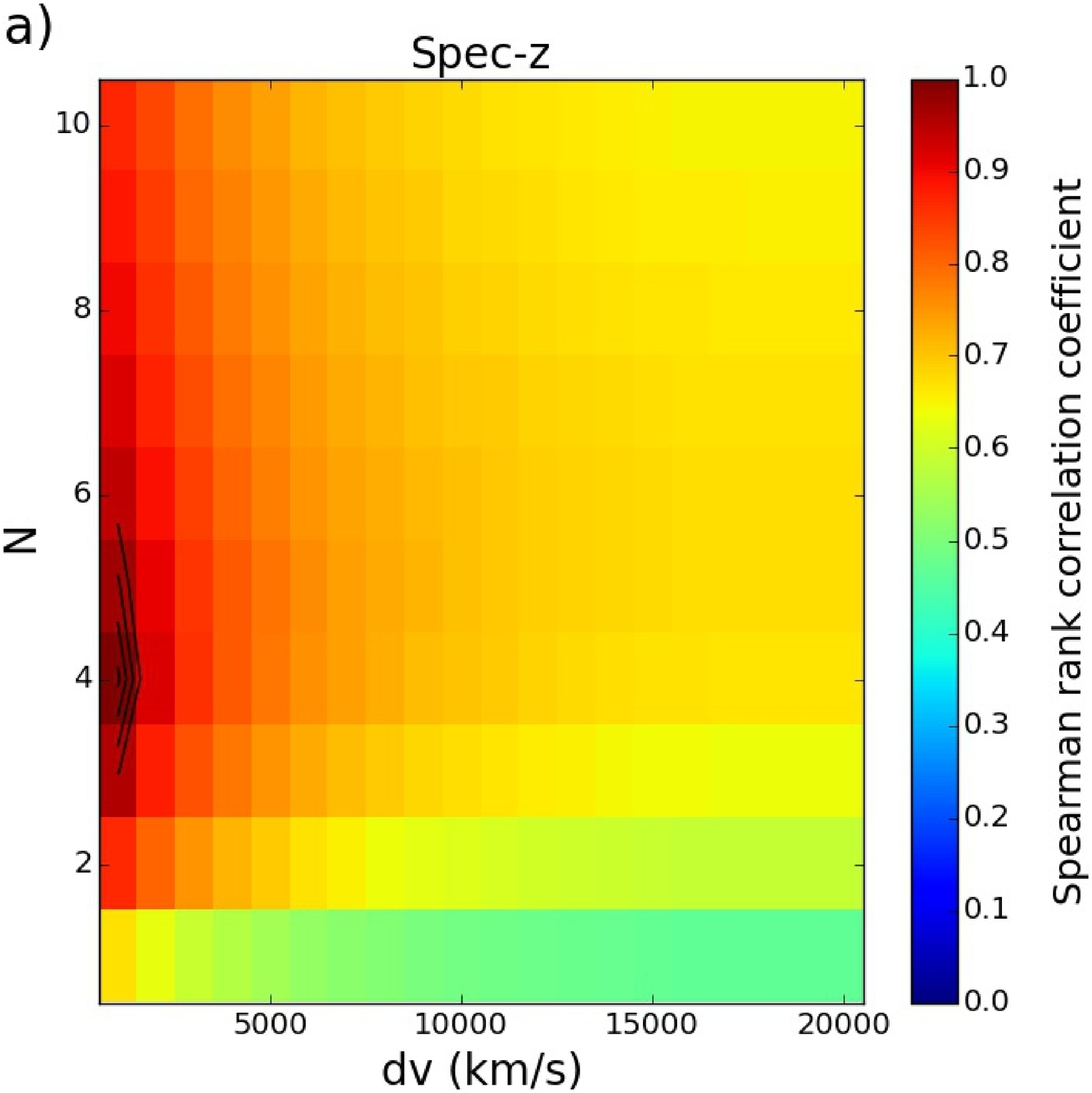}
\end{minipage}
\begin{minipage}[t]{0.33\linewidth}
  \centering  
  \includegraphics[width=.99\linewidth]{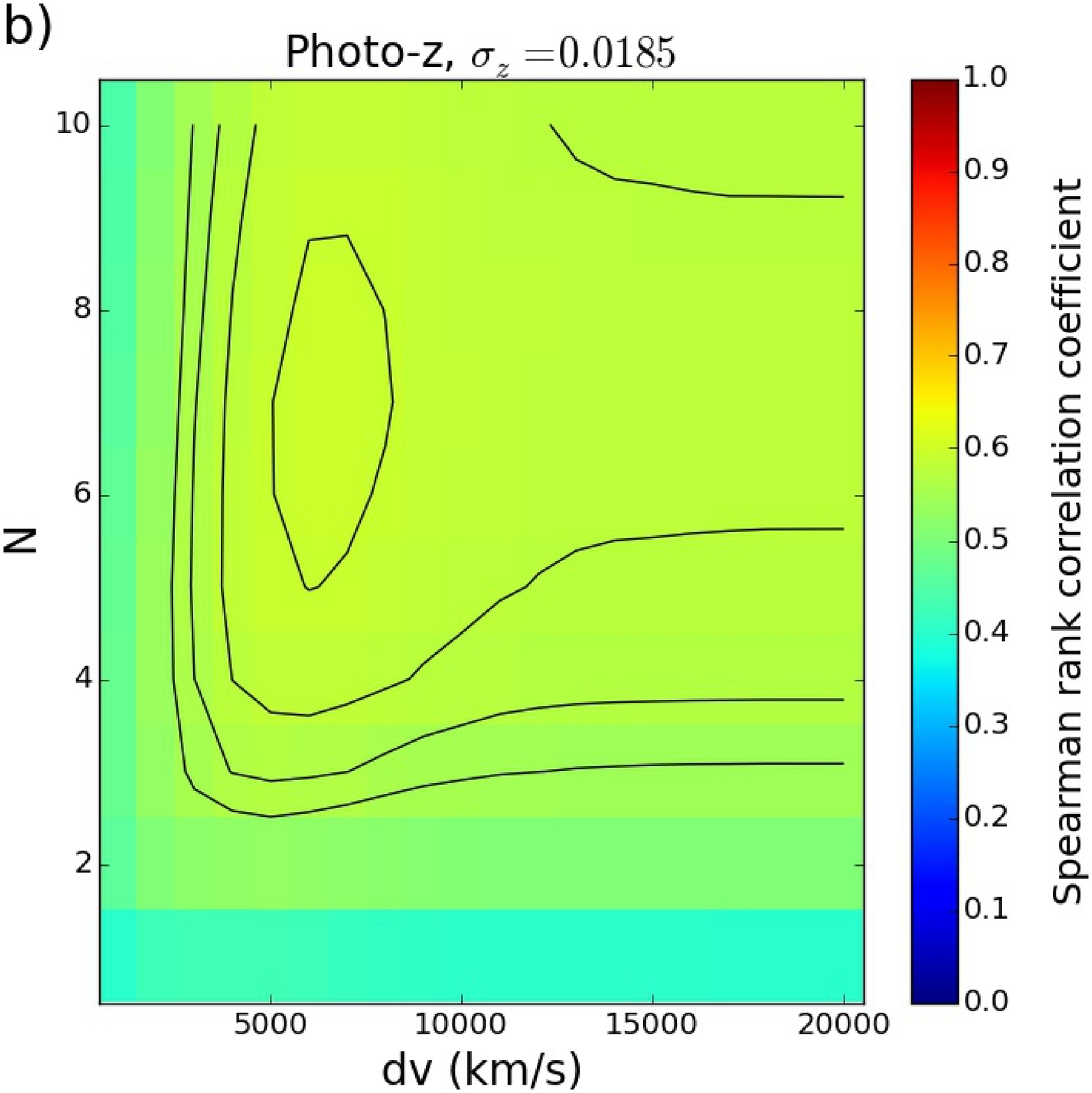}
\end{minipage}
\begin{minipage}[t]{0.33\linewidth}
  \centering  
  \includegraphics[width=.99\linewidth]{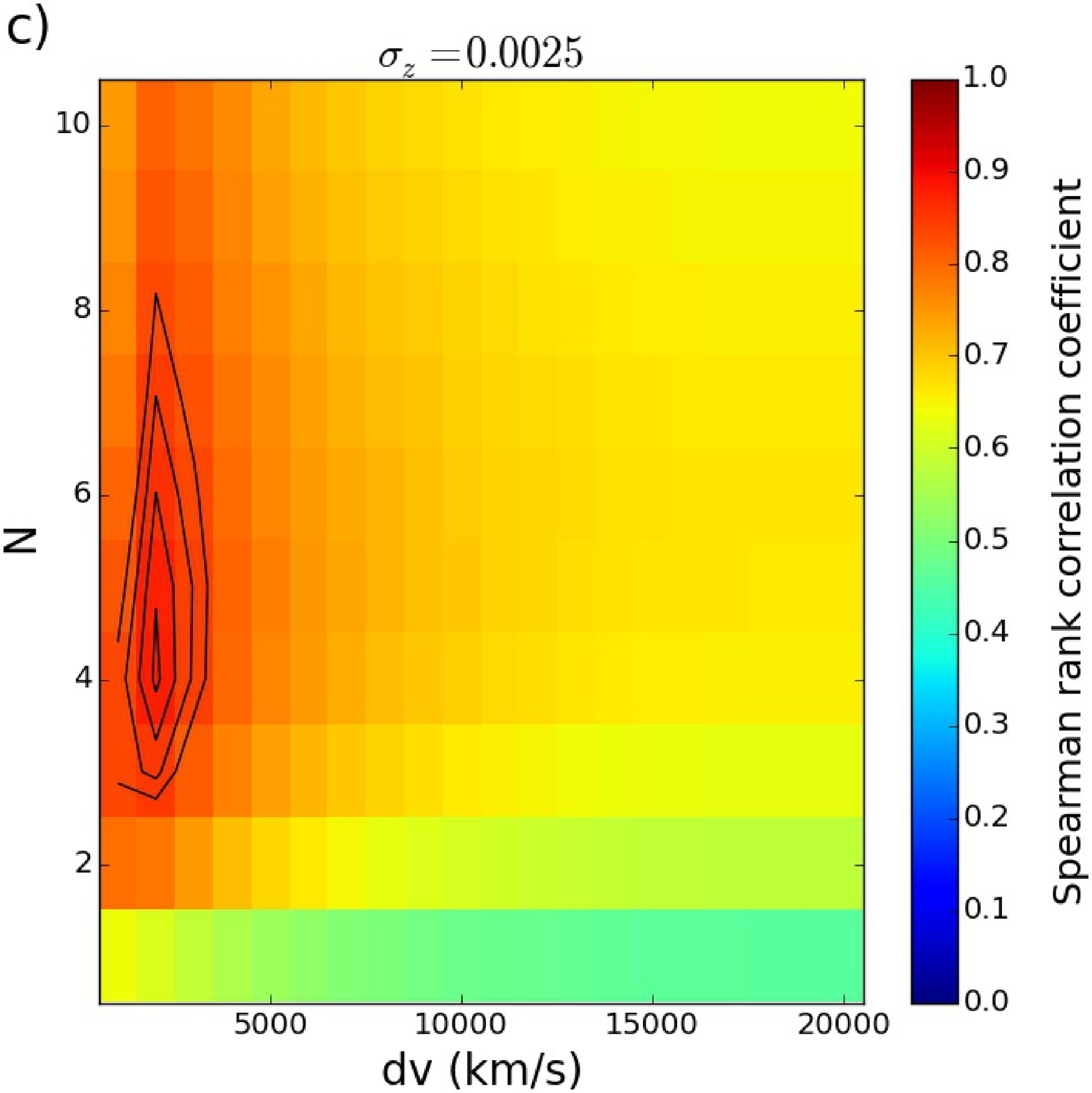}
\end{minipage}
\begin{minipage}[t]{0.33\linewidth}
   \centering 
   \includegraphics[width=.99\linewidth]{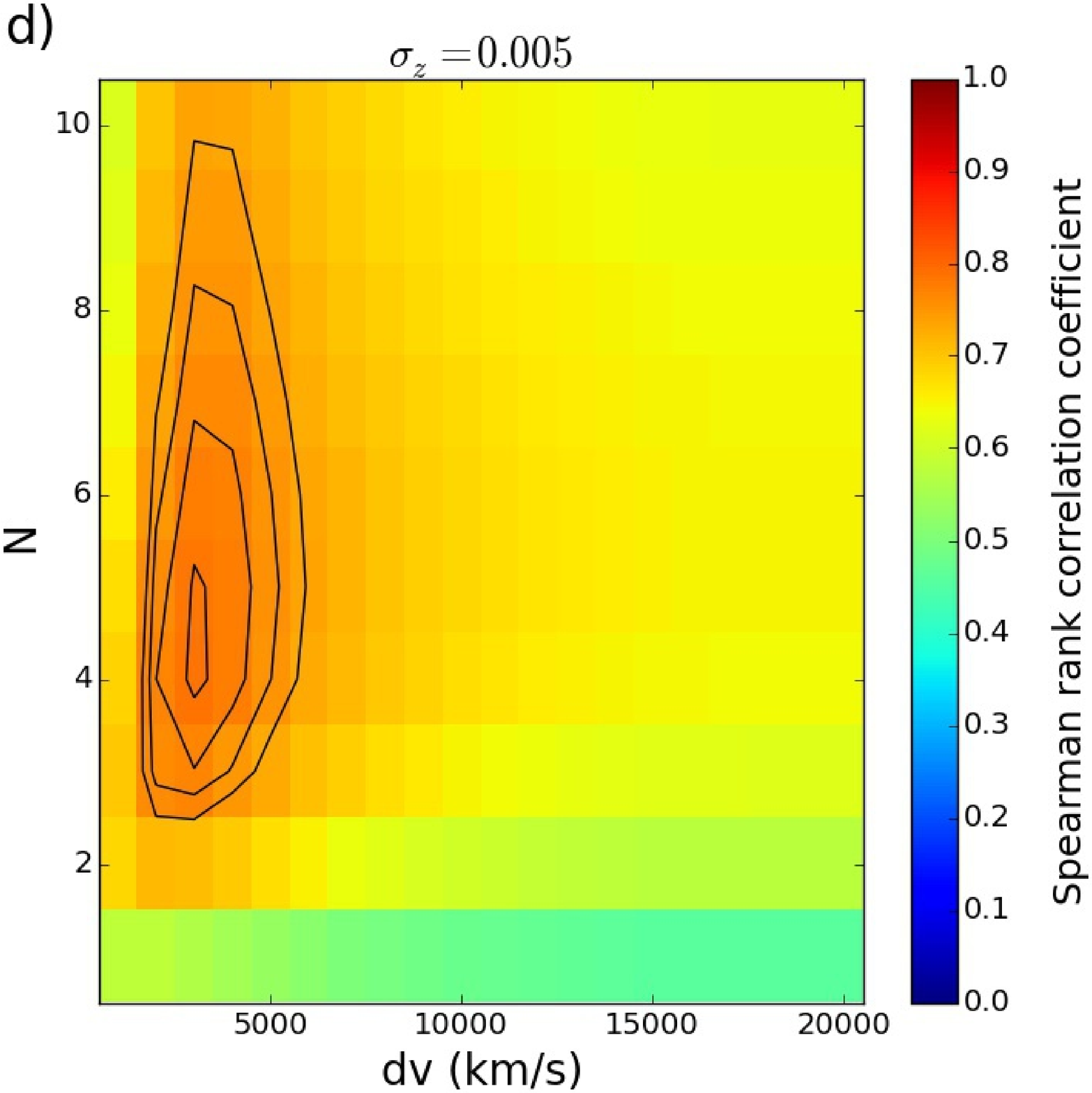}
\end{minipage}
\begin{minipage}[t]{0.33\linewidth}
  \centering  
  \includegraphics[width=.99\linewidth]{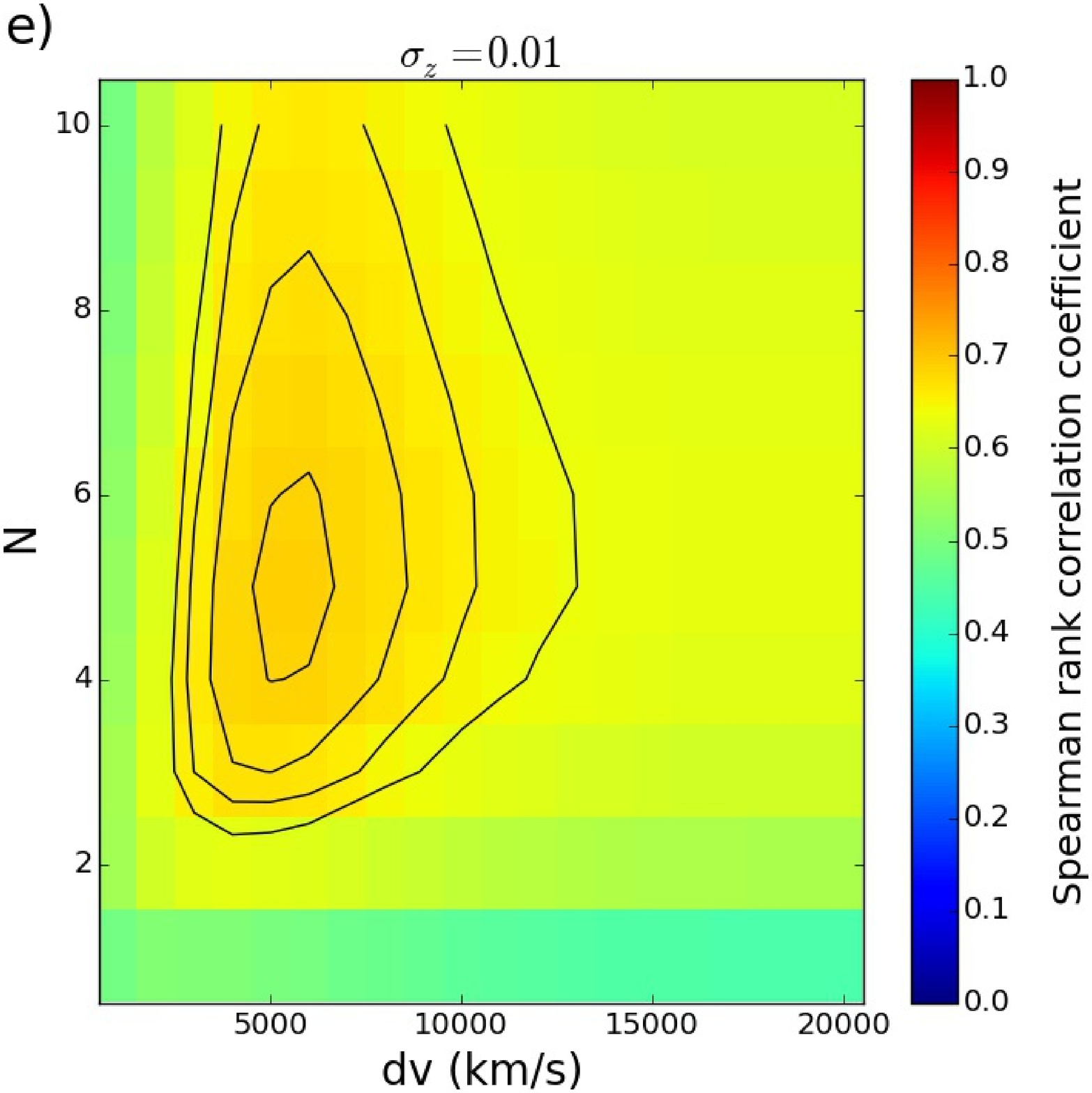}
\end{minipage}
\begin{minipage}[t]{0.33\linewidth}
  \centering  
  \includegraphics[width=.99\linewidth]{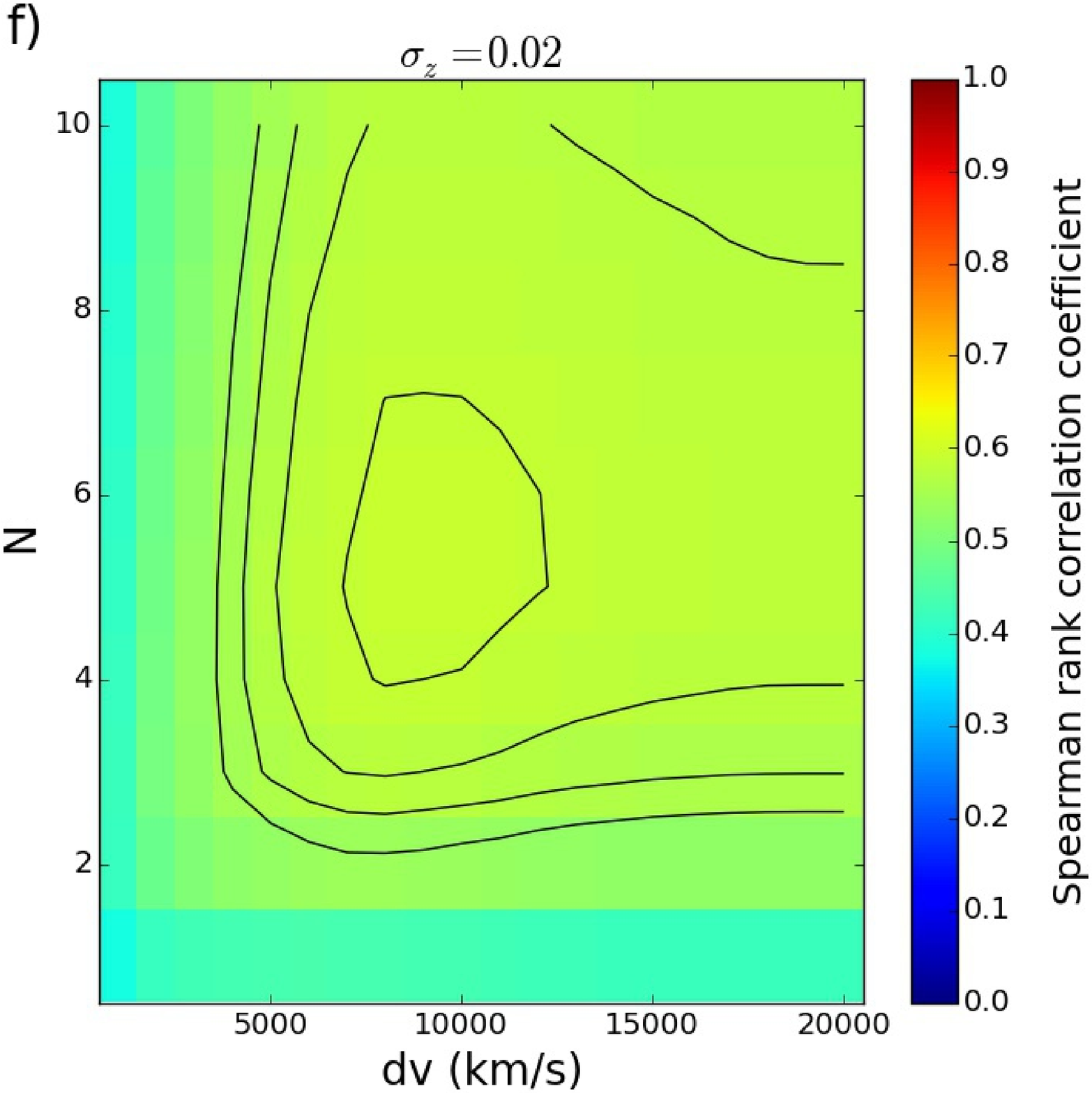}
\end{minipage}
\begin{minipage}[t]{0.33\linewidth}
   \centering 
   \includegraphics[width=.99\linewidth]{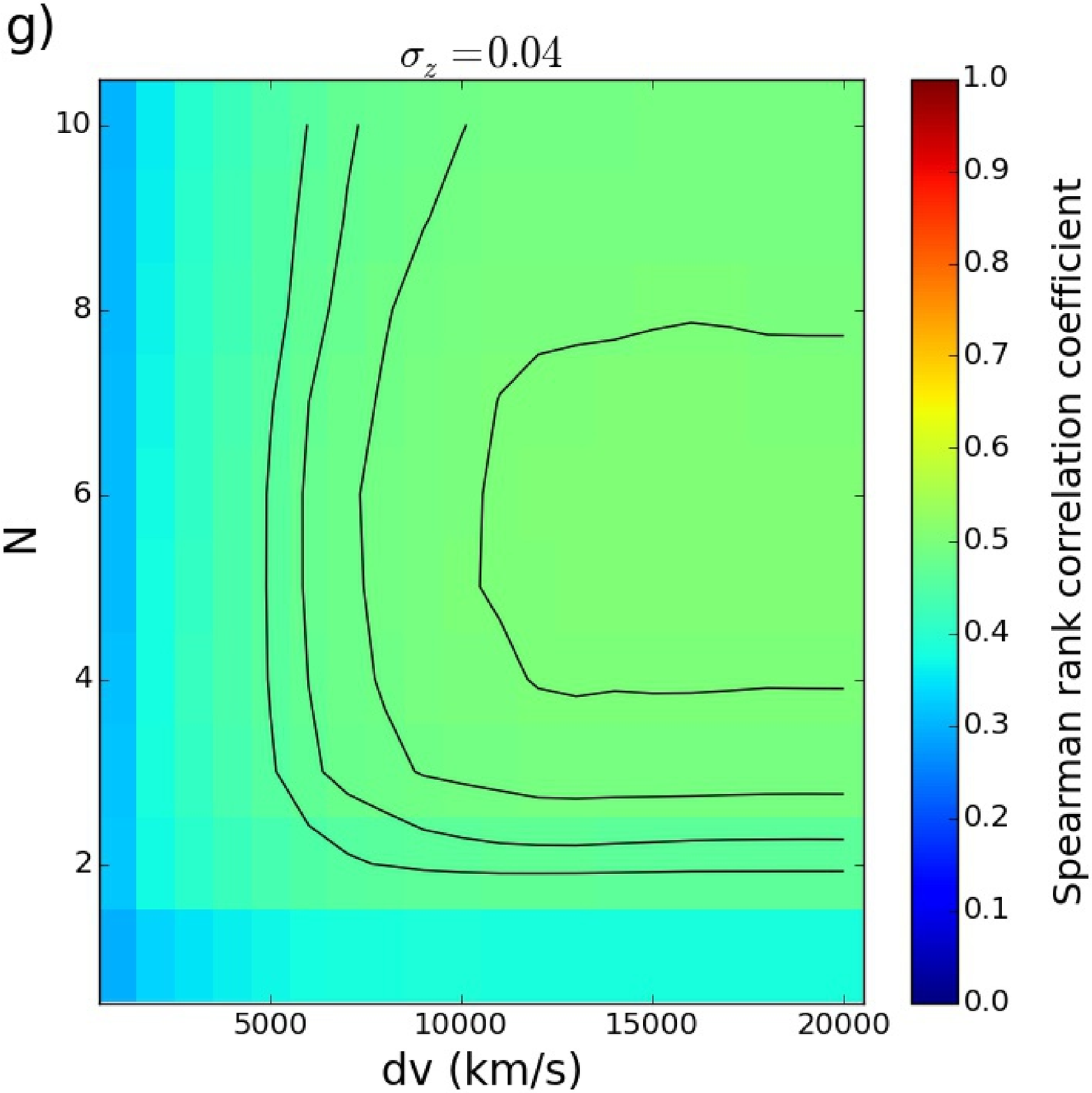}
\end{minipage}
\begin{minipage}[t]{0.33\linewidth}
  \centering  
  \includegraphics[width=.99\linewidth]{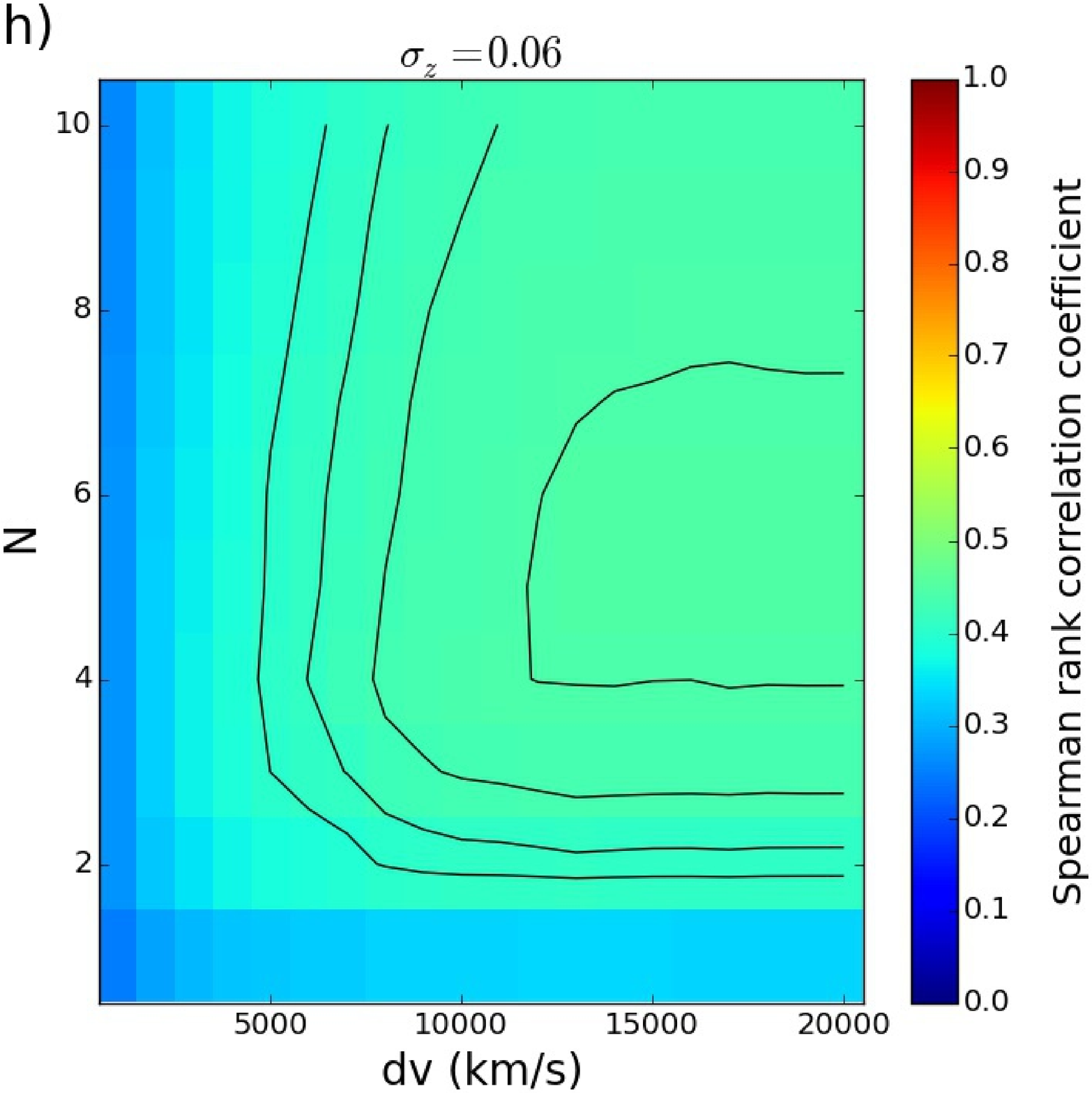}
\end{minipage}
\begin{minipage}[t]{0.33\linewidth}
  \centering  
  \includegraphics[width=.99\linewidth]{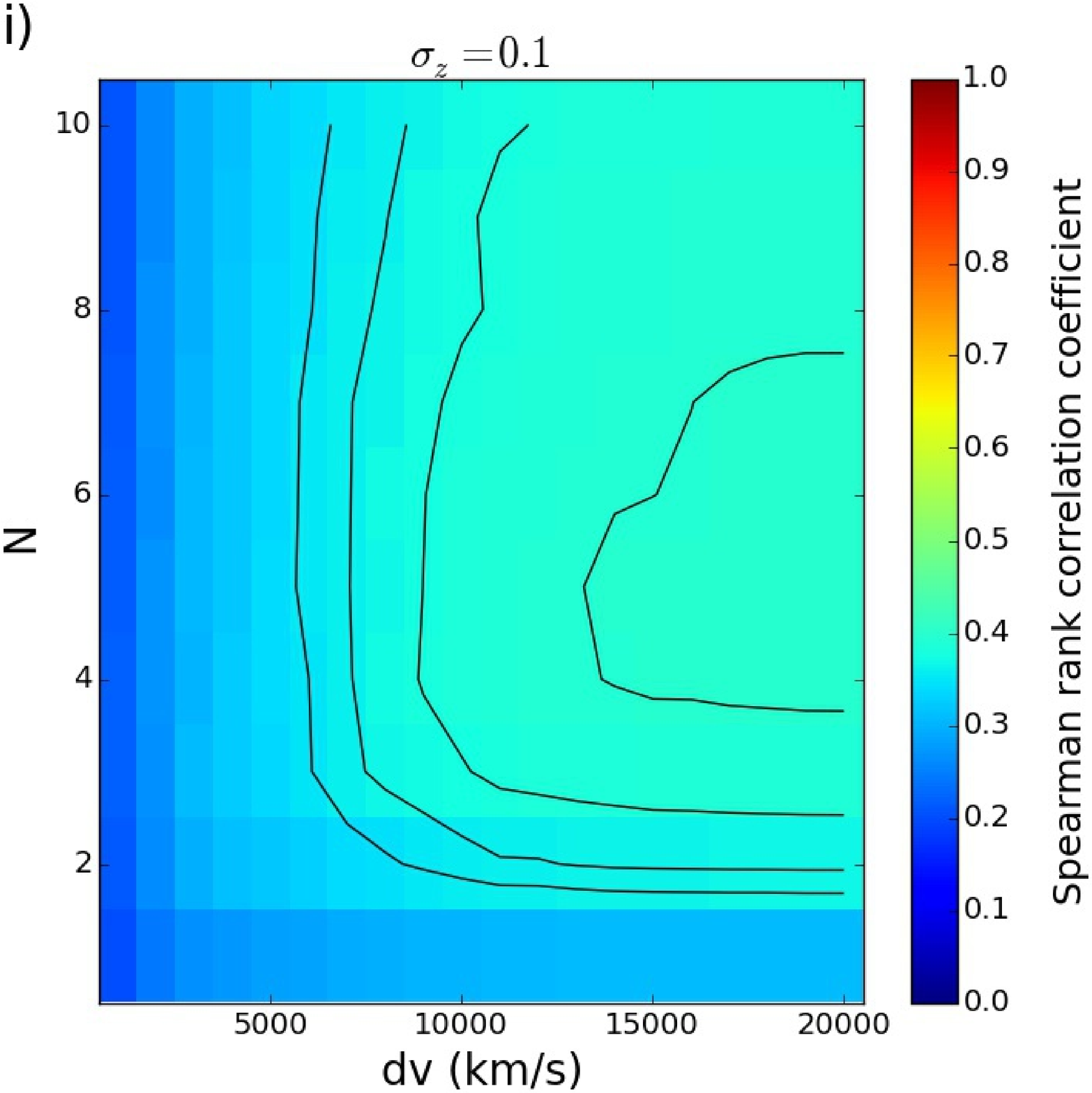}
\end{minipage}
\caption{Shows the Spearman rank correlation coefficients between the $N$\textsuperscript{th} nearest neighbour environments and the spectroscopic benchmark environments (where $N=4$ and $dv=1000\;$km/s) for nine catalogues of galaxies as a function of the aperture parameters: $N$ and the velocity cut ($dv$). a) Shows the correlations for the spectroscopic redshift catalogue. b) Shows the correlations for the photometric redshift catalogue. c) to i) show the correlations for simulated photometric redshift catalogues with redshift uncertainties of $0.0025$, $0.005$, $0.01$, $0.02$, $0.04$, $0.06$ and $0.1$ respectively. The correlation is represented with a colour in each grid cell. The equally spaced contours in the logarithm of the correlation highlight the location of the peak in the parameter space on each grid.}  
\label{fig:env_nnn_benchmark_correlation}
\end{figure*}

\begin{figure*}
\begin{minipage}[t]{0.33\linewidth}
   \centering 
   \includegraphics[width=.99\linewidth]{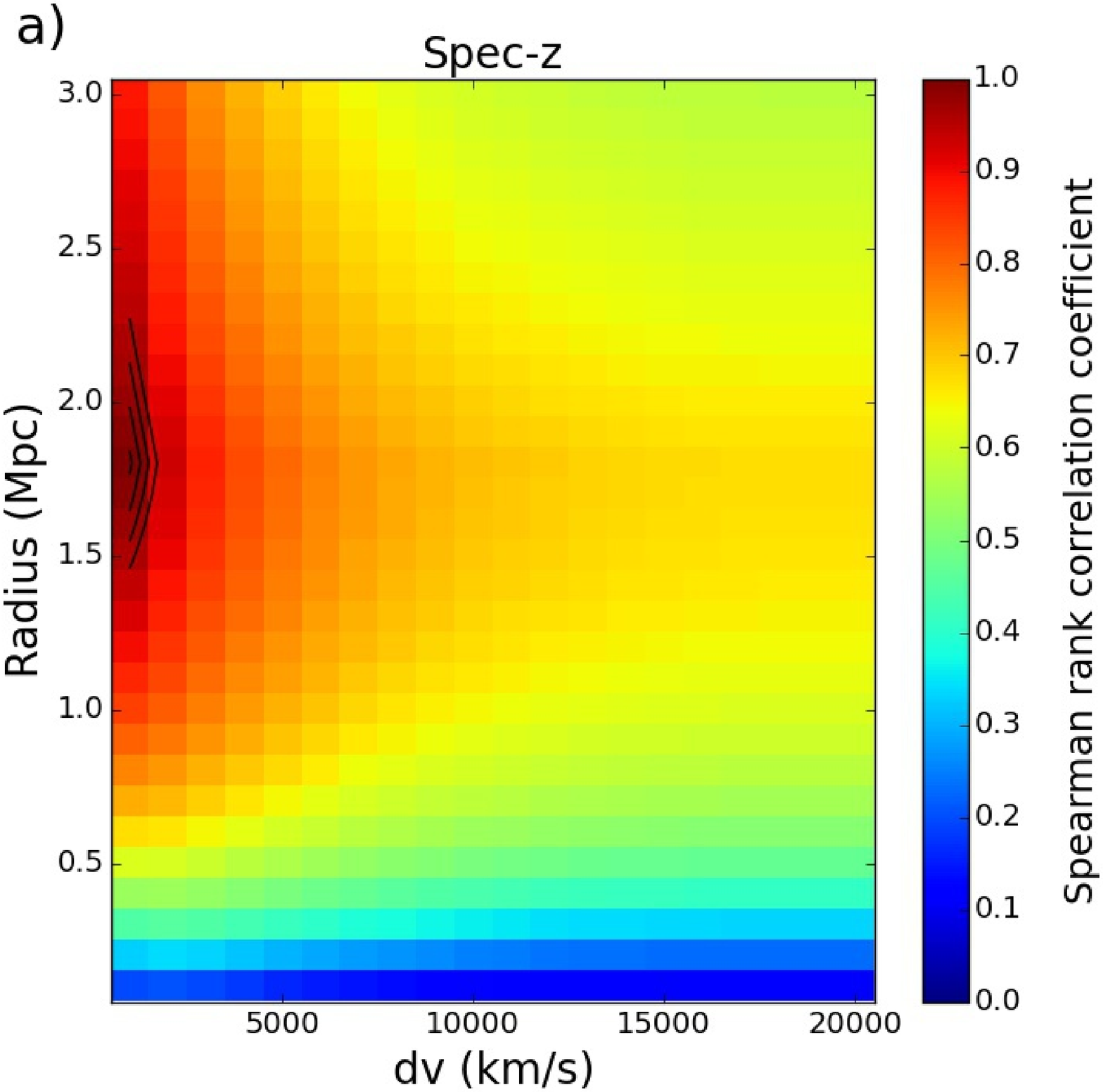}
\end{minipage}
\begin{minipage}[t]{0.33\linewidth}
  \centering  
  \includegraphics[width=.99\linewidth]{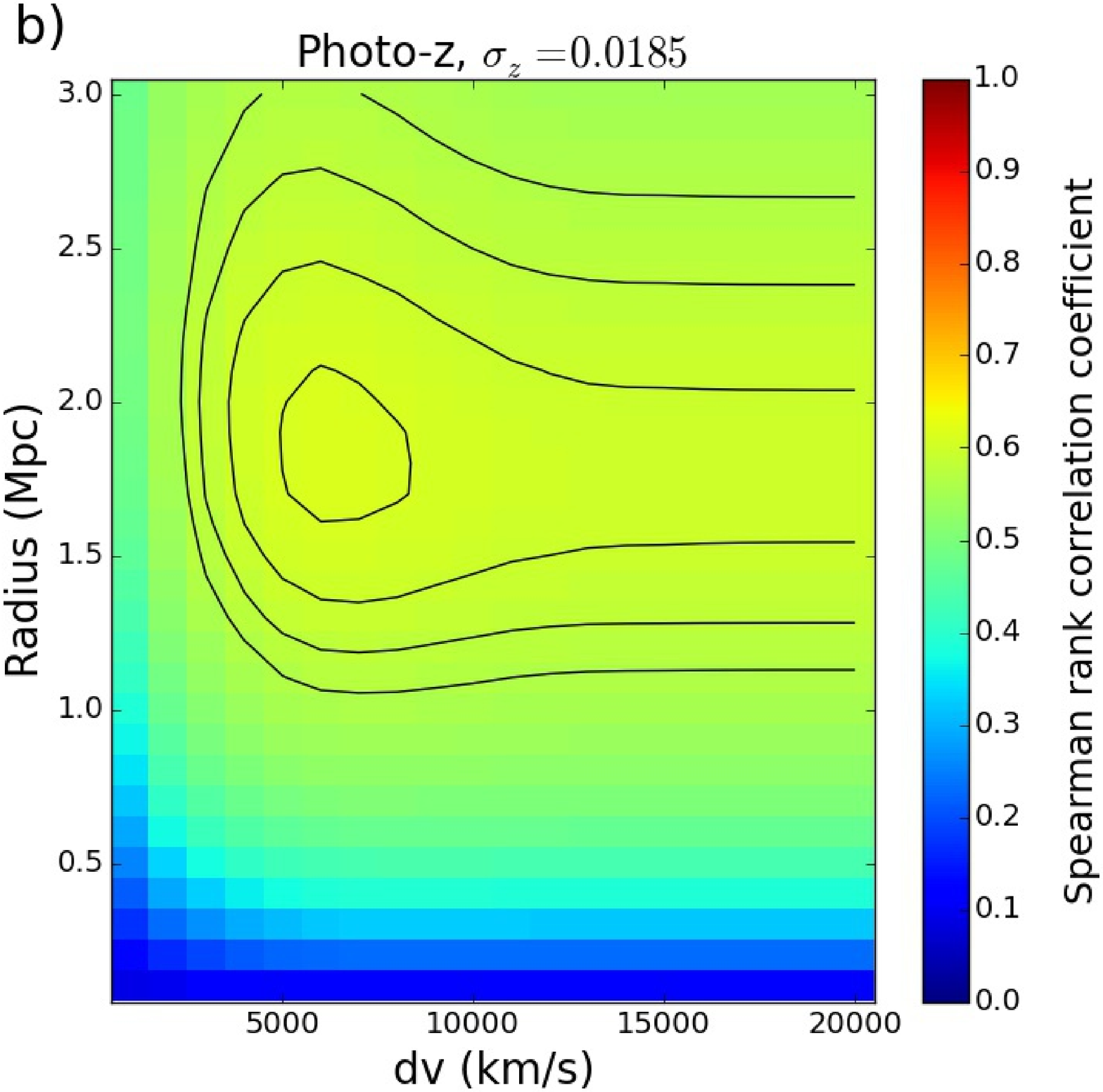}
\end{minipage}
\begin{minipage}[t]{0.33\linewidth}
  \centering  
  \includegraphics[width=.99\linewidth]{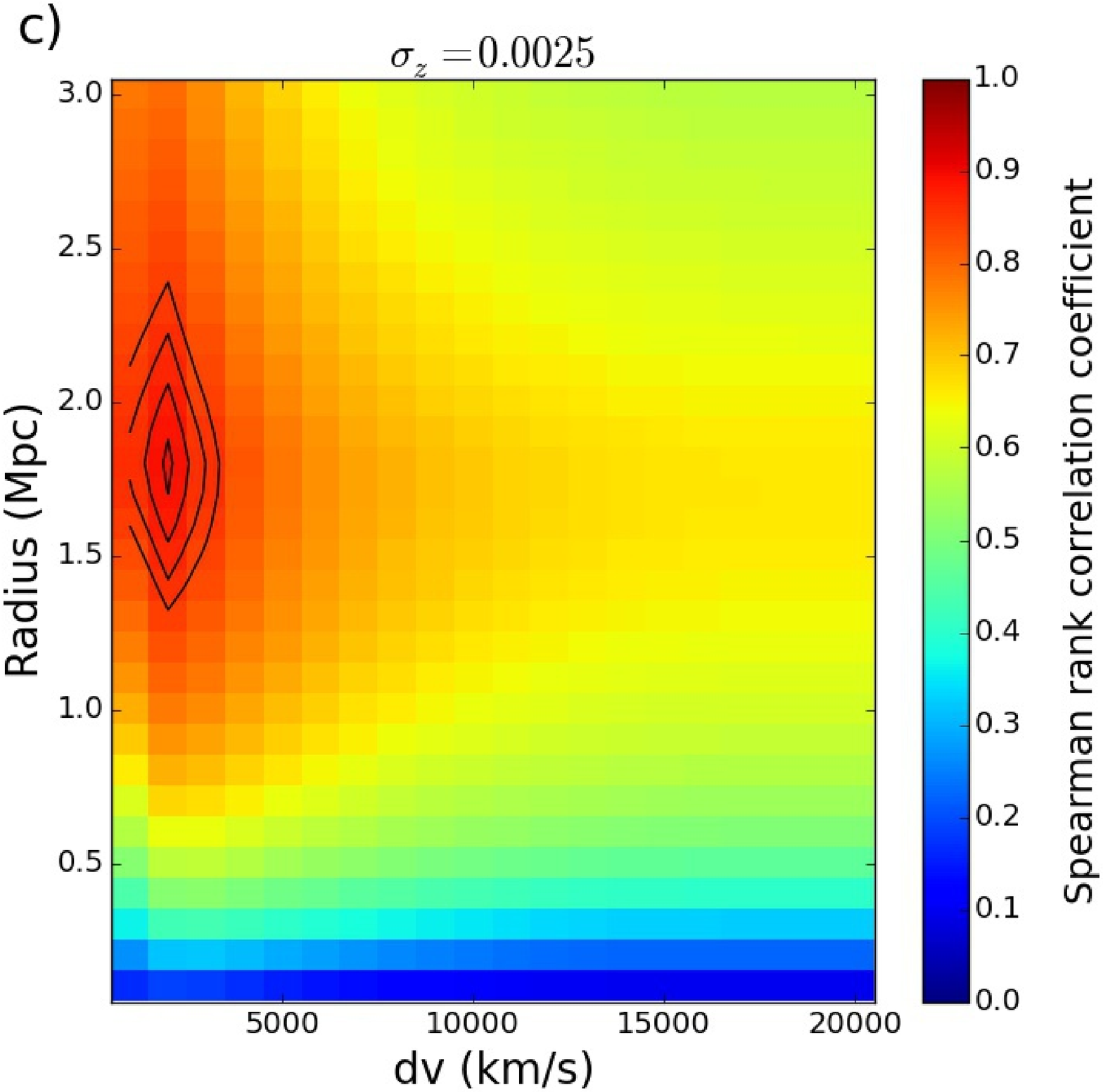}
\end{minipage}
\begin{minipage}[t]{0.33\linewidth}
   \centering 
   \includegraphics[width=.99\linewidth]{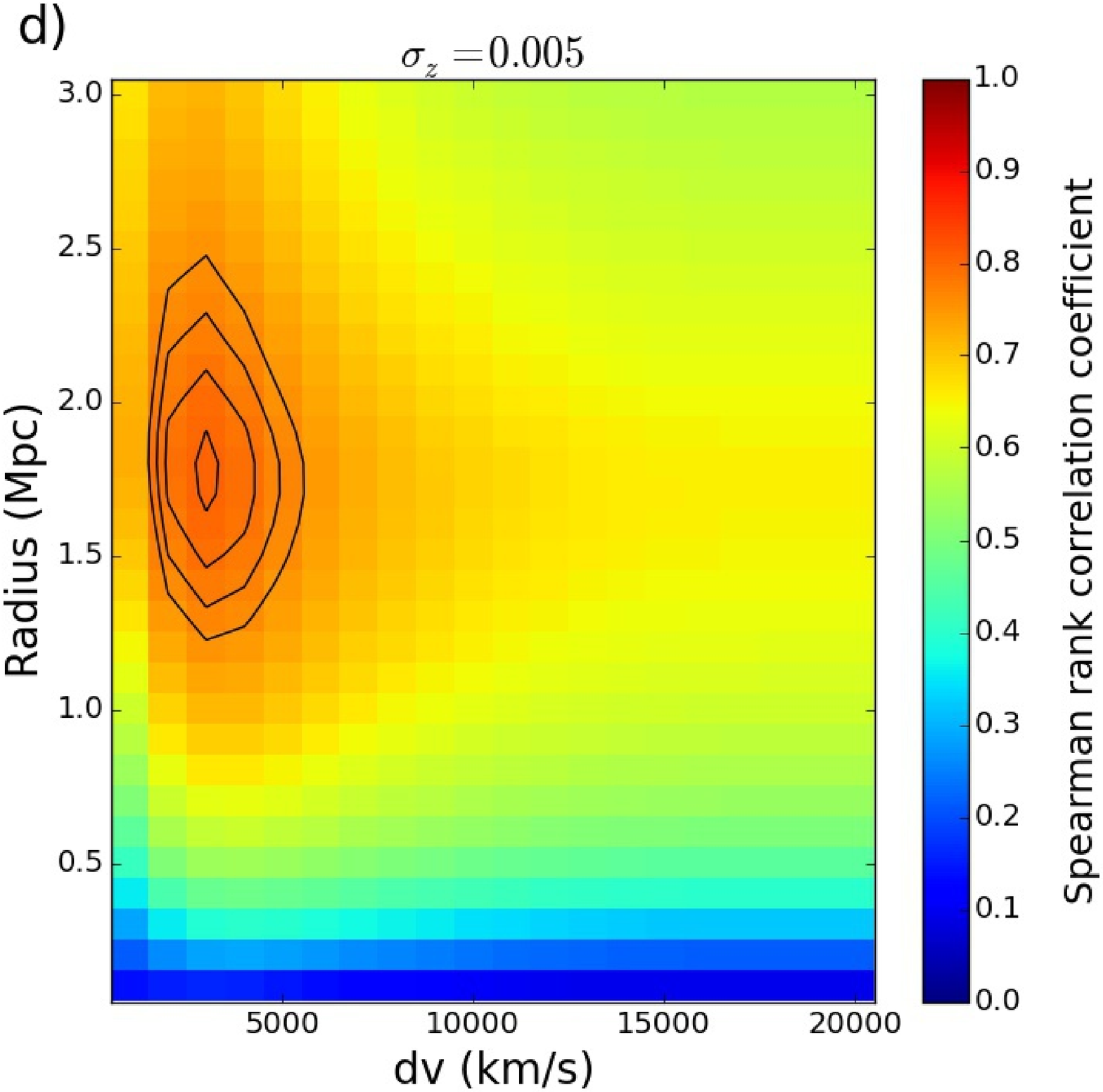}
\end{minipage}
\begin{minipage}[t]{0.33\linewidth}
  \centering  
  \includegraphics[width=.99\linewidth]{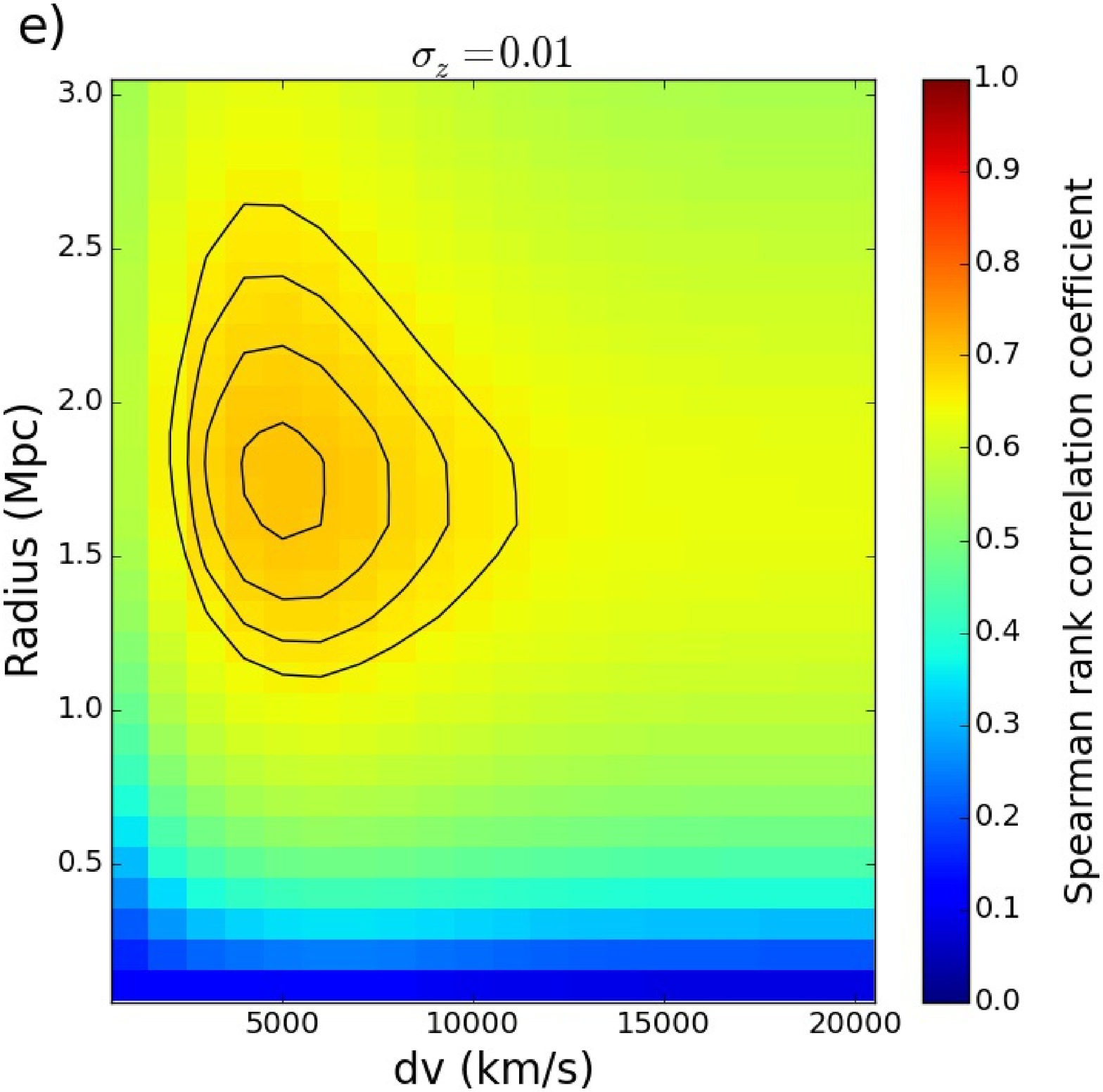}
\end{minipage}
\begin{minipage}[t]{0.33\linewidth}
  \centering  
  \includegraphics[width=.99\linewidth]{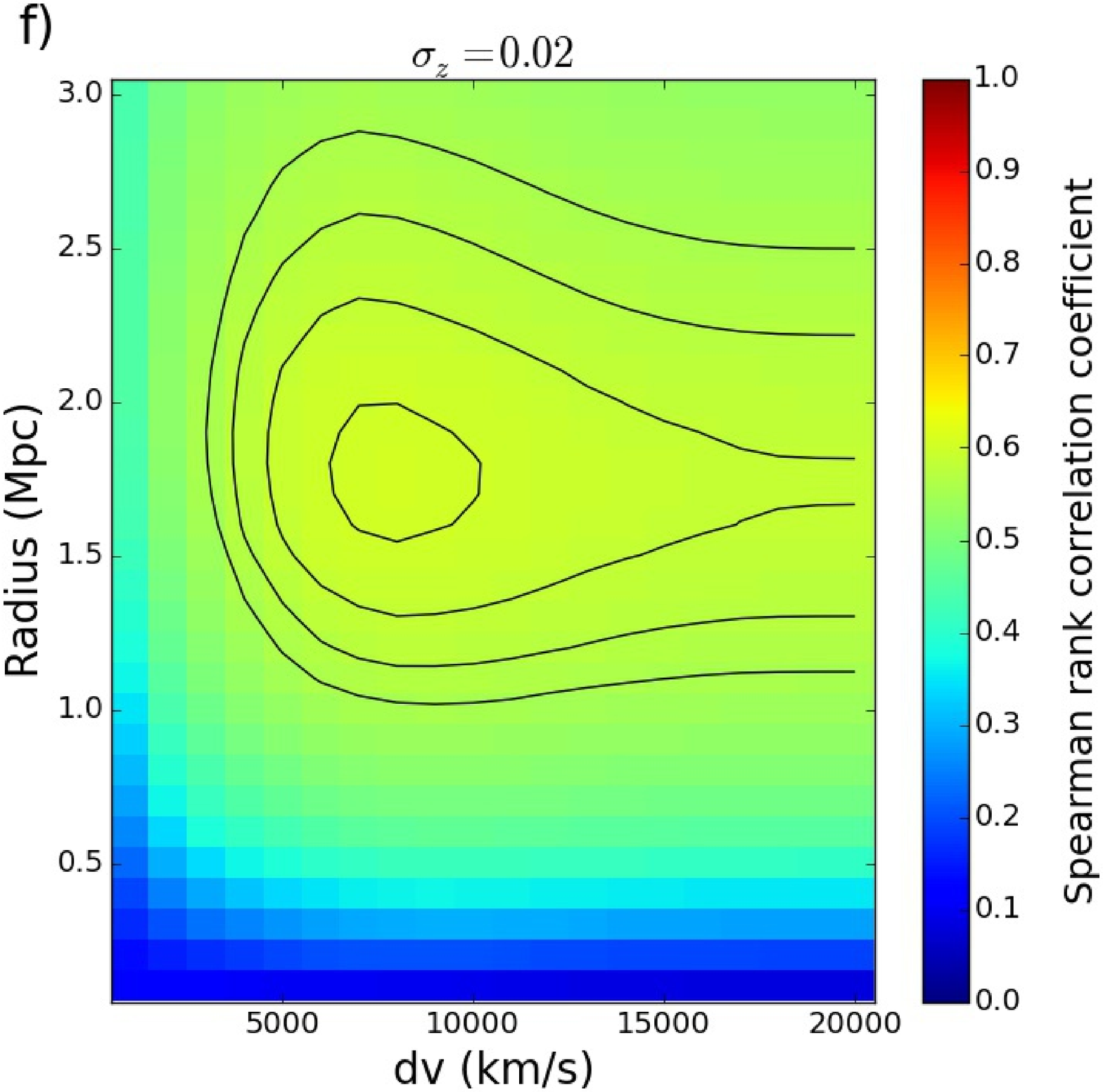}
\end{minipage}
\begin{minipage}[t]{0.33\linewidth}
   \centering 
   \includegraphics[width=.99\linewidth]{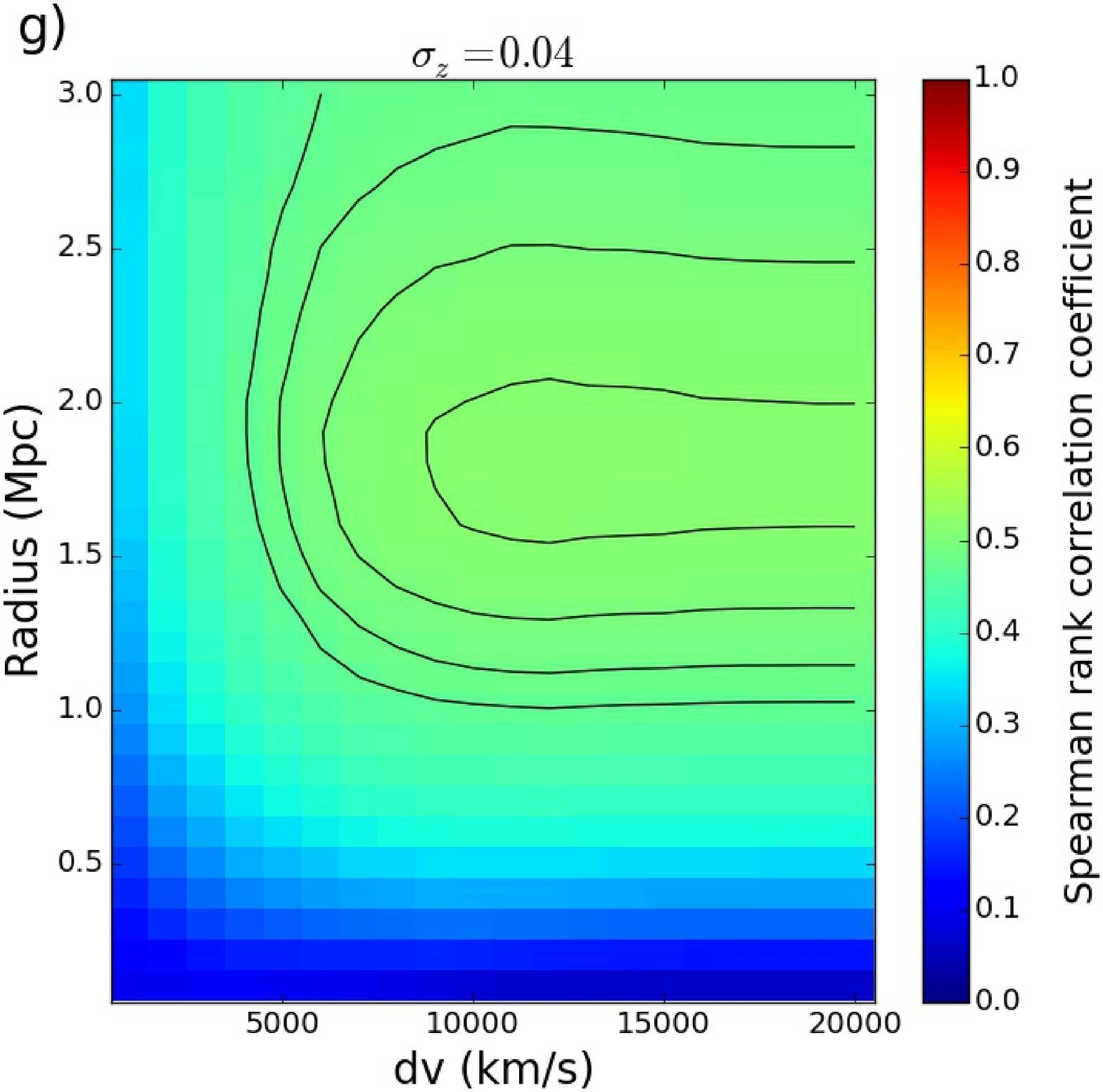}
\end{minipage}
\begin{minipage}[t]{0.33\linewidth}
  \centering  
  \includegraphics[width=.99\linewidth]{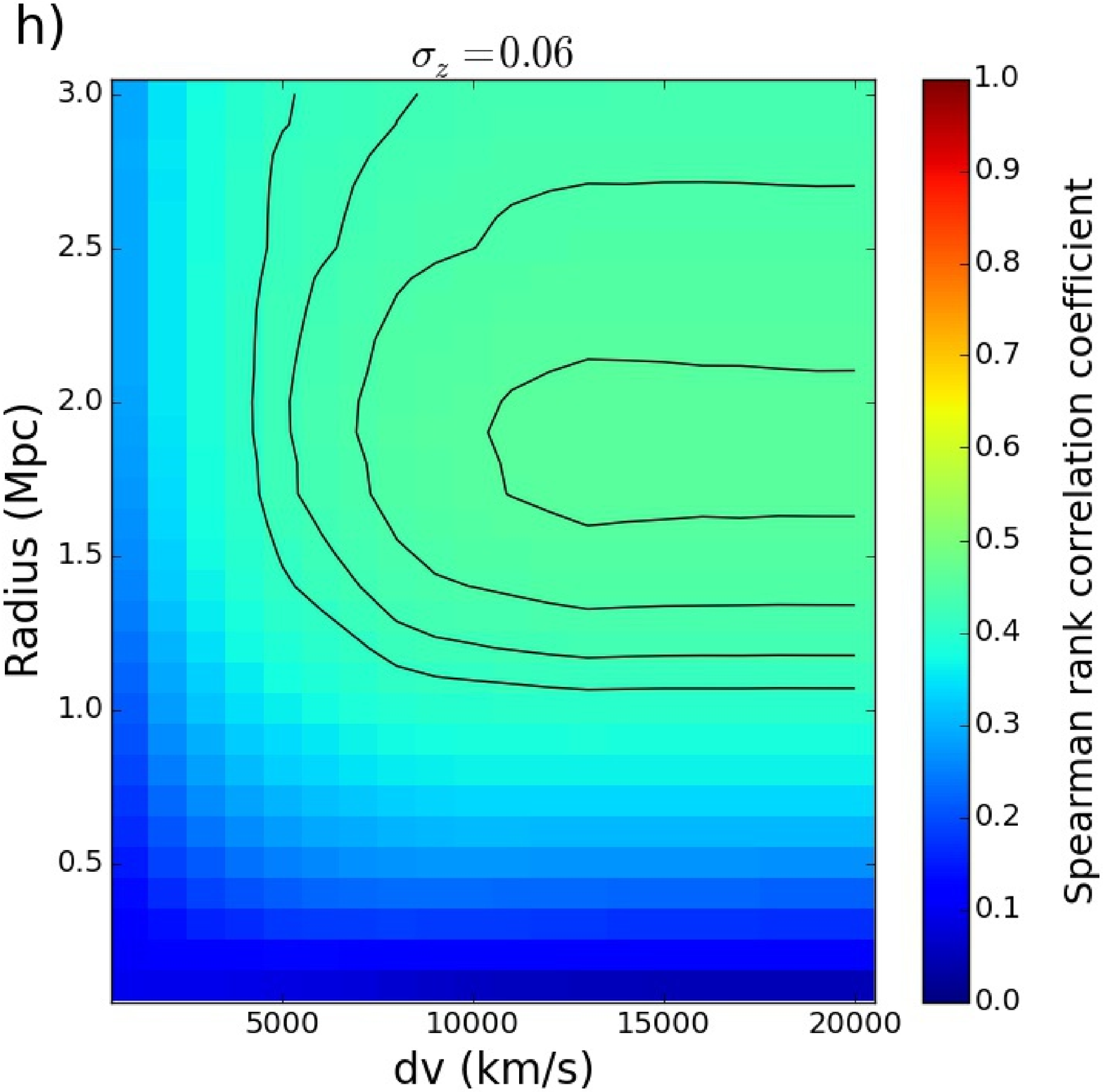}
\end{minipage}
\begin{minipage}[t]{0.33\linewidth}
  \centering  
  \includegraphics[width=.99\linewidth]{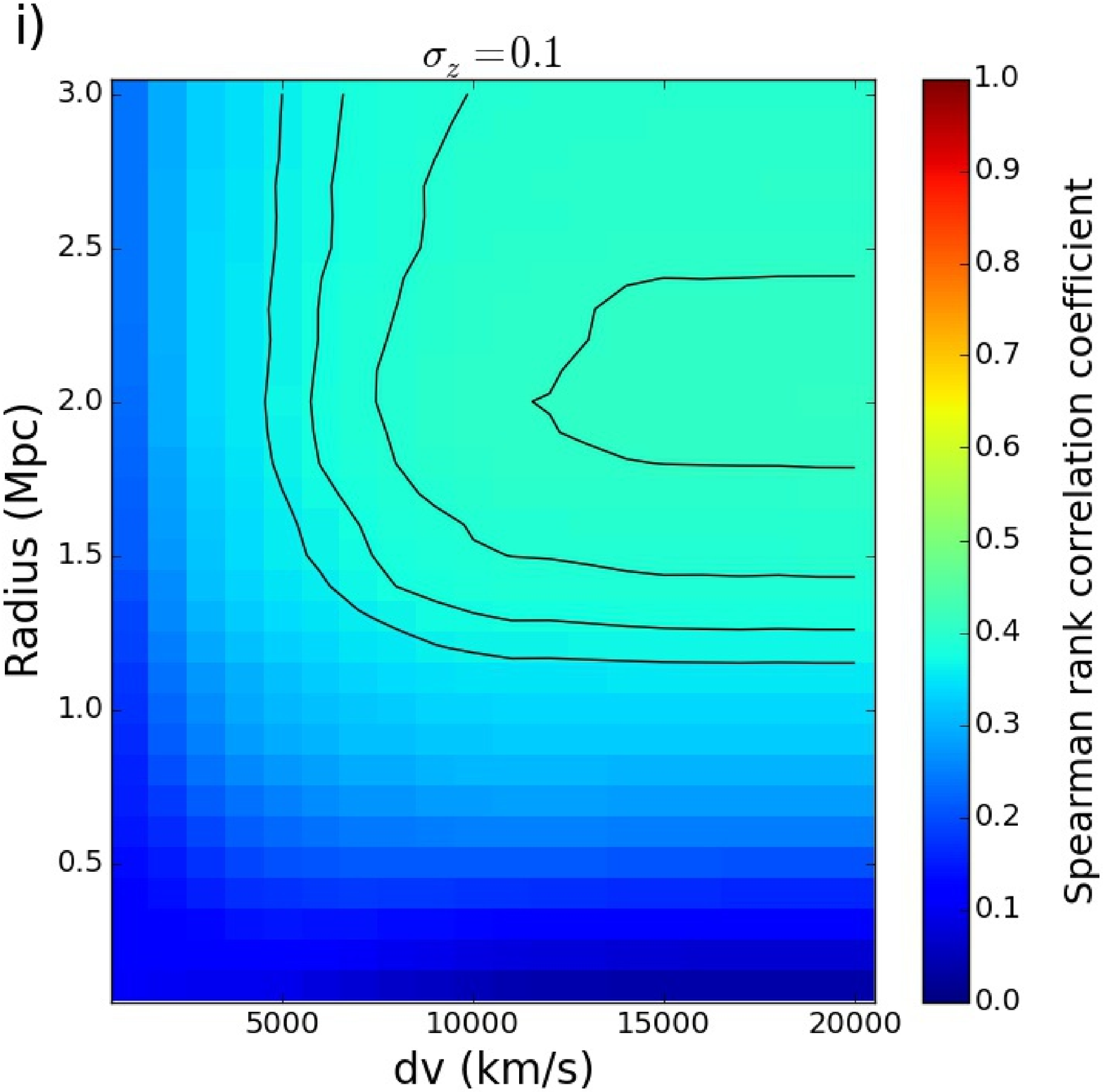}
\end{minipage}
\caption{Shows the Spearman rank correlation coefficients between the fixed aperture environments and the spectroscopic benchmark environments (where $r=1.8$ and $dv=1000\;$km/s) for nine catalogues of galaxies as a function of the aperture parameters: $r$ and the velocity cut ($dv$). a) Shows the correlations for the spectroscopic redshift catalogue. b) Shows the correlations for the photometric redshift catalogue. c) to i) show the correlations for simulated photometric redshift catalogues with redshift uncertainties of $0.0025$, $0.005$, $0.01$, $0.02$, $0.04$, $0.06$ and $0.1$ respectively. The correlation is represented with a colour in each grid cell. The equally spaced contours in the logarithm of the correlation highlight the location of the peak in the parameter space on each grid.}
\label{fig:env_conical_benchmark_correlation}
\end{figure*}

\begin{figure*}
\begin{minipage}[t]{0.33\linewidth}
   \centering 
   \includegraphics[width=.99\linewidth]{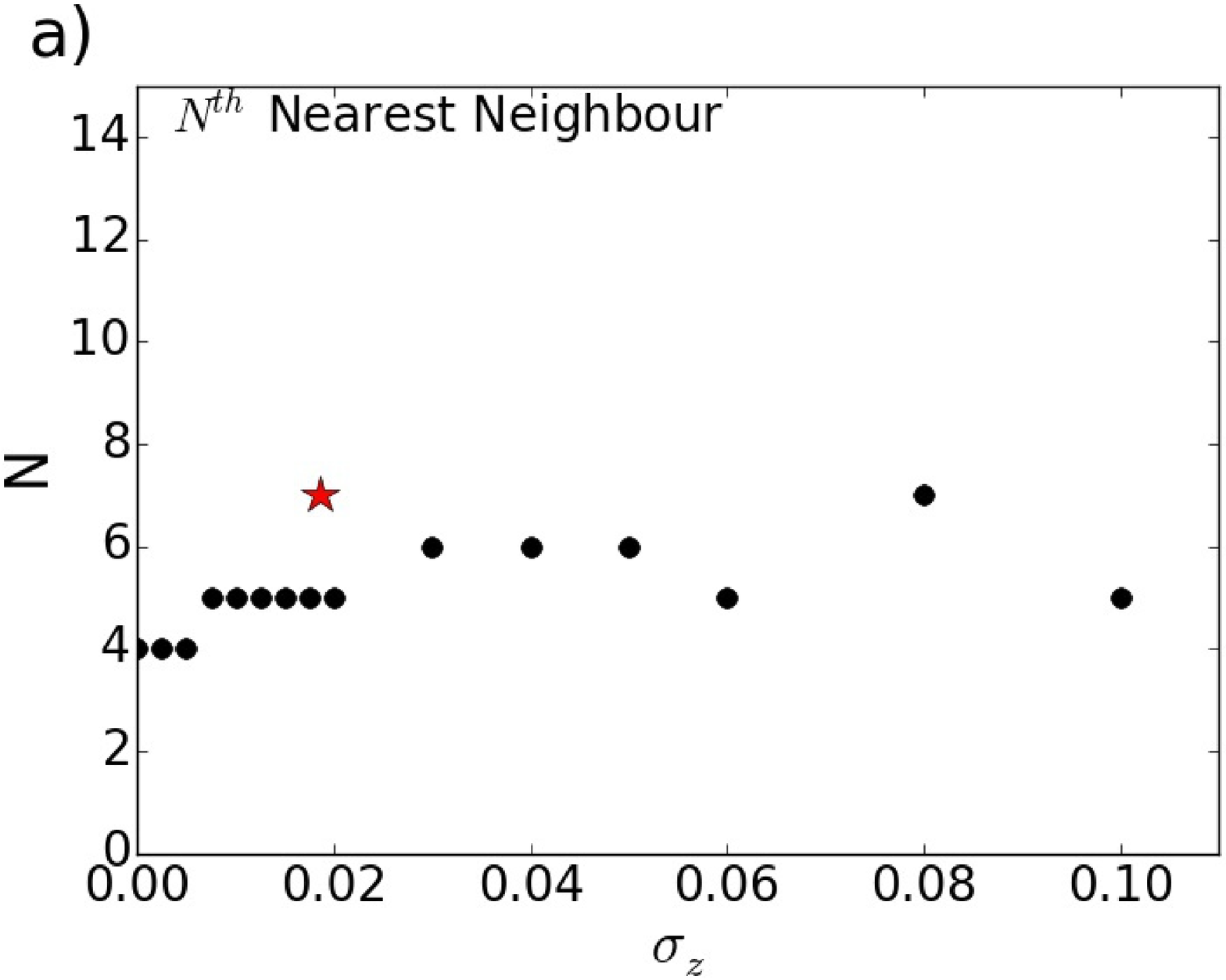}
\end{minipage}
\begin{minipage}[t]{0.33\linewidth}
  \centering  
  \includegraphics[width=.99\linewidth]{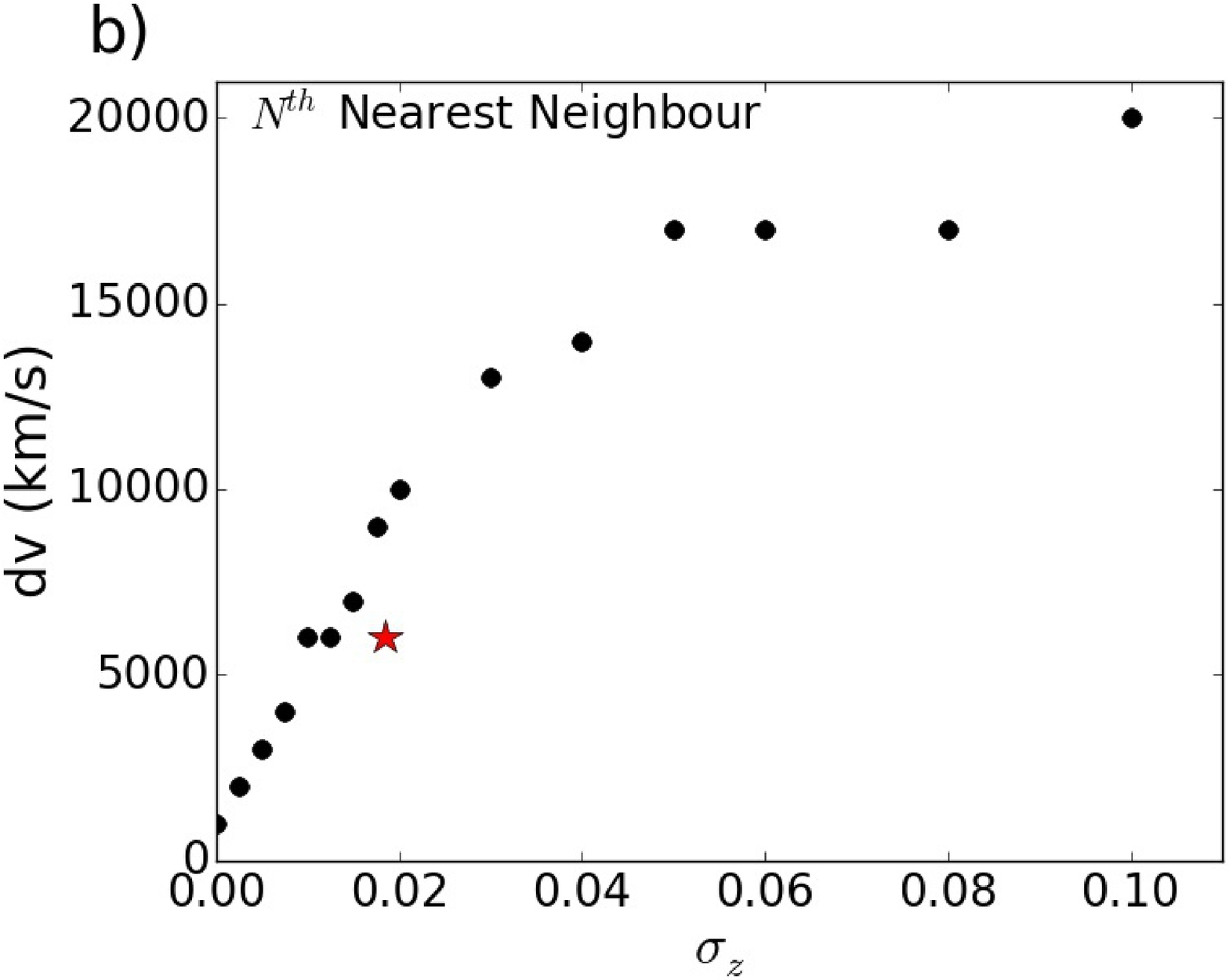}
\end{minipage}
\begin{minipage}[t]{0.33\linewidth}
  \centering  
  \includegraphics[width=.99\linewidth]{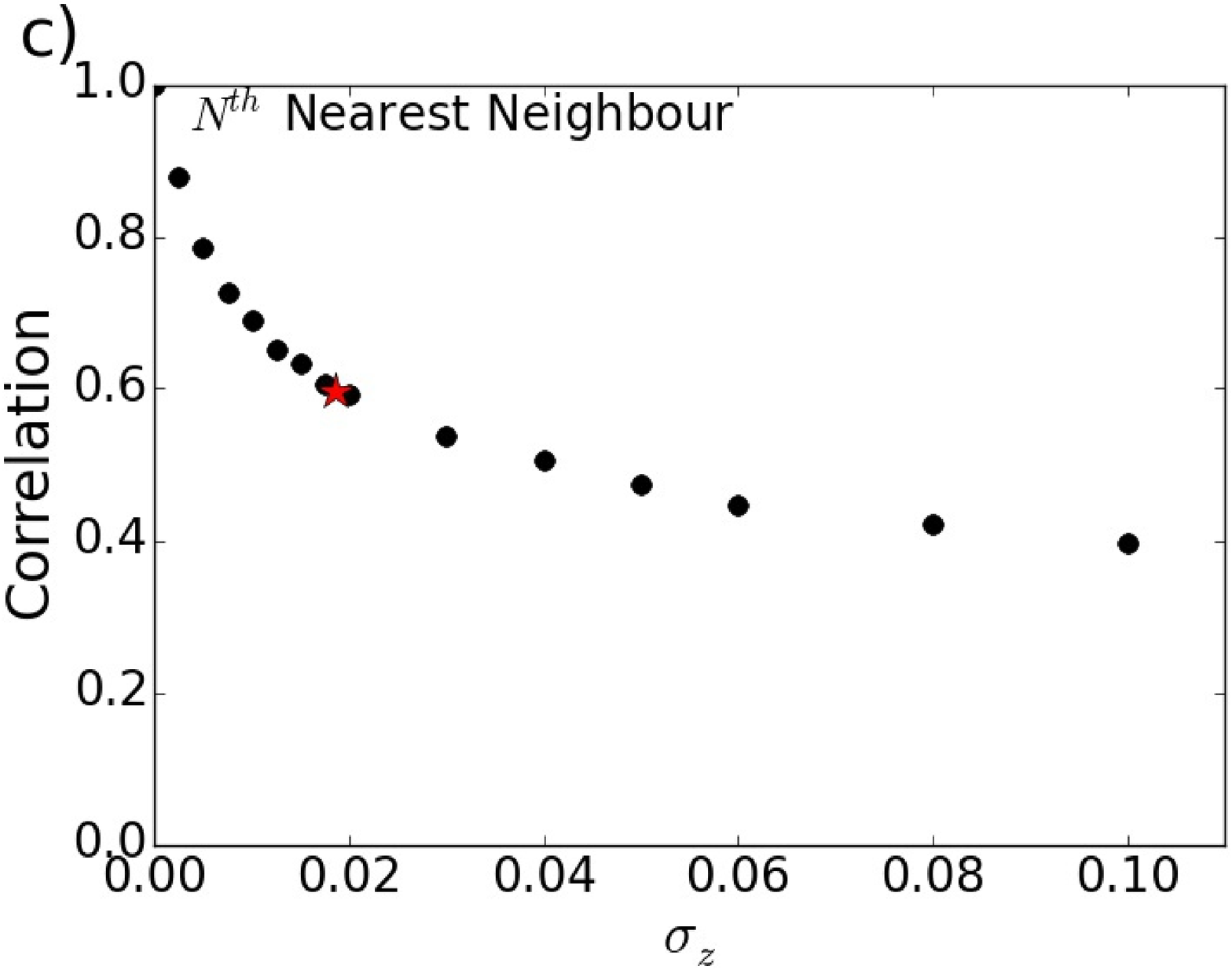}
\end{minipage}
\begin{minipage}[t]{0.33\linewidth}
   \centering 
   \includegraphics[width=.99\linewidth]{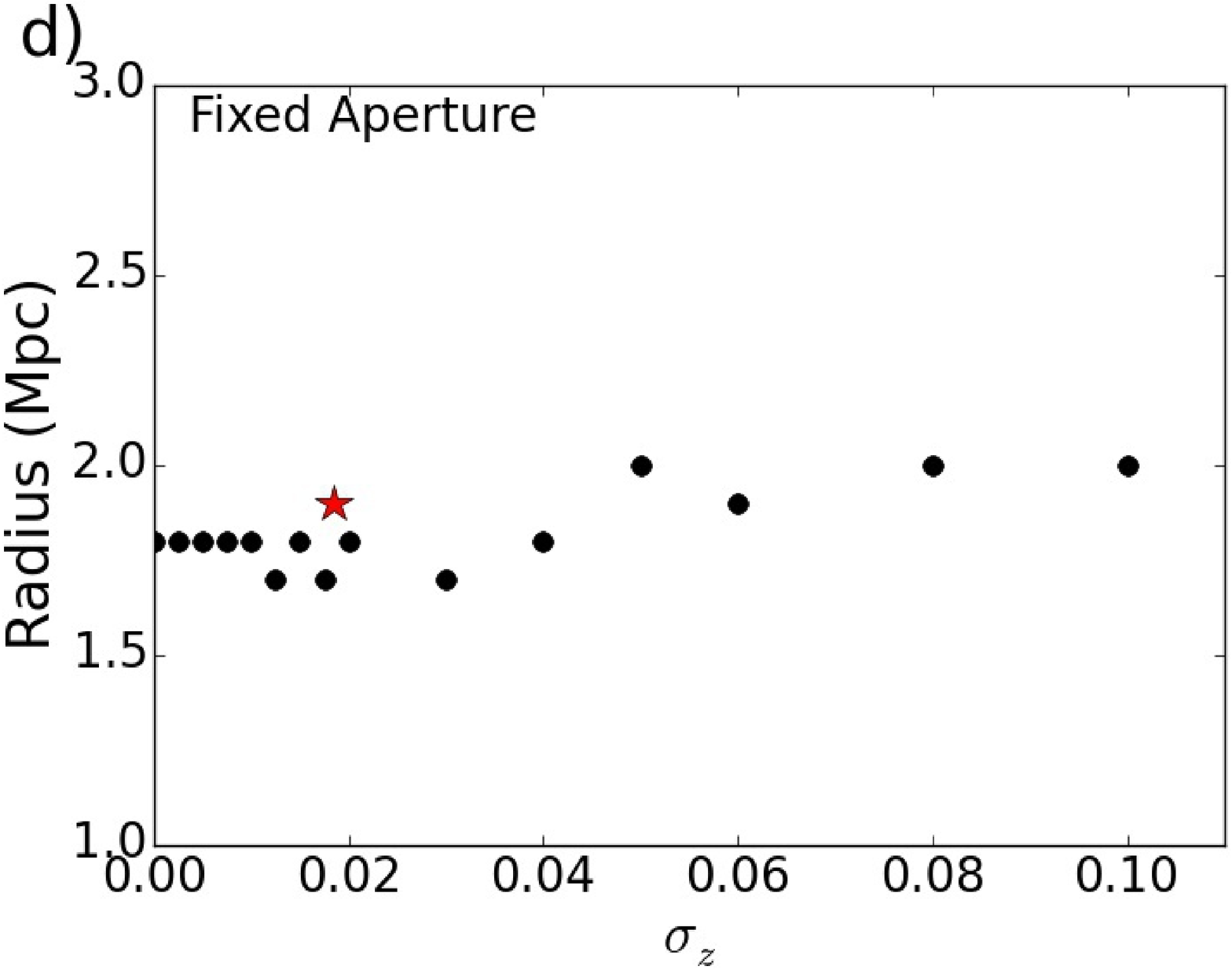}
\end{minipage}
\begin{minipage}[t]{0.33\linewidth}
  \centering  
  \includegraphics[width=.99\linewidth]{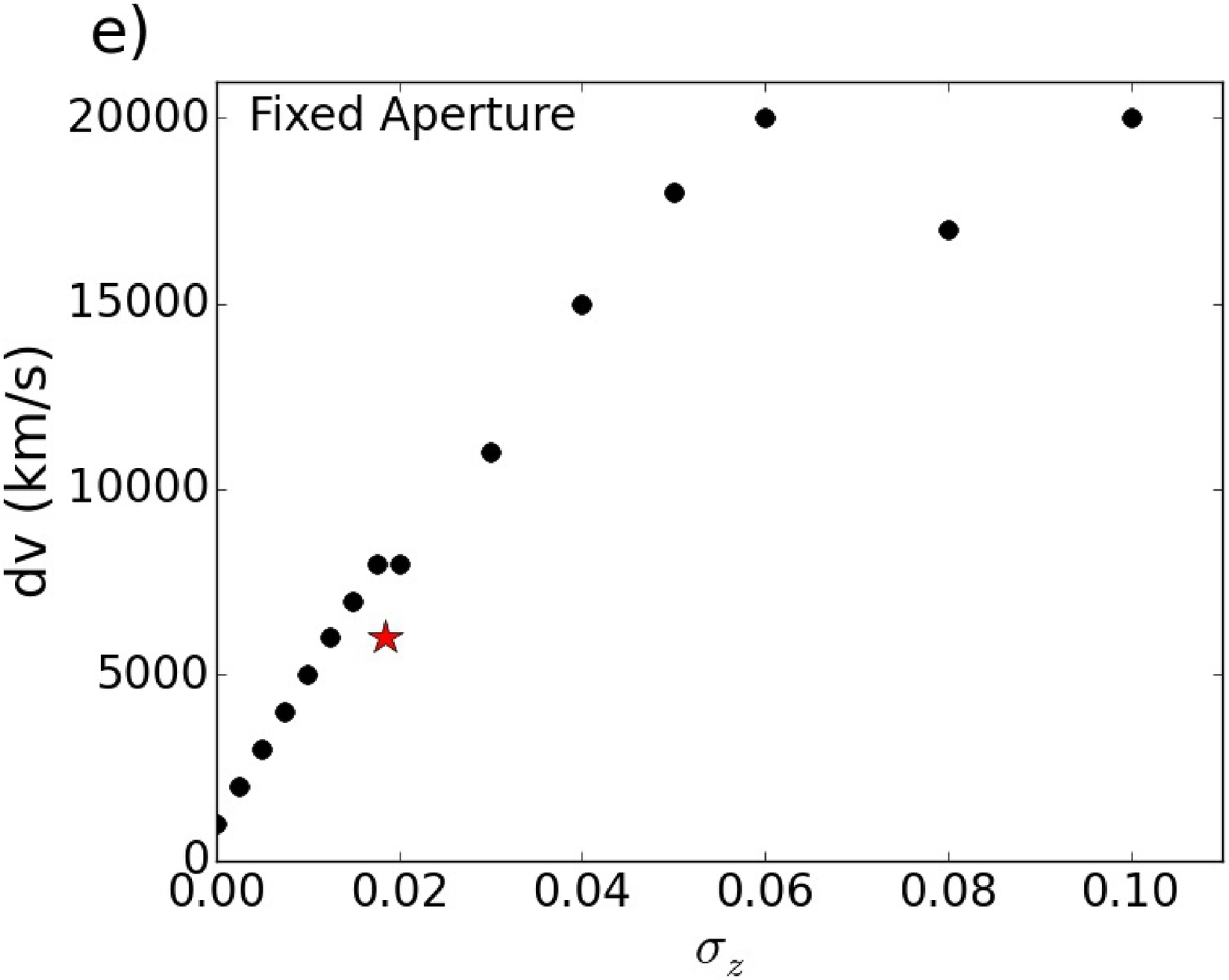}
\end{minipage}
\begin{minipage}[t]{0.33\linewidth}
  \centering  
  \includegraphics[width=.99\linewidth]{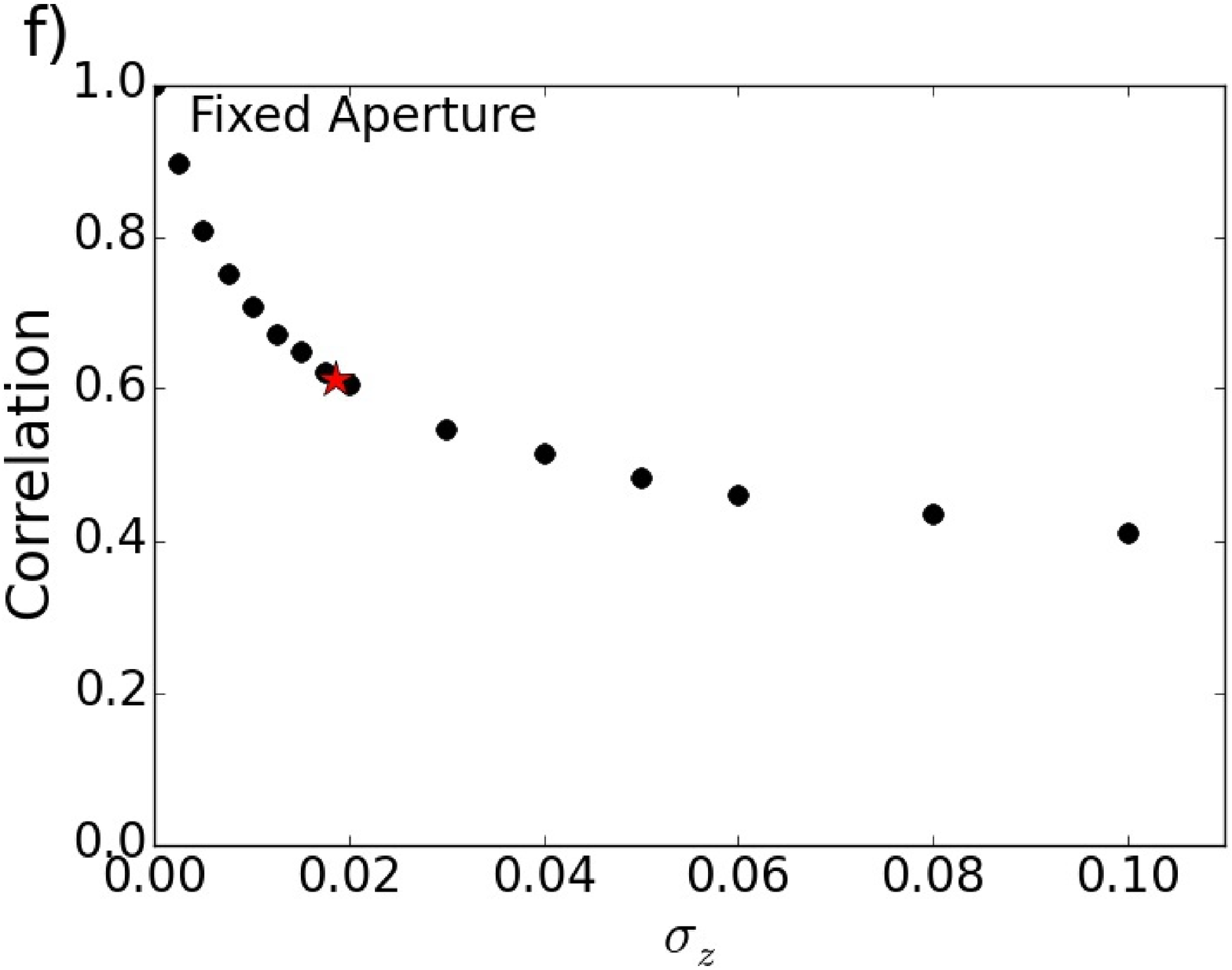}
\end{minipage}
\caption{Left column shows the optimal values for the parameters that control the projected size of the apertures as a function of redshift uncertainty. Middle column shows the optimal velocity cut as a function of redshift uncertainty. Right column shows the Spearman rank correlation coefficient between the environments and the spectroscopic benchmark environments achieved using the optimal parameter values. The top row is for the $N$\textsuperscript{th} nearest neighbour method and the bottom row is for the fixed aperture method. The results for the SDSS photometric redshift are shown with the star symbols.}  
\label{fig:optimal_aperture_parms}
\end{figure*}

\section{Results} \label{sec:results}

We present the results in four sections. In Section \ref{subsec:spec_benchmarks} we compare the scales probed by the $N$\textsuperscript{th} nearest neighbour and fixed aperture methods and establish spectroscopic benchmark environments. In Section \ref{subsec:spec_method_compare} we compare the spectroscopic environment measurements from $N$\textsuperscript{th} nearest neighbour and fixed aperture methods. In Section \ref{subsec:spec_vs_photo_env} we compare the spectroscopic and photometric environment measurements for the two methods. In Section \ref{subsec:corr_vs_parms_uncertainty} we investigate how the Spearman rank correlation coefficient between environment measurements and the spectroscopic benchmark environment measurements varies with the aperture parameters and redshift uncertainty. We determine the aperture parameters that give the strongest correlations as a function of redshift uncertainty. 

\subsection{Spectroscopic benchmark environments} \label{subsec:spec_benchmarks}

To establish spectroscopic benchmark environments to facilitate comparisons between environment measurements as a function of the aperture parameter values and redshift uncertainty we examine the comoving scales that the methods probe.

In both methods the depth of the aperture is controlled by the same parameter: the velocity cut. In this study we consider a range of velocity cuts: $1000-20,000\;$km/s. A velocity cut of $20,000\;$km/s gives an aperture that extends through the entire redshift range: $0.02-0.085$. The possible comoving aperture depths at the median redshift are: $27.9-271.7\;$Mpc.

Next we examine the size of the aperture perpendicular to the line of sight for the $N$\textsuperscript{th} nearest neighbour method. Figure \ref{fig:nth_dist_scales} shows the median $N$\textsuperscript{th} nearest neighbour distance as a function of the velocity cut using spectroscopic measurements for values of $N$ in the range: $1-10$. The plot illustrates the obvious points that as $N$ increases the median distance to the $N$\textsuperscript{th} neighbour also increases and as the velocity cut increases the median distance to the $N$\textsuperscript{th} nearest neighbour decreases. More importantly Figure \ref{fig:nth_dist_scales} shows that the median $N$\textsuperscript{th} nearest neighbour distance for this range of $N$ is {\raise.17ex\hbox{$\scriptstyle\sim$}}$0.3-3.4\;$Mpc. In this study we choose to calculate the $N$\textsuperscript{th} nearest neighbour environments using $N=1-10$ and fixed aperture environments using $r=0.1-3.0\;$Mpc so that the range of scales probed by the two methods are matched.

We establish two sets of spectroscopic benchmark environments: one set for the fixed aperture method and one set for the $N$\textsuperscript{th} nearest neighbour method. To strike a balance between probing small scales ({\raise.17ex\hbox{$\scriptstyle\sim$}}$1.0\;$Mpc) where environment effects are reported to be important \citep{Blanton2007, Wilman2010}, ensuring we have a large dynamic range of environments and we do not have too many apertures devoid of galaxies we choose an aperture scale of $1.8\;$Mpc perpendicular to the line of sight and a velocity cut of $1000\;$km/s for the fixed aperture spectroscopic benchmark measurements. Referring to Figure \ref{fig:nth_dist_scales} to ensure that the aperture scales for the two methods are well matched we choose the aperture parameters $N=4$ and $dv=1000\;$km/s for the $N$\textsuperscript{th} nearest neighbour spectroscopic benchmark measurements.

\subsection{Comparison of spectroscopic environments} \label{subsec:spec_method_compare}

Figure \ref{fig:spec_methods_comparison} shows the $N$\textsuperscript{th} nearest neighbour spectroscopic benchmark environments versus the fixed aperture benchmark spectroscopic environments. $N=4$ and $dv=1000\;$km/s are adopted for the $N$\textsuperscript{th} nearest neighbour method and $r=1.8\;$Mpc and $dv=1000\;$km/s are adopted for the fixed aperture method. The dynamic range for the fixed aperture method is smaller than the $N$\textsuperscript{th} nearest neighbour method and so the best linear fit (gradient of $0.56$) relationship shown with the solid black line deviates from the 1-to-1 line shown with the dashed black symbols. Although we show a linear fit on the plot the relationship between the environments is non-linear. For large and small environments the $N$\textsuperscript{th} nearest neighbour method tends to `stretch out' the fixed aperture environments to higher and lower values respectively. Nevertheless there is a strong correlation between the spectroscopic environments measured with the two methods. The Spearman rank correlation coefficient is $0.91$. 

\subsection{Spectroscopic vs photometric environments} \label{subsec:spec_vs_photo_env}

Figure \ref{fig:spec_vs_photo_envs} shows density plots of the spectroscopic benchmark environments versus the SDSS photometric environments on the left and the spectroscopic (red) and photometric (blue) environment distributions on the right for the $N$\textsuperscript{th} nearest neighbour (top) and fixed aperture method (bottom). The benchmark parameters are adopted ($N=4$ and $dv=1000\;$km/s for the $N$\textsuperscript{th} nearest neighbour method and $r=1.8\;$Mpc and $dv=1000\;$km/s for the fixed aperture method) for the spectroscopic measurements whereas we choose $N=7$ and $dv=7000\;$km/s for the $N$\textsuperscript{th} nearest neighbour method and $r=1.9\;$Mpc and $dv=6000\;$km/s for the fixed aperture method for the photometric measurements as we find these parameters lead to the strongest correlations (see Section \ref{subsec:corr_vs_parms_uncertainty}). Included on the density plots are the 1-to-1 lines (dashed) and the linear best fit lines (solid). The colour of each grid cell on the density plots represents the number of galaxies in that cell. 

Plots a) and b) show the relatively smooth and continuous nature of the $N$\textsuperscript{th} nearest neighbour environments. Conversely the underdense, fixed aperture, spectroscopic environments shown in plots c) and d) are noticeably discretized. The fixed aperture method produces forbidden environments because this method is based on counting the number of \textit{whole} galaxies within a volume. The spectroscopic fixed aperture distribution does not fall to zero in these forbidden ranges, however, for two reasons: (i) \raise.17ex\hbox{$\scriptstyle\sim$}10 percent of the galaxies are weighted by the target sampling rate correction and (ii) densities are measured relative to the mean densities at the redshifts of the target galaxies. This discretization effect is unnoticeable in the photometric fixed aperture environment distribution because a larger velocity cut was adopted leading to more galaxies on average being found within the apertures.  

Plots b) and d) show that the photometric environment distributions are more peaked and their ranges are smaller than the corresponding spectroscopic distributions. Quantitatively the standard deviations of the environment distributions also shows this is the case. The standard deviation of the $N$\textsuperscript{th} nearest neighbour method for the spectroscopic benchmark environment distribution is $0.7$ whereas the standard deviation for the photometric environment distribution is $0.4$. The standard deviation for the fixed aperture method reduces from $0.44$ to $0.31$. The most dense environments measured with spectroscopy are found to be less dense with photometry. This is because imprecise redshifts tend to have the effect of moving galaxies away from the centres of dense regions. Conversely, low density regions in the spectroscopic sample tend to be contaminated, in the photometric sample, by galaxies that have scattered into them from higher or lower redshifts.

We illustrate this with Figures \ref{fig:density_example} and \ref{fig:wedge}. Figure \ref{fig:density_example} shows two typical target (marked in red) galaxies: one in a low density region (top row) and one in a high density region (bottom row). The redshifts of the galaxies are measured with spectroscopy in the plots on the left and photometry in the plots on the right. The plots show the galaxies that are constrained by velocity cuts of $\pm 1000\;$km/s. The low density environment that was measured with spectroscopy is measured to be more dense with photometry. The number of density defining galaxies found within $1.8\;$Mpc of the target (marked by the blue circle) increases from $1$ to $5$. The high density environment that was measured with spectroscopy is measured to be less dense when measured with photometry. The number of density defining galaxies found within $1.8\;$Mpc of the target decreases from $24$ to $13$. 

Figure \ref{fig:wedge} shows the redshift and declination of galaxies within a wedge of volume using spectroscopic measurements (a) and photometric measurements (b). Clustering can easily be seen in the spectroscopic plot whereas in the photometric plot the distribution of galaxies looks more random. It is difficult to see the clustering (from this perspective) in the photometric plot because of the redshift errors. High density regions are smoothed into the surrounding low density regions leading to a more homogenous density field. 

In Figure \ref{fig:spec_vs_photo_envs} we also computed the gradients of the linear best fit lines and Spearman rank correlation coefficients for plots a) and c). The gradient of the linear best fit lines between the spectroscopic and photometric environment measurements are $0.38$ for the $N$\textsuperscript{th} nearest neighbour method and $0.45$ for the fixed aperture method. We note that although we have shown linear best fit lines on the plots the relationships between the spectroscopic and photometric environments are nonlinear. The high density photometric environments for both methods deviate away from the linear fits to larger values. We find in agreement with previous studies that the high density environment are recovered better than the low density environments \citep[e.g][]{Cooper2005}.

The Spearman rank correlation coefficients for the $N$\textsuperscript{th} nearest neighbour method are $0.60$ and $0.62$ for the fixed aperture method. These large correlations suggest that, although it might not be possible to use photometry to measure galaxy environment for individual galaxies, with a high degree of accuracy, it should be possible to extract an environment signal from a sufficiently large sample of galaxies.

\subsection{Correlation vs aperture parameters and redshift uncertainty} \label{subsec:corr_vs_parms_uncertainty}

Figure \ref{fig:env_nnn_benchmark_correlation} and Figure \ref{fig:env_conical_benchmark_correlation} show how the Spearman rank correlation coefficient between environments and the spectroscopic benchmark environments vary as a function of the aperture parameters and redshift uncertainty for the $N$\textsuperscript{th} nearest neighbour and fixed aperture methods respectively. Each figure consists of a sequence of nine coloured grids. Each grid shows the results for a different redshift uncertainty. The results for the SDSS spectroscopic redshift sample are shown in plot a). The results for the SDSS photometric redshift sample are shown in plot b) and the results for the simulated photometric samples with $\sigma_z=0.0025, 0.005, 0.01, 0.02, 0.04, 0.06$ and $0.1$ are shown in plots c) to i) respectively. In Figure \ref{fig:env_nnn_benchmark_correlation} the aperture parameter: $N$ varies along the vertical axis whereas in Figure \ref{fig:env_conical_benchmark_correlation} the radius of the fixed aperture varies along the vertical axis. The velocity cut ($dv$) used to constrain the depth of the apertures varies along the horizontal axis of each grid. The colour of each grid cell represents the Spearman rank correlation coefficient between the environments and the spectroscopic benchmark environments. The smallest correlations are shown in blue and the largest correlations are shown in red. The equally spaced contours in the logarithm of the correlation highlight the location of the peak in the parameter space on each grid.

From visual inspection of Figure \ref{fig:env_nnn_benchmark_correlation} and Figure \ref{fig:env_conical_benchmark_correlation} the correlation between environments and the spectroscopic benchmark environments for the fixed aperture method and $N$\textsuperscript{th} nearest neighbour method behave in a similar way. As the redshift uncertainty increases the coloured grids transition from red to blue. The Spearman rank correlation coefficient between the environments and the spectroscopic benchmark environments therefore decreases as the redshift uncertainty increases. 

The Spearman rank correlation coefficient between the environments and the spectroscopic benchmark environments depends not only on the redshift uncertainty but also on the aperture parameters. For low redshift uncertainties ($\sigma_z < 0.02$) there exists a `sweet spot' (e.g. see plot e) in either Figure). A prudent choice for the aperture parameter values enables the strongest environment signal to be extracted. At larger redshift uncertainties providing the aperture parameter values are large enough (e.g. $N>3$, $r>1.2\;$Mpc, $dv>10,000\;$km/s and smaller than the maximum values we study) the correlation is relatively insensitive to the aperture parameter values.

Figure \ref{fig:optimal_aperture_parms} shows the parameter values that give the largest Spearman rank correlation coefficient between the environments and spectroscopic benchmark environments as a function of redshift uncertainty. The figure consists of six plots. The top row (a, b and c) are for the $N$\textsuperscript{th} nearest neighbour method and bottom row (d, e and f) are for the fixed aperture method. The left hand column (a and d) shows the optimal parameter values for the parameters that control the projected sizes of the apertures. The best values of $N$ are shown in plot a) and the best values of $r$ are shown in plot d) as a function of redshift uncertainty. The middle column (b and e) shows the optimal velocity cut as a function of redshift uncertainty for the two methods and the right hand column (c and e) shows the the Spearman rank correlation coefficient obtained with the optimal parameter values as a function of redshift uncertainty. The results for the SDSS photometric redshifts are shown with the star symbols. 

The optimal parameter values for the parameters that control the projected size of the apertures are aligned with the values chosen for the spectroscopic benchmark measurements. Both $N$ and $r$ are fairly insensitive to the redshift uncertainty (there appears to be a mild trend with increasing redshift uncertainty). $N=5$ and $r=1.8\;$Mpc are good choices across the entire range of redshift uncertainties. 

The optimal velocity cut for both methods however is dependent on the redshift uncertainty. To obtain the optimal correlation with the spectroscopic benchmark measurements it appears that the $N$\textsuperscript{th} nearest neighbour method requires a slightly deeper aperture than the fixed aperture method. Nevertheless for both methods as the redshift uncertainty increases a larger velocity cut should be employed. This ensures that galaxies displaced from their spectroscopic positions along the line of sight by uncertain redshift measurements are captured by the apertures. 

The optimal parameter values and correlations obtained using the SDSS photometric redshift sample (which has an uncertainty of $0.0185$) are consistent with the values obtained with the simulated photometric redshifts.

The best Spearman rank correlation coefficient between the environments and the spectroscopic benchmark environments decreases from $1.0$ (i.e. when comparing the benchmark environments with themselves) down to {\raise.17ex\hbox{$\scriptstyle\sim$}}$0.4$ with simulated redshifts with an uncertainty of $0.1$ for both the $N$\textsuperscript{th} nearest neighbour and fixed aperture methods. The correlation decays rapidly for small redshift uncertainties ($\sigma_z < 0.02$) and then more gradually for larger redshift uncertainties ($\sigma_z > 0.02$). Although the correlation at a redshift uncertainty of $0.1$ is reduced it is significantly larger than zero. It therefore should be possible to extract an environment signal from large scale photometric surveys such as DES.

\begin{figure*}
\centering
\includegraphics[width=0.49\linewidth]{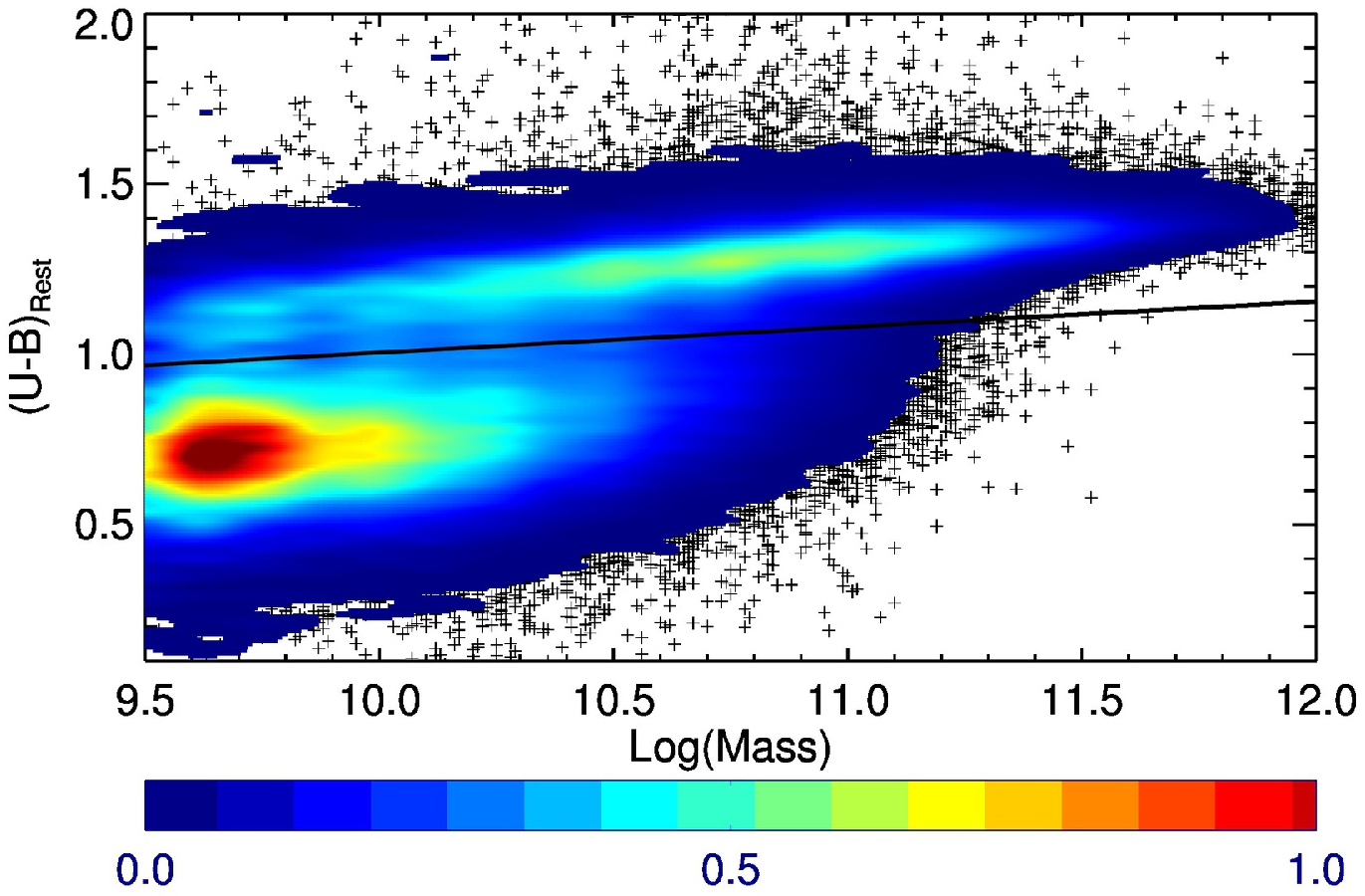}
\includegraphics[width=0.49\linewidth]{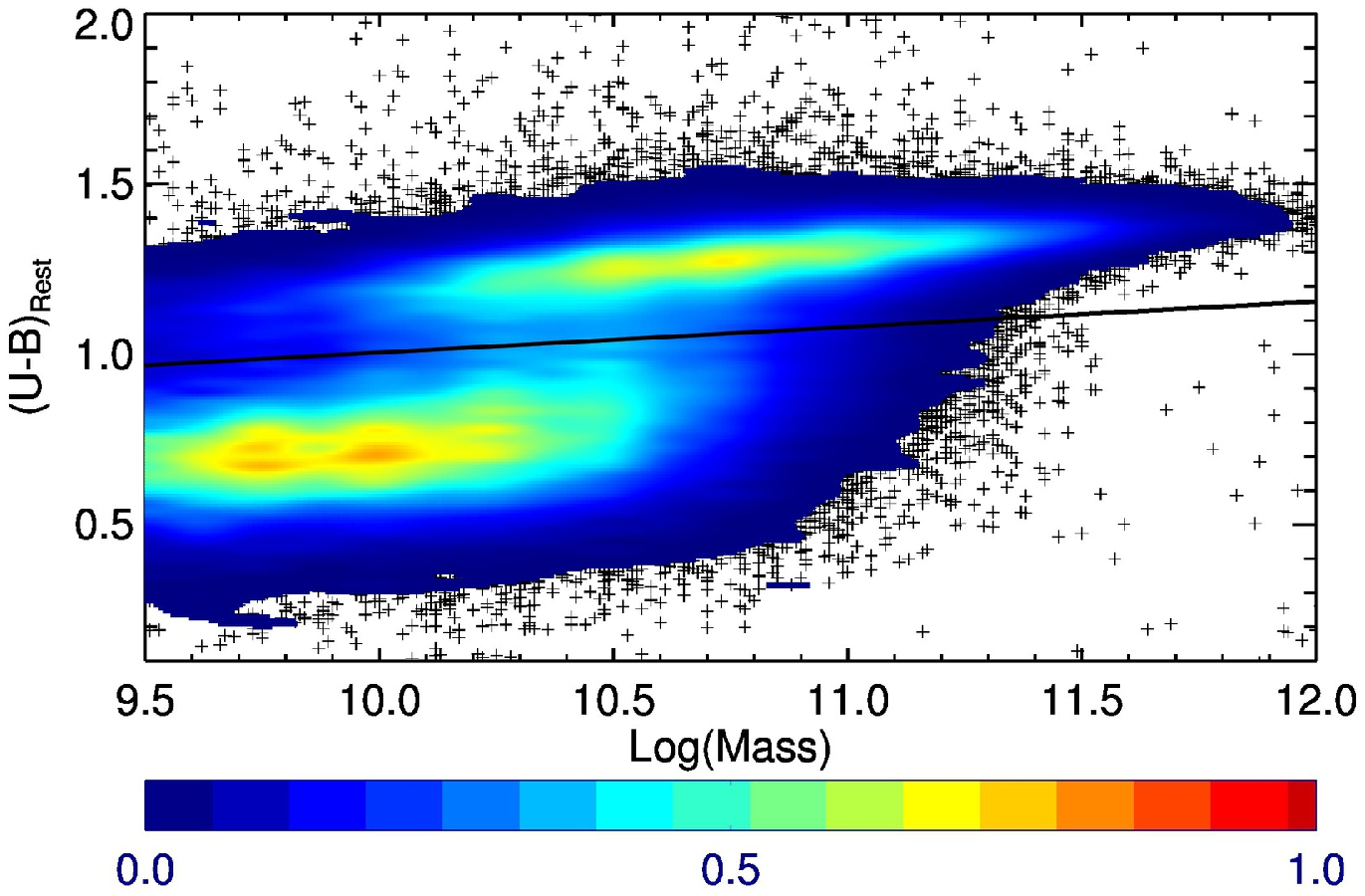}
\caption{(U-B) rest frame colour as a function of the log(mass) for the spectroscopic selection (left) and the photometric selection (right). The colour represents the relative density of the galaxies in the parameter space with the target sampling rate and volume corrections applied. Beneath a threshold the colour scheme is not used but the individual galaxies are overplotted with the ``+'' symbol. The black dividing line (specified in the main text) splits the red sequence and blue cloud populations.}  
\label{fig:color_mass}
\end{figure*}

\begin{figure*}
\begin{minipage}[t]{0.33\linewidth}
   \centering 
   \includegraphics[width=.99\linewidth]{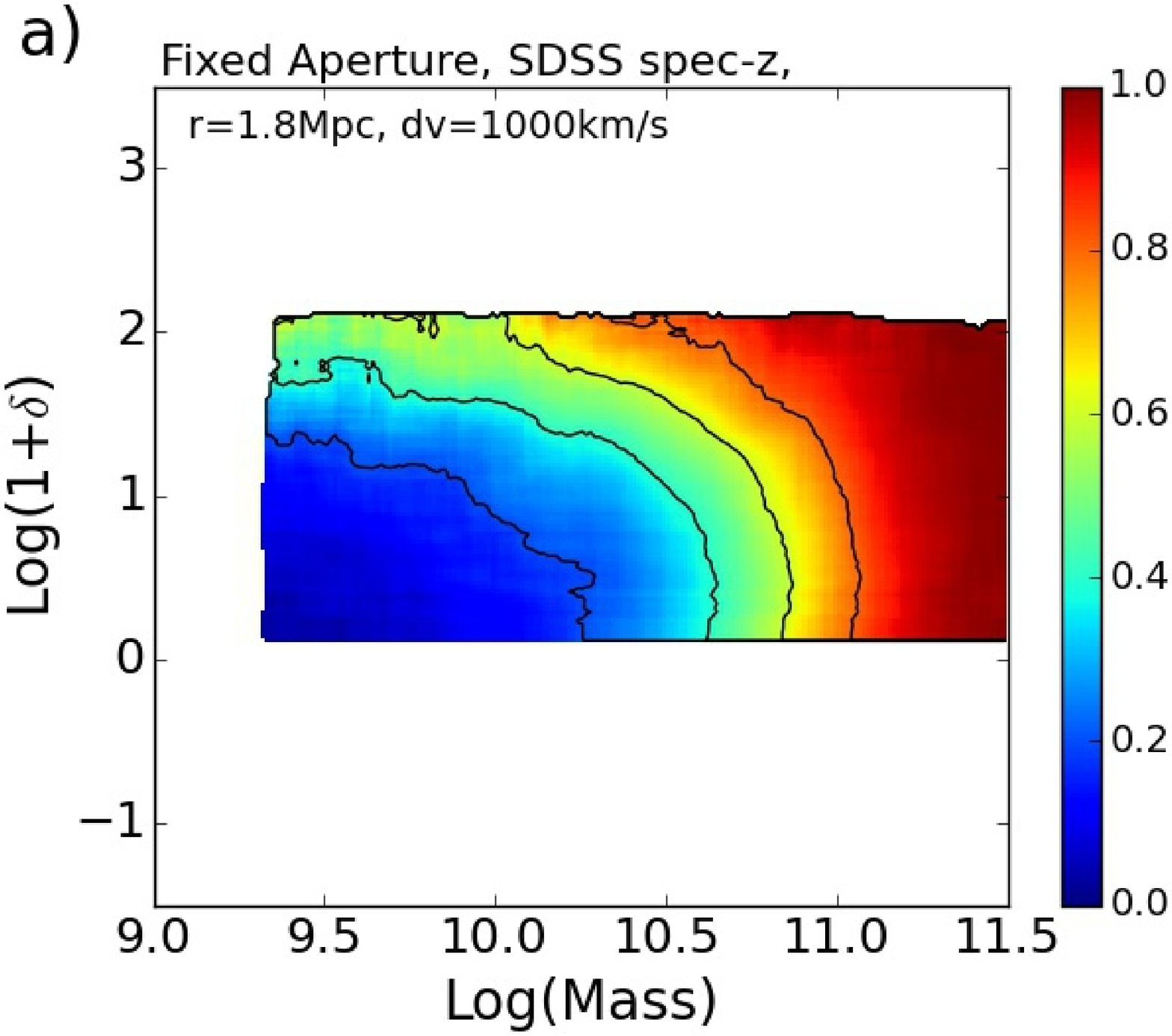}
\end{minipage}
\begin{minipage}[t]{0.33\linewidth}
  \centering  
  \includegraphics[width=.99\linewidth]{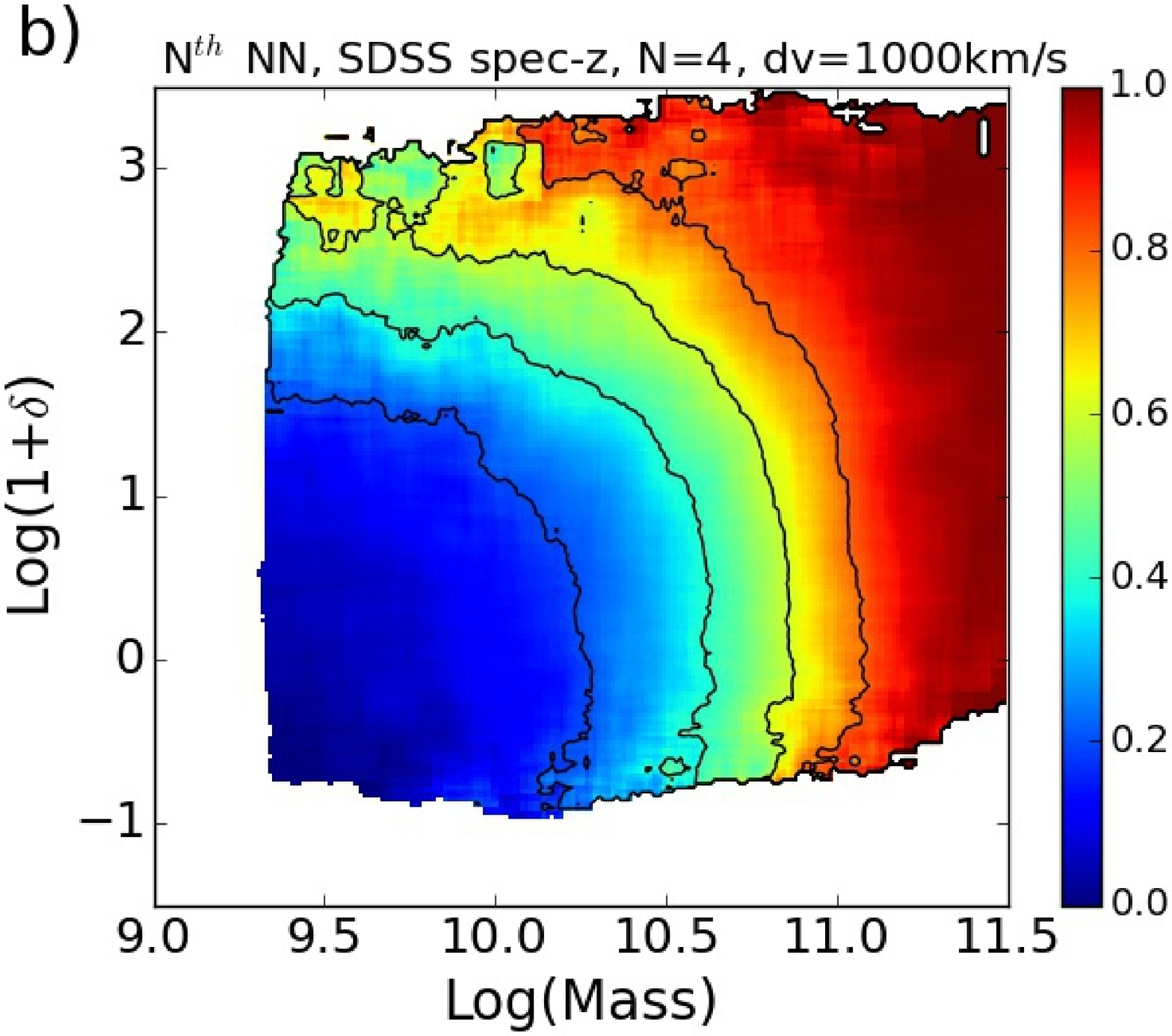}
\end{minipage}
\begin{minipage}[t]{0.33\linewidth}
  \centering  
  \hfill
\end{minipage}
\begin{minipage}[t]{0.33\linewidth}
   \centering 
   \includegraphics[width=.99\linewidth]{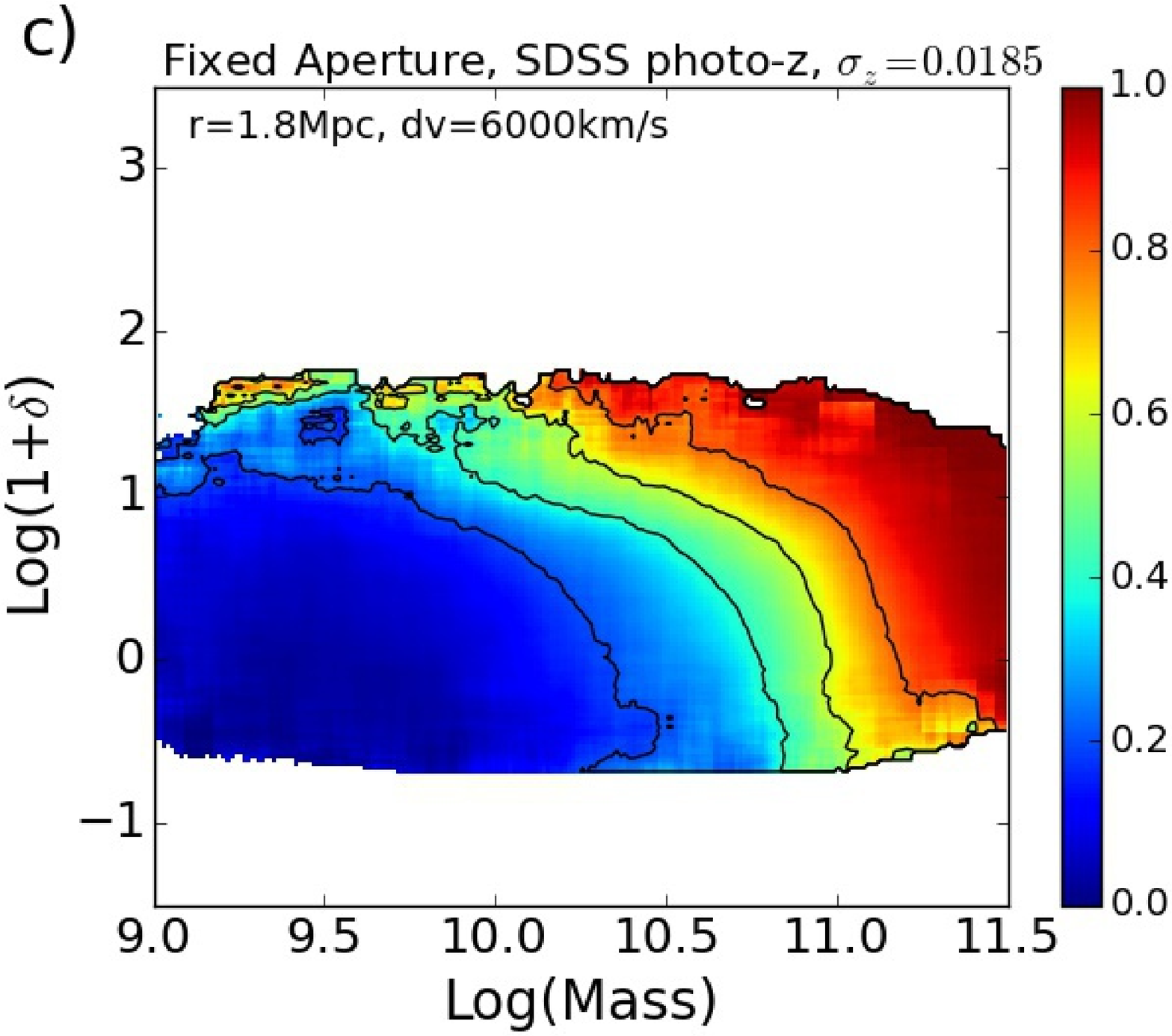}
\end{minipage}
\begin{minipage}[t]{0.33\linewidth}
  \centering  
  \includegraphics[width=.99\linewidth]{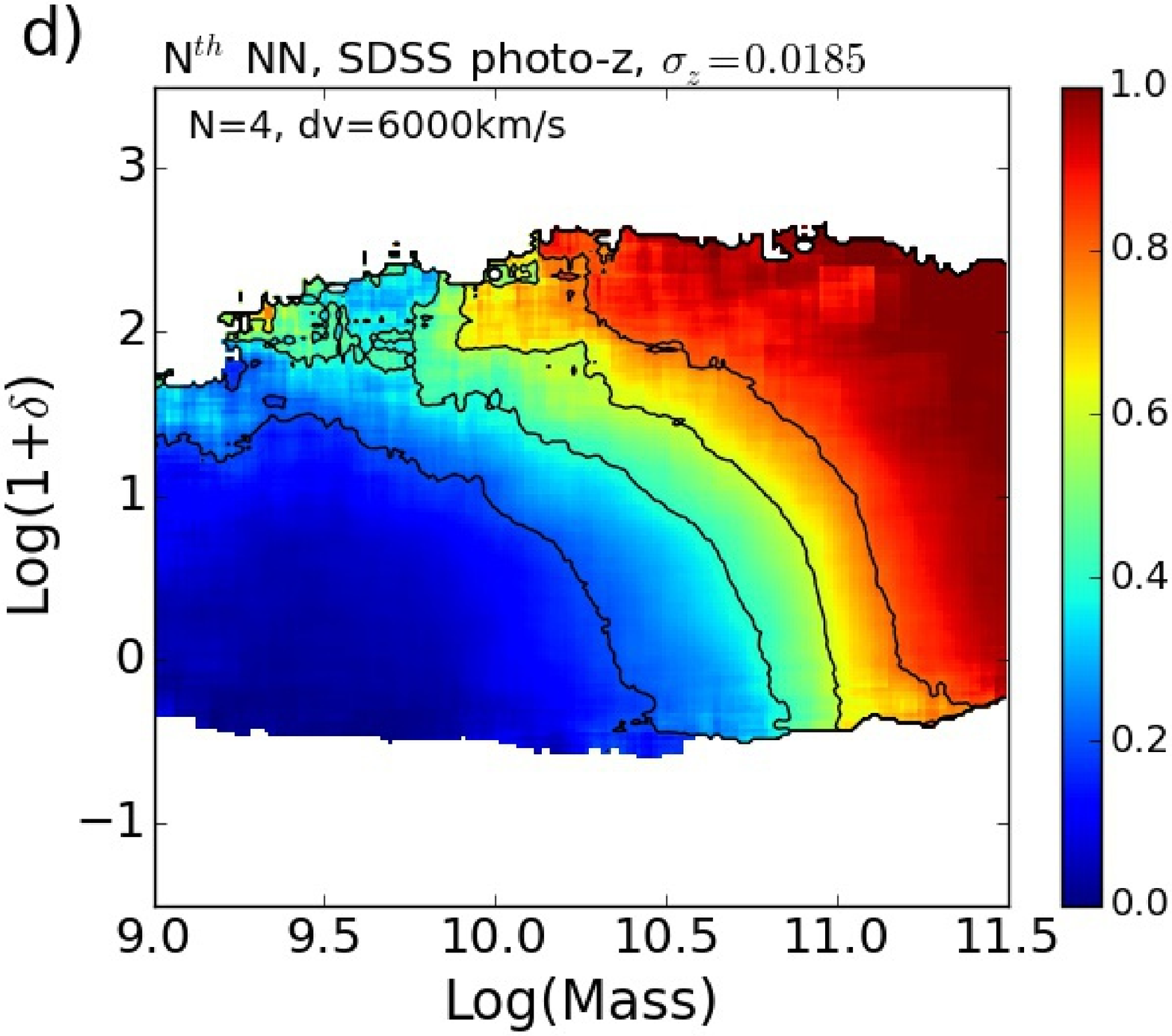}
\end{minipage}
\begin{minipage}[t]{0.33\linewidth}
  \centering  
  \includegraphics[width=.99\linewidth]{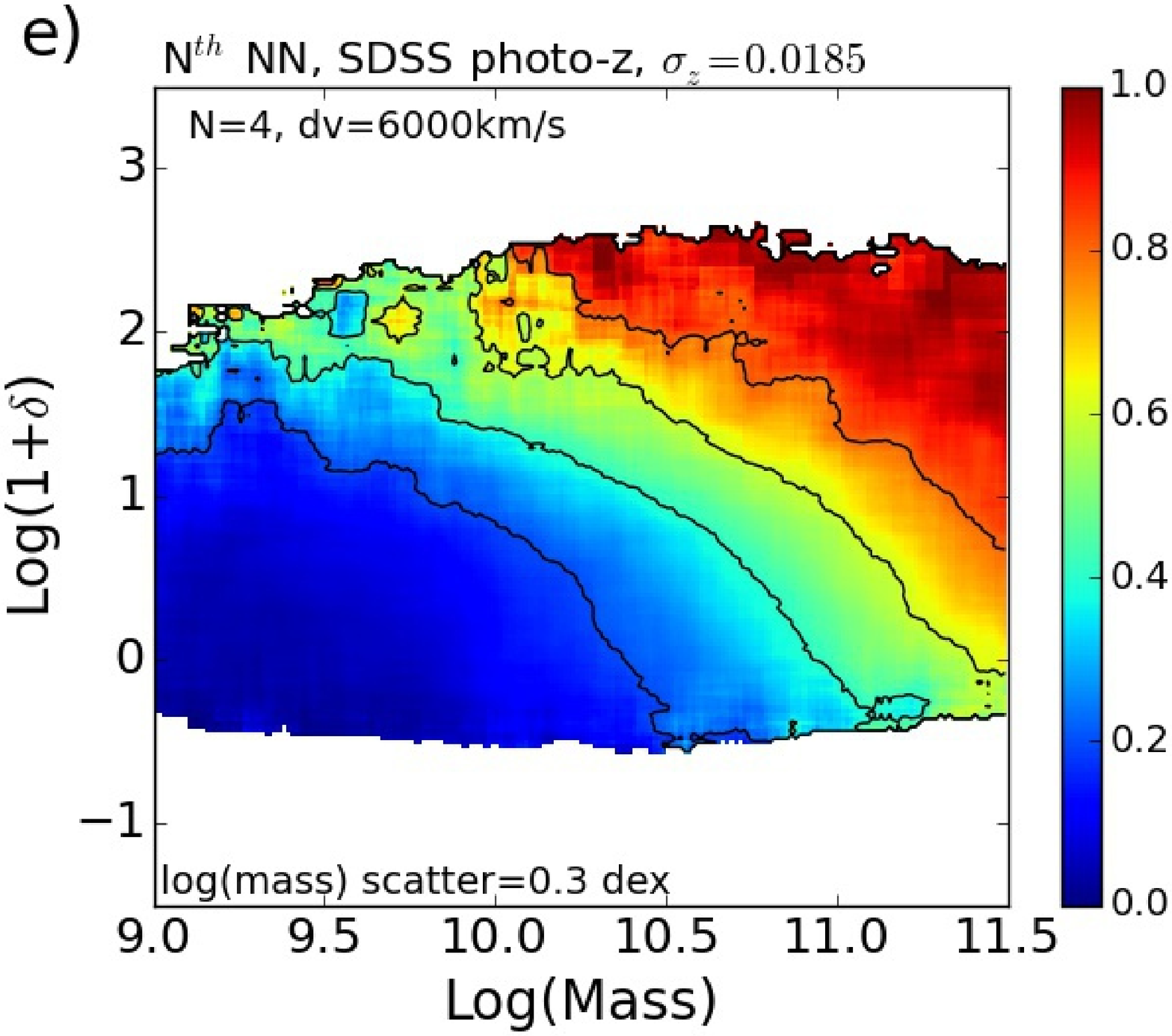}
\end{minipage}
\begin{minipage}[t]{0.33\linewidth}
  \centering  
  \includegraphics[width=.99\linewidth]{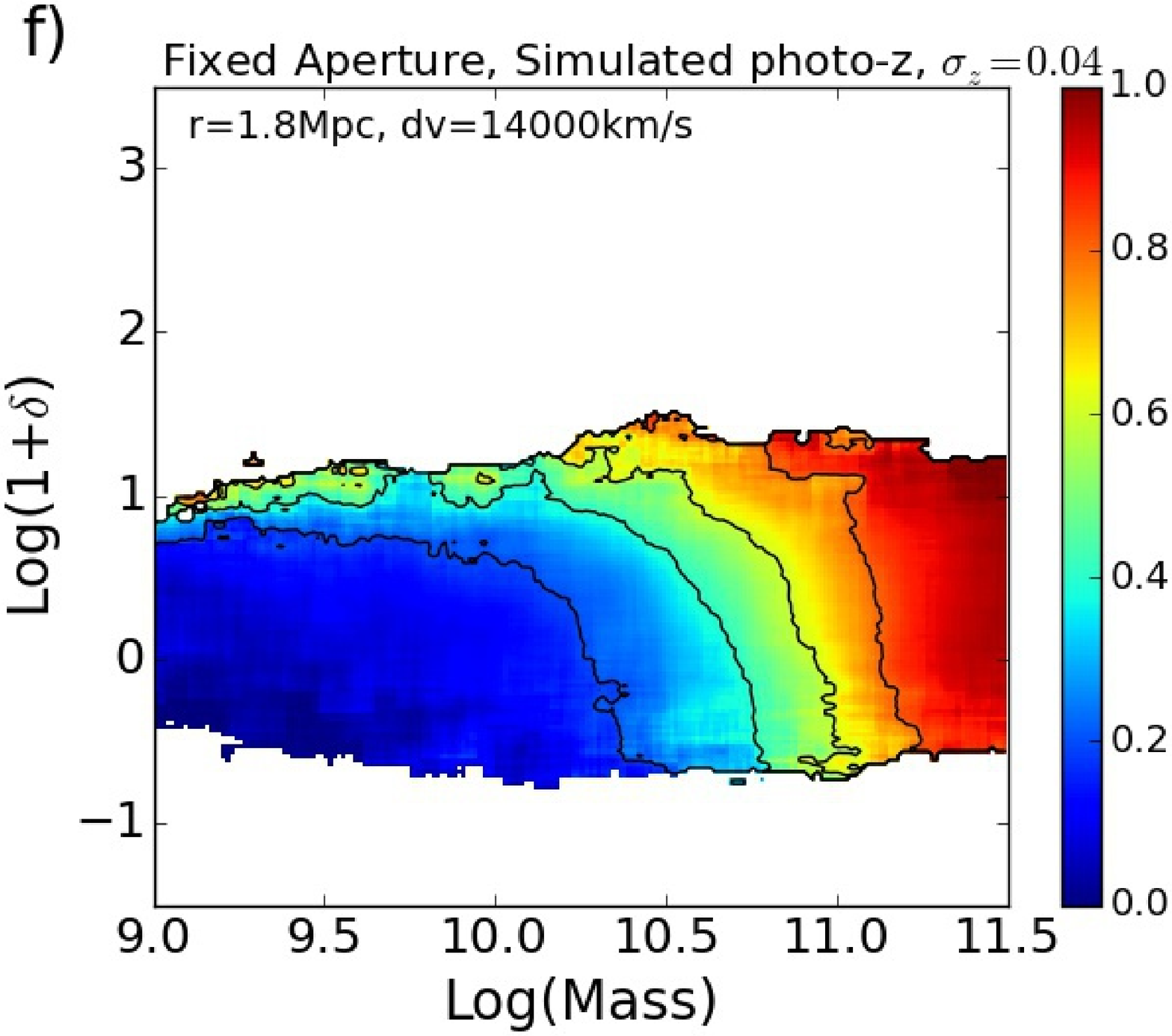}
\end{minipage}
\begin{minipage}[t]{0.33\linewidth}
  \centering  
  \includegraphics[width=.99\linewidth]{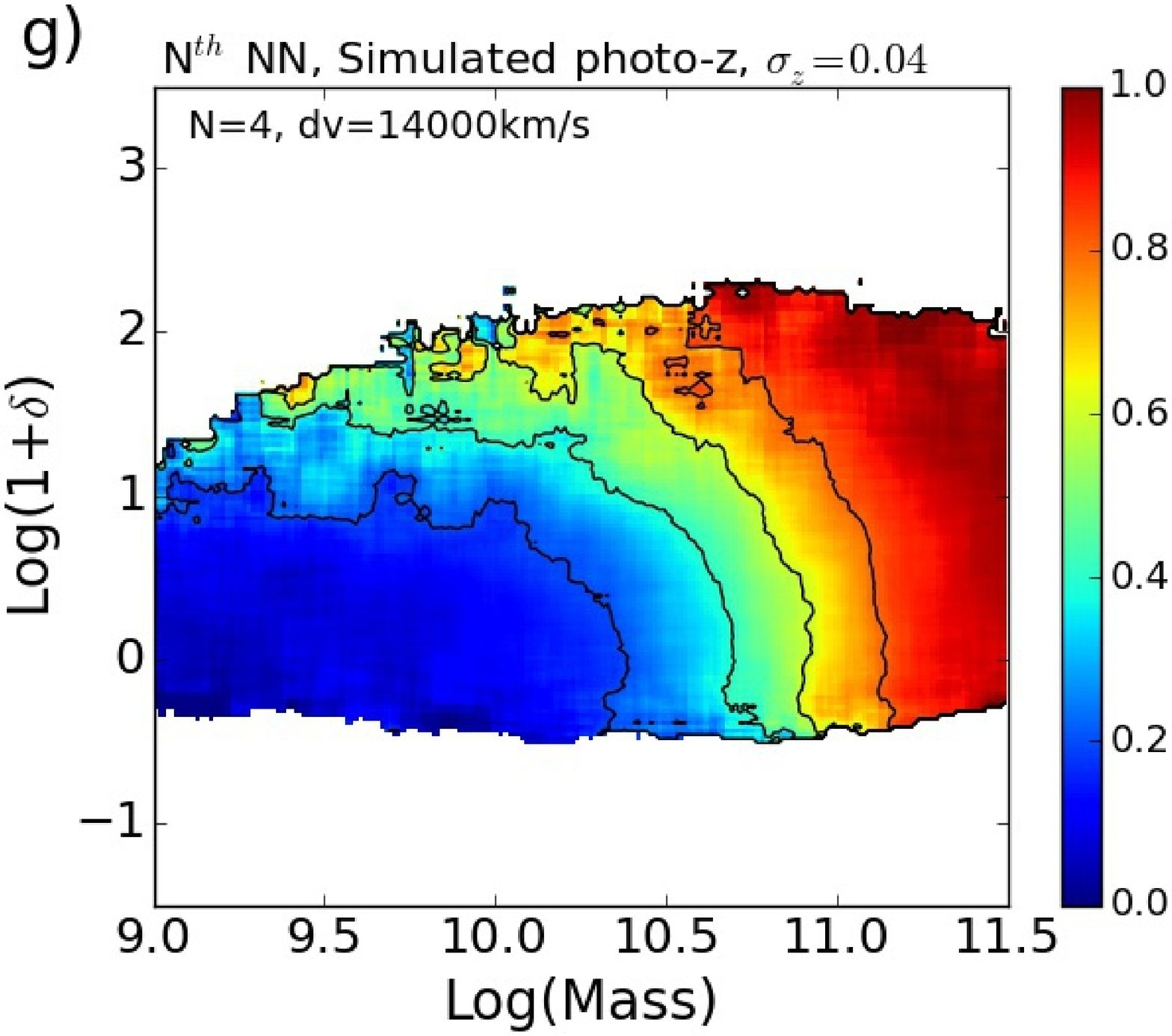}
\end{minipage}
\begin{minipage}[t]{0.33\linewidth}
   \centering 
  \includegraphics[width=.99\linewidth]{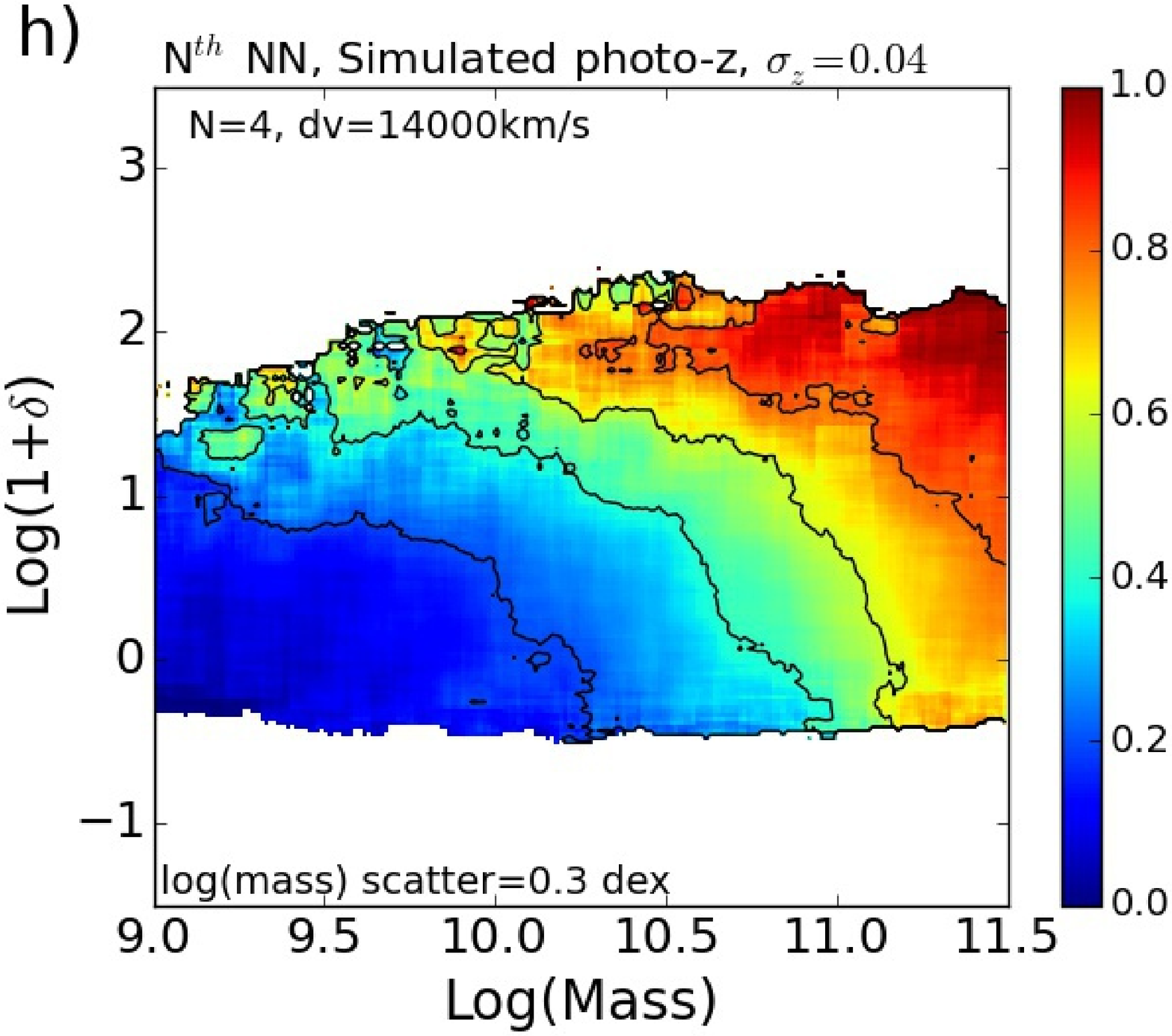}
\end{minipage}
\begin{minipage}[t]{0.33\linewidth}
  \centering  
  \includegraphics[width=.99\linewidth]{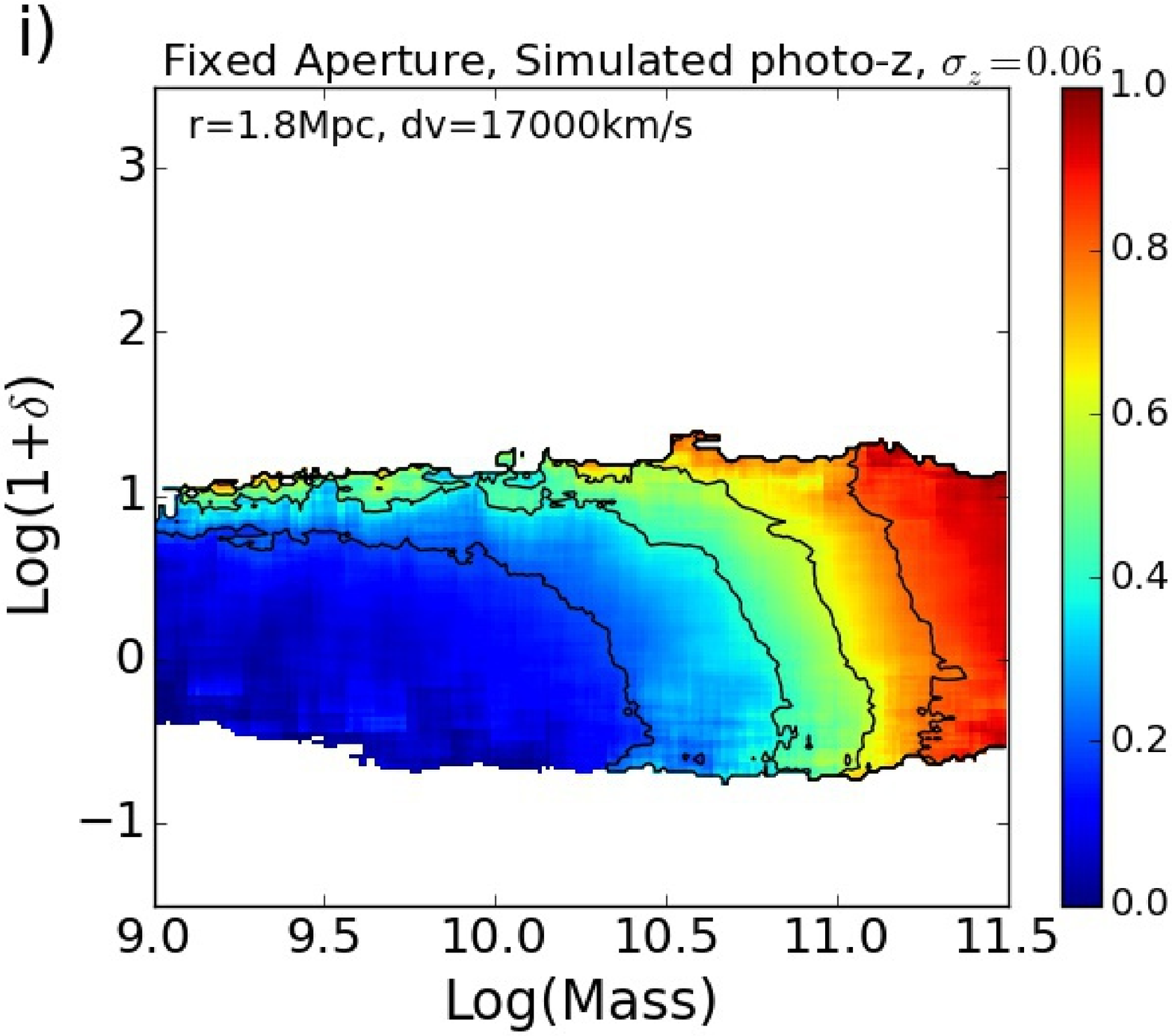}
\end{minipage}
\begin{minipage}[t]{0.33\linewidth}
  \centering  
  \includegraphics[width=.99\linewidth]{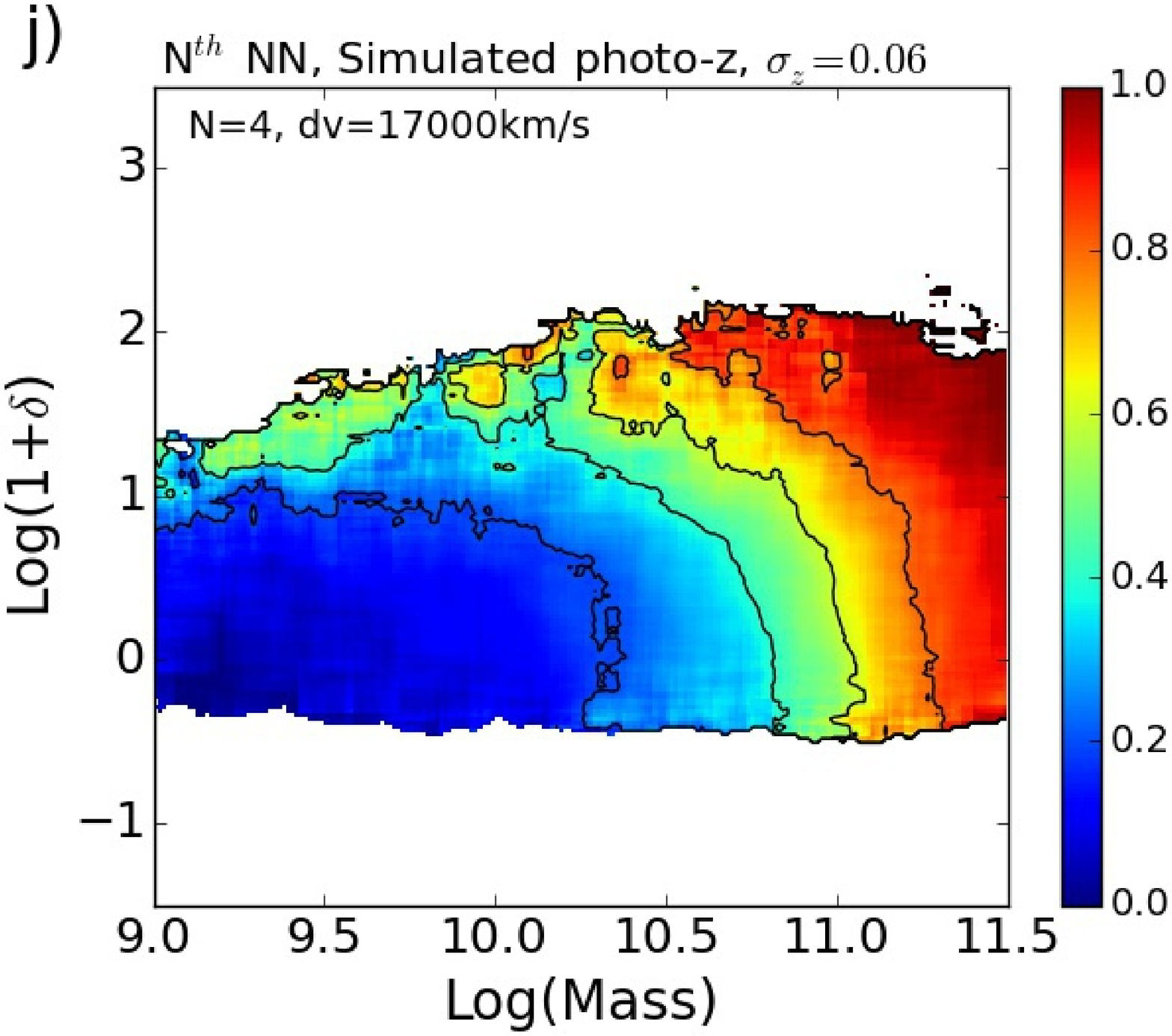}
\end{minipage}
\begin{minipage}[t]{0.33\linewidth}
   \centering 
   \includegraphics[width=.99\linewidth]{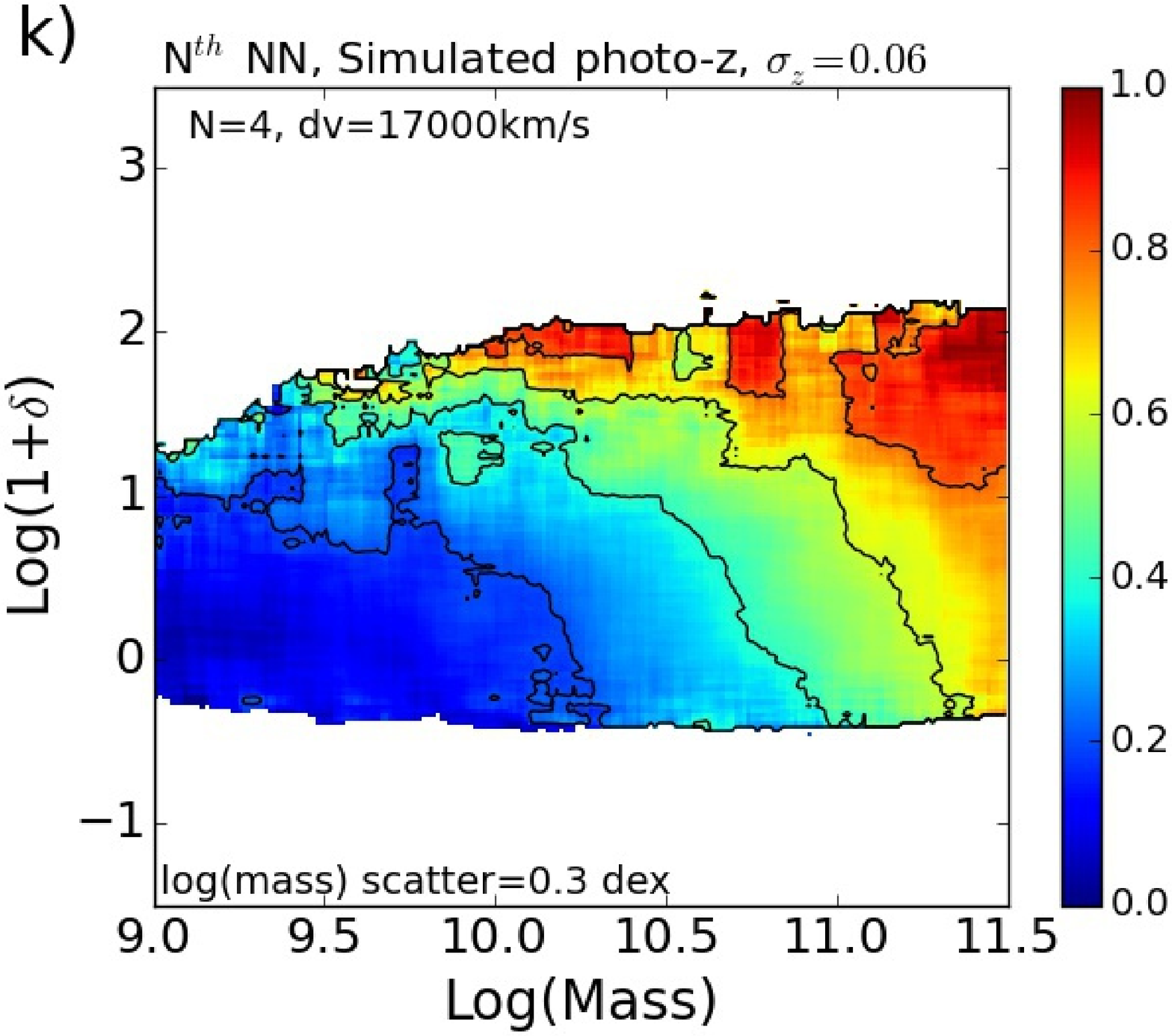}
\end{minipage}
\caption{Shows the red fraction as a function of Log(mass) and environment for the fixed aperture method (left column) and the $N$\textsuperscript{th} nearest neighbour method (middle and right columns). The derived masses are based on the spectroscopic redshifts in the left and middle columns. The masses of the galaxies in the plots in the right hand column have been convolved with a Gaussian distribution with a width of 0.3 dex to model the impact of the photometric redshifts on the mass estimates. The spectroscopic redshift sample is shown in the first row (a and b), the photometric redshift sample is shown in the second row (c, d and e) and simulated photometric redshift samples with redshift uncertainties of $0.04$ and $0.06$ are shown in the third (f, g and h) and fourth rows (i, j and k). The aperture parameter values used are stated on each plot. The contours mark the red fractions: $0.2$, $0.4$, $0.6$ and $0.8$. }
\label{fig:surface_plots}
\end{figure*}

\begin{figure*}
\begin{minipage}[t]{0.49\linewidth}
  \centering
  \includegraphics[width=0.95\linewidth]{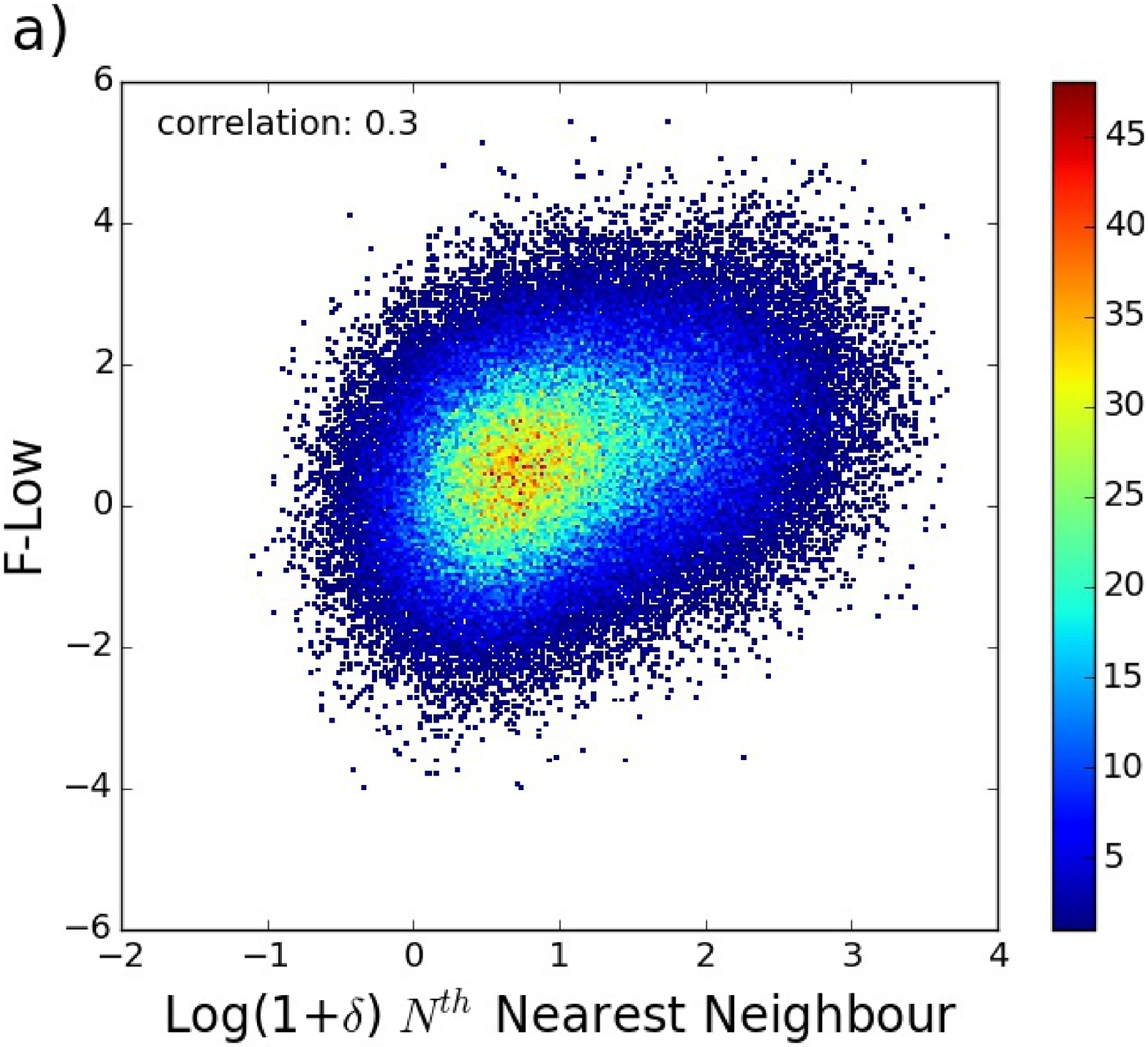}
\end{minipage}
\begin{minipage}[t]{0.49\linewidth}
  \centering
  \includegraphics[width=0.95\linewidth]{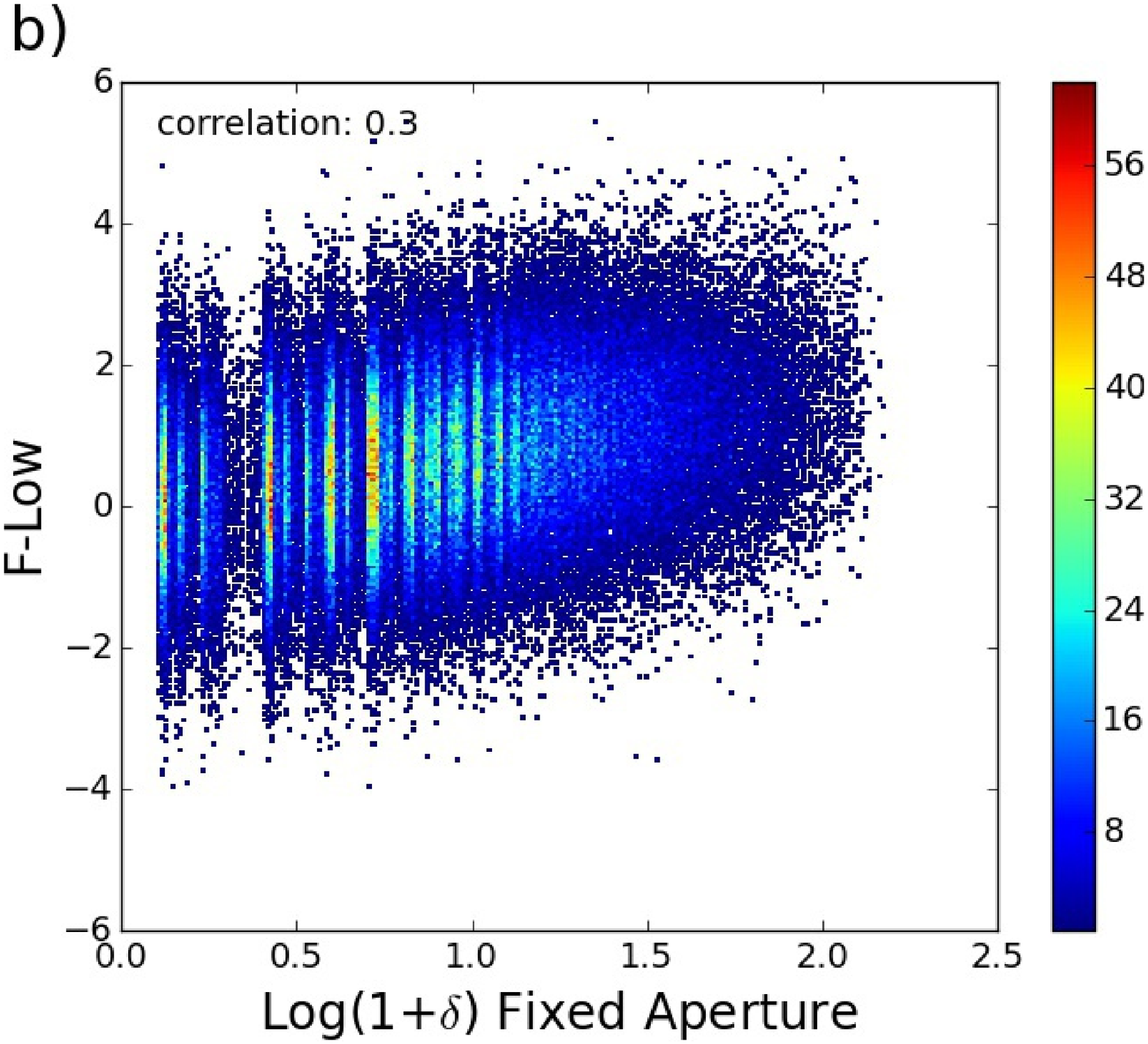}
\end{minipage}
\begin{minipage}[t]{0.49\linewidth}
   \centering 
  \includegraphics[width=0.95\linewidth]{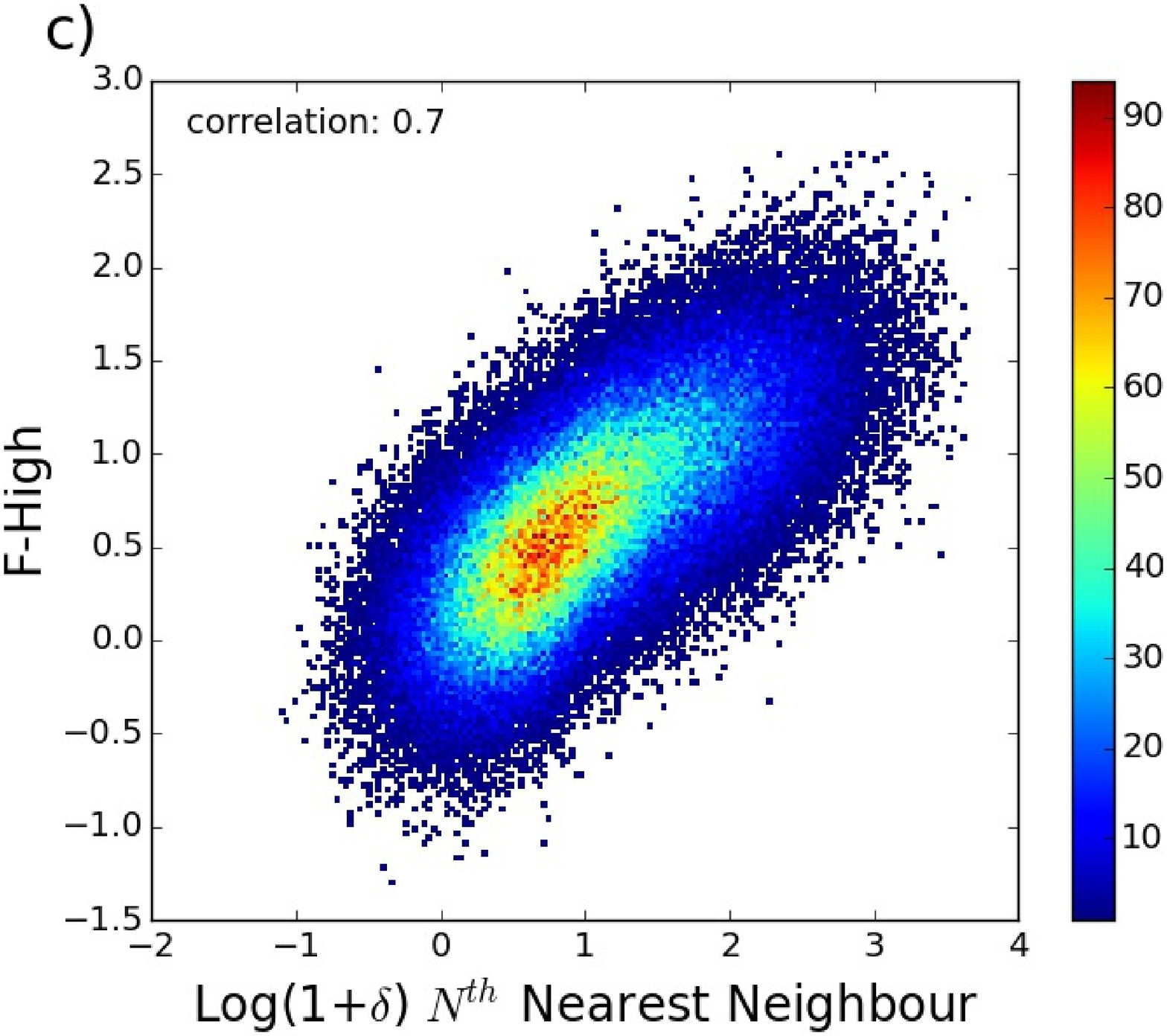}
\end{minipage}
\begin{minipage}[t]{0.49\linewidth}
   \centering 
   \includegraphics[width=0.95\linewidth]{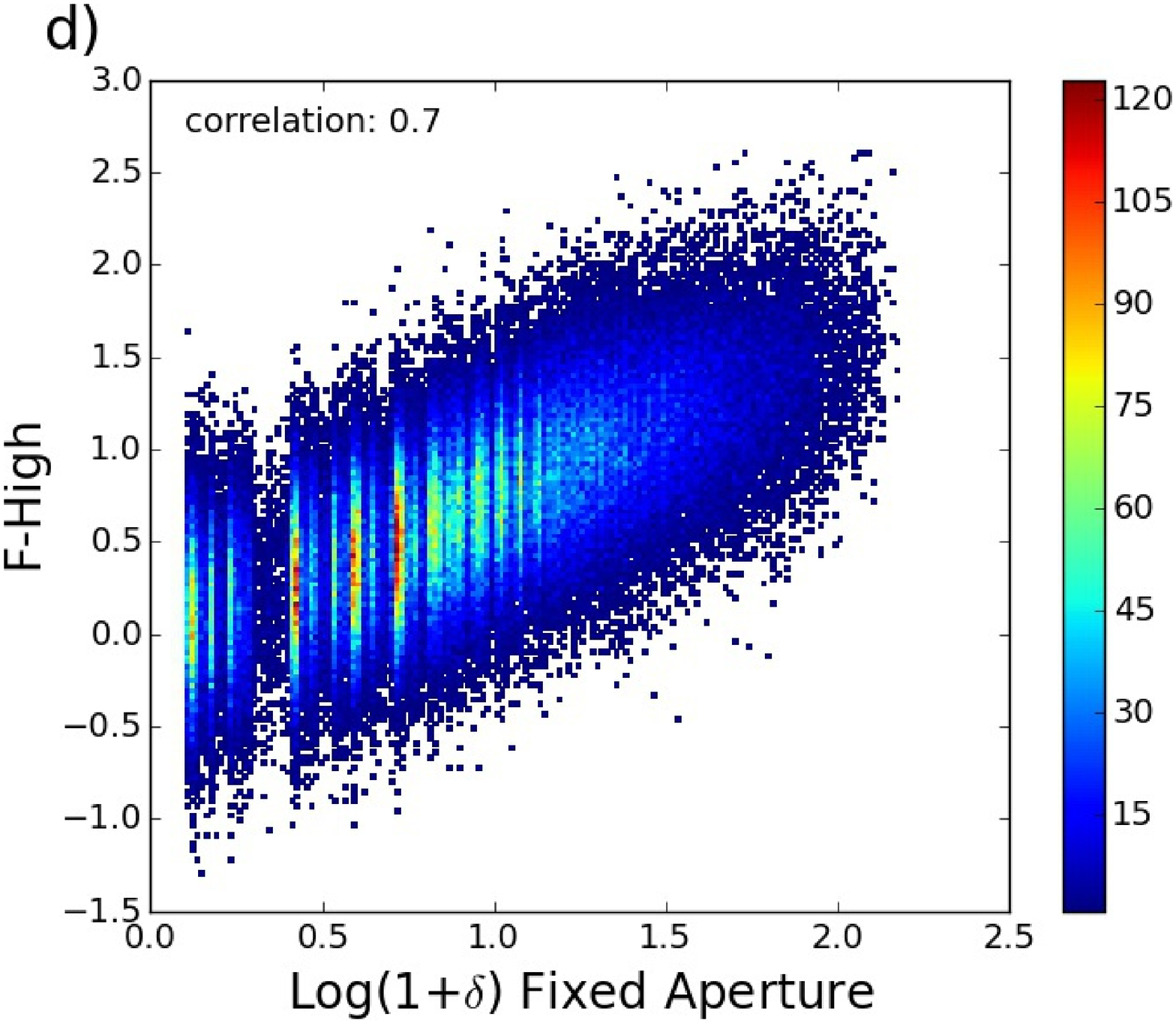}
\end{minipage}
\caption{Shows the spectroscopic benchmark environment measurements for the $N$\textsuperscript{th} nearest neighbour method (a and c) and the fixed aperture method (b and d) versus two sets of faked data. One set of faked data was constructed to have a low correlation (a and b) and the other a high correlation (c and d) with the spectroscopic benchmark environments. The aperture parameters for the spectroscopic benchmark measurements are $N=4$ and $dv=1000\;$km/s for the $N$\textsuperscript{th} nearest neighbour method and $r=1.8\;$Mpc and $dv=1000\;$km/s for the fixed aperture method.}  
\label{fig:property_env_trend}
\end{figure*}

\section{Discussion} \label{sec:discussion}

This section consists of three parts. In Section \ref{subsec:red_fraction} we examine the environmental dependence of the galaxy red fraction using photometric redshifts. In Section \ref{subsec:sample_size} we estimate how much larger, than spectroscopic samples, photometric samples must be to obtain equivalent measurements of environment correlations and in Section \ref{subsec:lit_compare} we briefly compare this work with others in the literature. 

\subsection{Red fraction: mass-environment dependence} \label{subsec:red_fraction}

In the local universe it has been shown that the fraction of red galaxies depends on stellar mass and environment \citep{Baldry2006, Peng2010}. As a case study we investigate the dependence of the galaxy red fraction on environment measured with photometric redshifts.

\cite{Peng2010} showed that in the local universe the red fraction is high for massive galaxies even in low density environments and the red fraction is high for galaxies in very dense environments even for galaxies with low masses. Recent work has shown that this dependence is establish at least at $z=0.7$ \citep{Kovac2014}. 

These results are consistent with the idea of two main channels to quench star formation: ``mass quenching'' and ``environment quenching''. These quenching modes act independently and facilitate the transition of galaxies from the ``blue cloud'' to the ``red sequence''. One candidate for ``mass quenching'' is AGN feedback. In this scenario the central black hole drives energetic outflows which can heat gas and hence inhibit star formation. ``Environment quenching'' could manifest itself when satellite galaxies fall into larger dark matter haloes. Ram pressure stripping or strangulation are physical mechanisms that could be responsible for this quenching mode.

To study the dependence of red fraction on mass and environment as a function of redshift uncertainty, here, we extend the range of masses down to $10^{9.5}$ $M_{\odot}$. To do this we calculate environments for galaxies that are intrinsically fainter than the absolute magnitude cut off used to construct the volume limited samples previously. These faint galaxies however do not become part of the density defining populations. To account for the fact that these galaxies are not detectable through the entire redshift range we determine a volume correction for these galaxy. The volume correction is the ratio of the total volume imposed by the redshift range to the detectable volume of the galaxy with a particular intrinsic brightness. The volume correction is used to weight galaxies (together with the target sampling correction) when representing the whole population of galaxies. 

We now outline how we determined the red fraction for the galaxies in the samples. For each galaxy the SDSS apparent magnitudes: u,g,r,i and z were adjusted to the AB system using small calibration offsets \citep{Eisenstein2006, Blanton2007b}. The (u-g)\textsubscript{Rest} colour for each galaxy was calculated by subtracting the K-correction (k\textsubscript{ug}) obtained from the stellar population fits.

\begin{equation}  \label{eq:rest_color}
(u-g)\textsubscript{Rest} = (u - k\textsubscript{u}) - (g - k\textsubscript{g}) = (u-g) - k\textsubscript{ug}
\end{equation}

The (U-B)\textsubscript{Rest} colour was then calculated using the transformation equation by Lupton (2005) found on the SDSS website:

\begin{equation}
(U-B)\textsubscript{Rest} = 0.8116((u-g)\textsubscript{Rest}) - 0.1313
\end{equation}

The colour-mass diagrams for the spectroscopic and photometric selections are shown in Figure \ref{fig:color_mass}. The bimodality in the galaxy population can clearly be seen in both samples. Following \cite{Peng2010} a dividing line was employed, restated here:

\begin{equation}  \label{eq:color_divide}
(U-B)\textsubscript{Rest} = 1.10 + 0.075log( m / 10^{10} M_{\odot}) - 0.18z
\end{equation}

The galaxies with a (U-B)\textsubscript{Rest} colour larger than the dividing value, for the mass and redshift, defined by equation (\ref{eq:color_divide}) were considered to be red and the galaxies with a (U-B)\textsubscript{Rest} colour smaller than the dividing value were marked as blue. This binary division facilitated a simple way to compute the fraction of red galaxies in any bin of mass and environment. The red fraction we used was simply the number of red galaxies, weighted by the target sampling rate and volume corrections, in the bin divided by the total weighted number of galaxies in the bin. 

Figure \ref{fig:surface_plots} shows plots of the galaxy red fraction as a function of log(mass) and environment for the fixed aperture method (left column) and the $N$\textsuperscript{th} nearest neighbour method (middle and right columns). In the left and middle columns the derived masses are based on the spectroscopic redshifts. In the right column the derived masses are also based on the spectroscopic redshifts but the masses have been convolved with a Gaussian distribution with a width of 0.3 dex. This is to account for errors in the mass estimates due to the photometric redshifts \citep[e.g.][]{Taylor2009, Fossati2015}. The SDSS spectroscopic redshift sample is shown in the first row (a and b), the SDSS photometric redshift sample is shown in the second row (c, d and e) and simulated photometric redshift samples with redshift uncertainties of $0.4$ and $0.6$ are shown in the third (f, g, h) and fourth (i, j, k) rows respectively. The aperture parameter values used are close to the optimal values determined in Section \ref{subsec:corr_vs_parms_uncertainty} and are stated on each plot. The contours mark the red fractions: $0.2$, $0.4$, $0.6$ and $0.8$

The first row of plots (a and b) for the spectroscopic redshift sample show that the the red fraction increases with both log(mass) and environment. There are clear contours of constant red fraction covering several orders of magnitude in log(mass) and environment for both the $N$\textsuperscript{th} nearest neighbour method (b) and the fixed aperture method (a). We reproduce the result presented by \cite{Peng2010} using spectroscopic redshifts using the $N$\textsuperscript{th} nearest neighbour method and in addition the fixed aperture method. The red fraction is high for massive galaxies even in low density environments and the effect of environment is most important for galaxies with low masses.

The second row of plots (c, d and e) for the SDSS photometric redshift sample also show contours of constant red fraction, albeit some deterioration. The information to separate the red fraction by log(mass) and environment is still present in the SDSS photometric redshift measurements. Galaxies with smaller masses than in the spectroscopic sample are found in the photometric samples because some galaxies with small spectroscopic redshifts ($z<0.02$) are scattered into the sample because of the uncertain redshift measurements. 

As noted earlier (in Section \ref{subsec:spec_vs_photo_env}) for fixed aperture parameters the range of environments for the photometric measurements are smaller than the range for the spectroscopic measurements. The reduction in dynamic range is particularly evident when comparing the red fraction surfaces for the SDSS spectroscopic and photometric redshift samples for the $N$\textsuperscript{th} nearest neighbour method (plots b and d). The red fraction surface for the SDSS photometric redshift sample is shrunk along the environment axis compared with the spectroscopic redshift sample. The dynamic ranges for the SDSS photometric redshift and spectroscopic redshift samples for the fixed aperture method (plots a and c) are however similar. In this method the much larger velocity cut (x6) adopted for the SDSS photometric redshift (plot c) is able to compensate for the reduction in dynamic range due to the photometric redshifts.

Increasing the redshift uncertainty to $0.4$ and $0.6$ (third and fourth rows) results in more deteriorated red fraction surfaces and smaller dynamic ranges. Nevertheless, particularly for the $N$\textsuperscript{th} nearest neighbour method the red fraction contours still behave in a similar manner to the SDSS spectroscopic sample. The red fraction dependence on galaxy environment does not break down even for samples with large redshift uncertainties. The $N$\textsuperscript{th} nearest neighbour method appears to fare better than the fixed aperture method as the redshift uncertainty increases. This is because the $N$\textsuperscript{th} nearest neighbour method tends to probe a larger dynamic range of environments than the fixed aperture method. 

To model the impact of the photometric redshifts on the mass estimates we have applied an additional uncertainty of 0.3 dex to the mass estimates in the right hand column. Both the middle and right hand columns employ the $N$\textsuperscript{th} nearest neighbour method so comparing the red fraction surfaces in these columns shows the impact of the additional mass uncertainty. The additional uncertainty in the mass estimates tends to mix up the red and blue galaxies resulting in more homogenous red fractions. However even with this additional uncertainty the general trends we described above are still present. 

\begin{figure}
\includegraphics[width=0.95\linewidth]{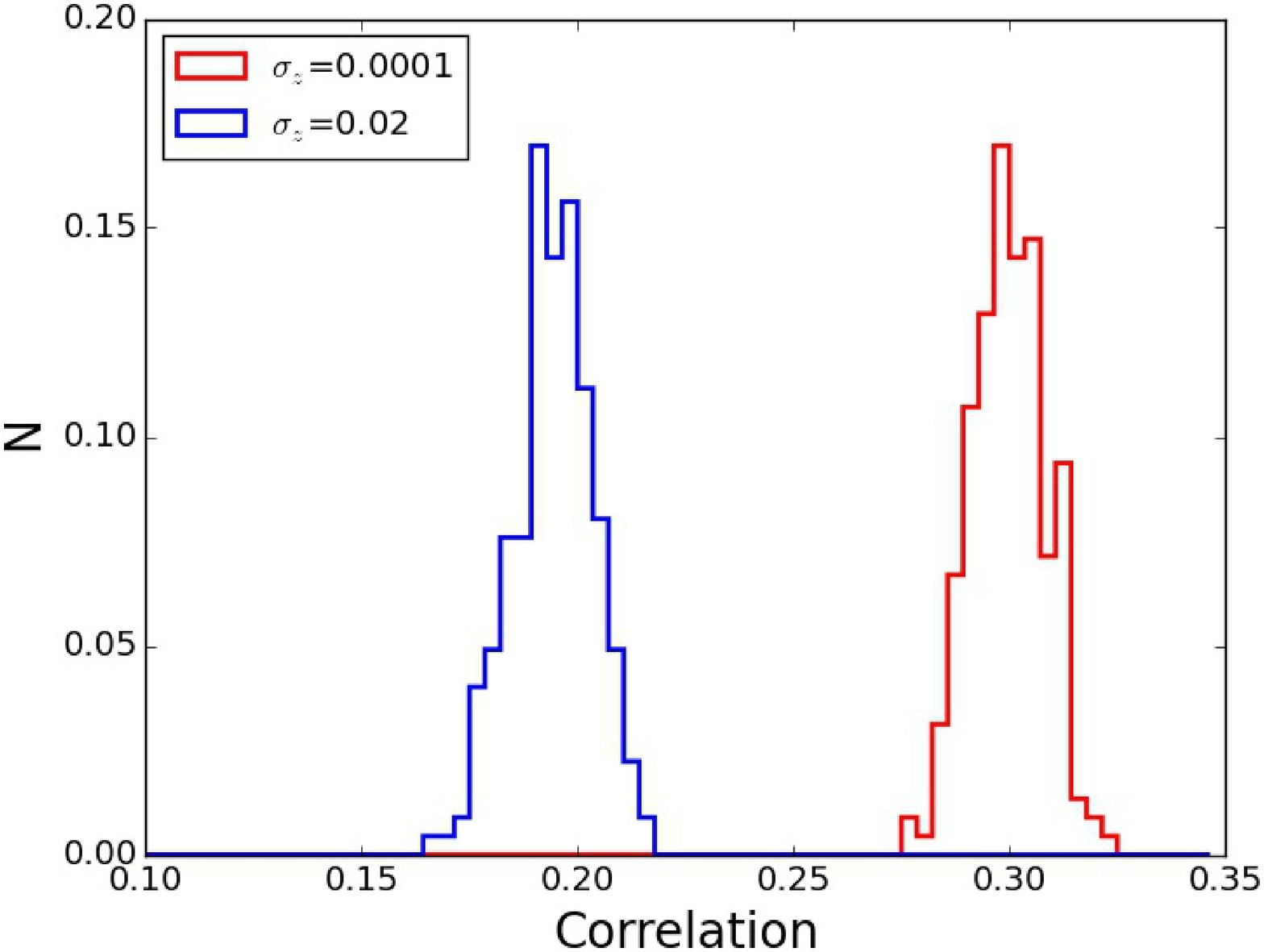}
\caption{Histograms of the Spearman rank correlation coefficients between the optimal $N$\textsuperscript{th} nearest neighbour environments and the faked data (constructed to have a correlation of $0.3$ with the spectroscopic benchmark environments) for $224$ realizations of the spectroscopic catalogue (red) with $\sigma_z=0.0001$ and $224$ realizations of the photometric catalogue (blue) with $\sigma_z=0.02$. The Spearman rank correlation coefficient is calculated using $10,000$ objects from each realization.}  
\label{fig:nn_correlation_hist}
\end{figure}

\begin{figure*}
\begin{minipage}[t]{0.49\linewidth}
   \centering 
  \includegraphics[width=0.95\linewidth]{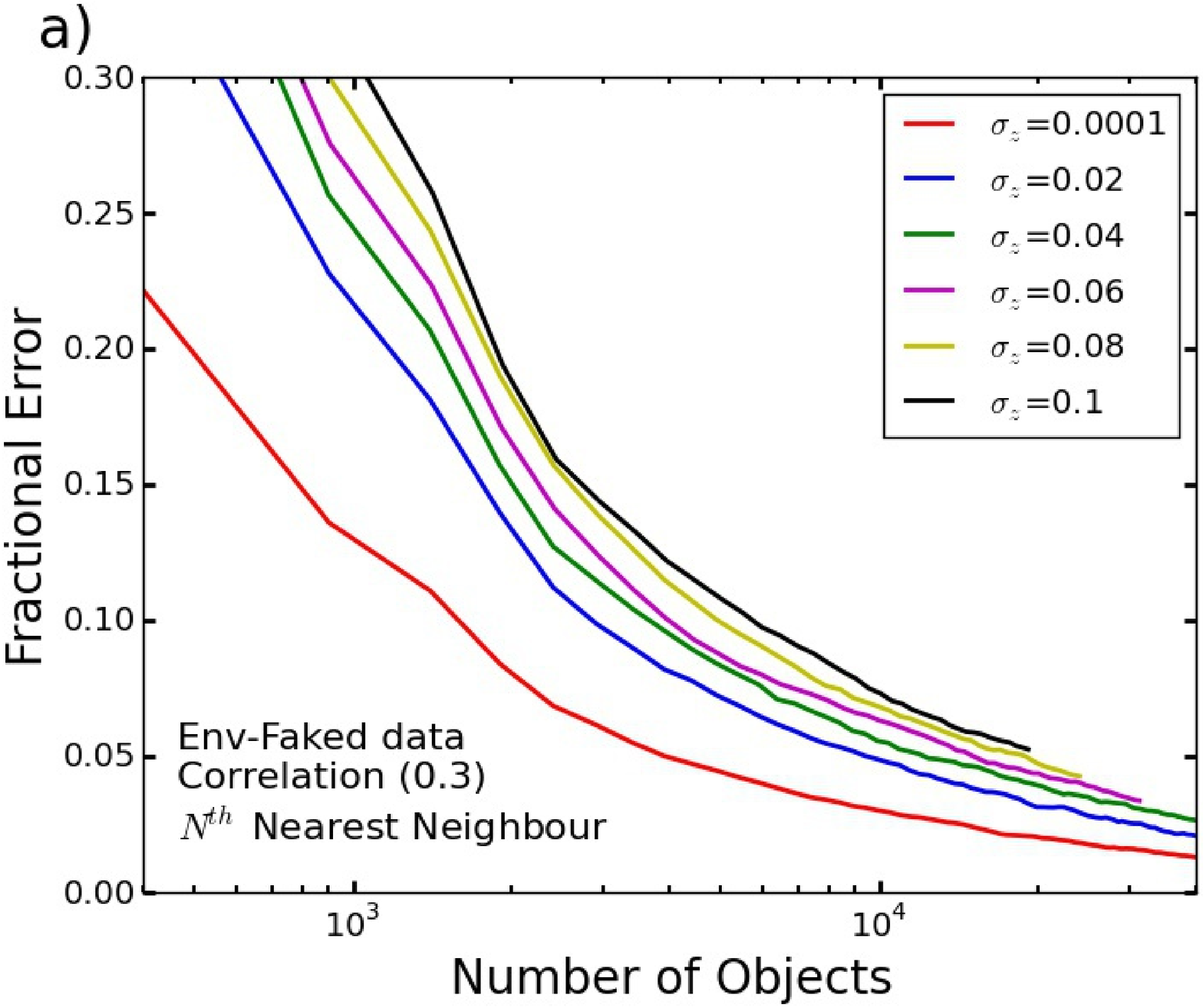}
\end{minipage}
\begin{minipage}[t]{0.49\linewidth}
   \centering 
   \includegraphics[width=0.95\linewidth]{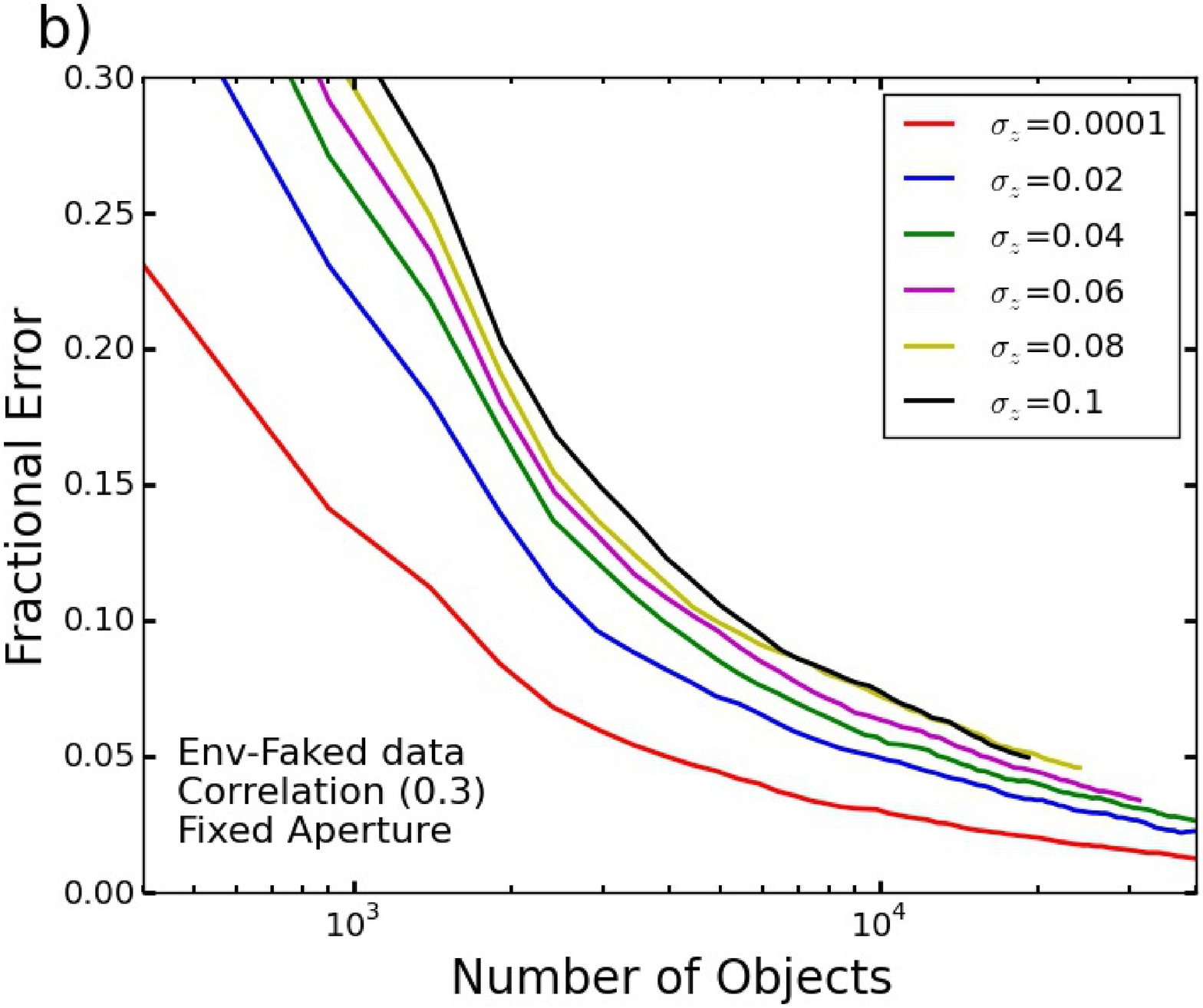}
\end{minipage}
\begin{minipage}[t]{0.49\linewidth}
   \centering 
  \includegraphics[width=0.95\linewidth]{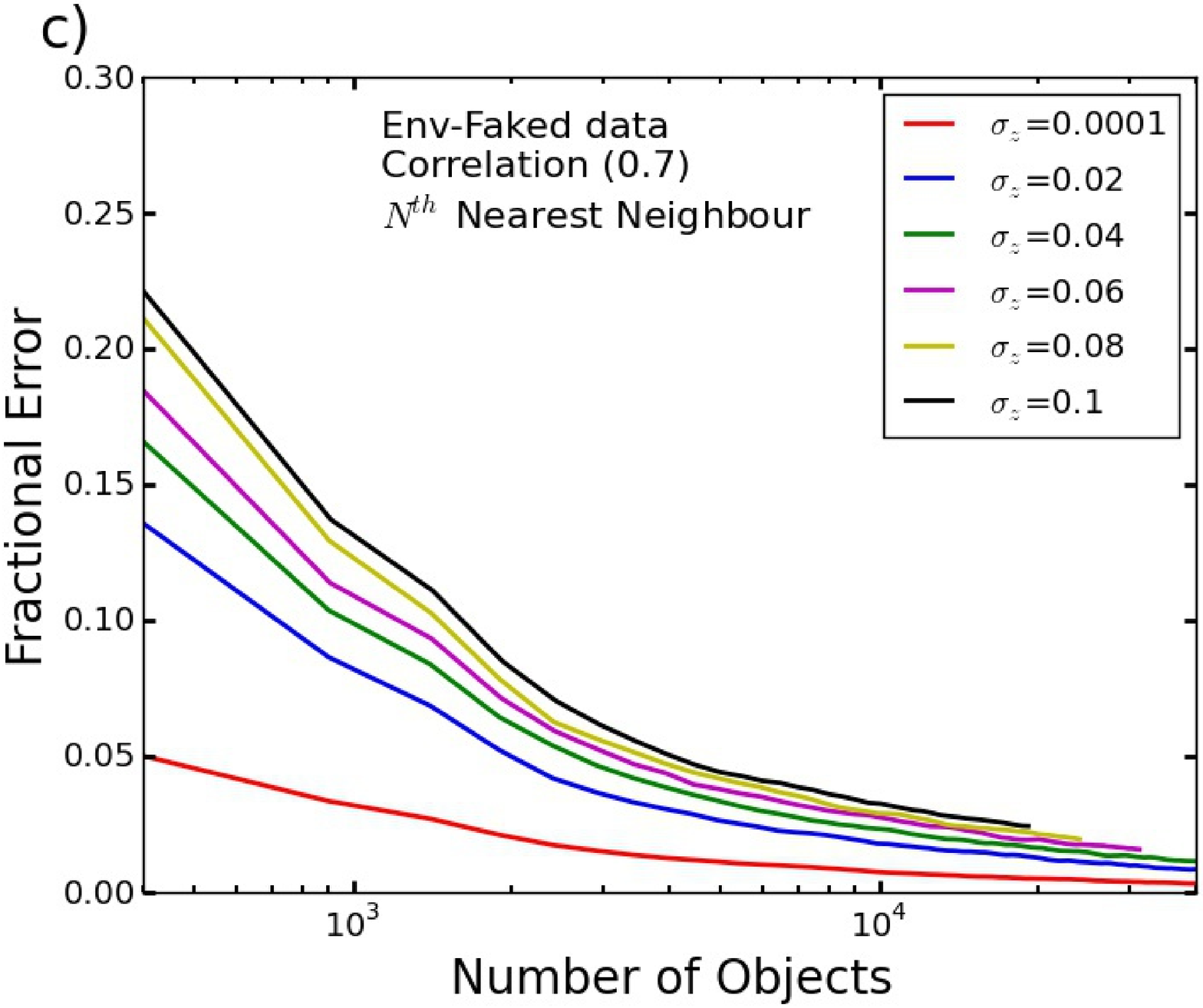}
\end{minipage}
\begin{minipage}[t]{0.49\linewidth}
   \centering 
   \includegraphics[width=0.95\linewidth]{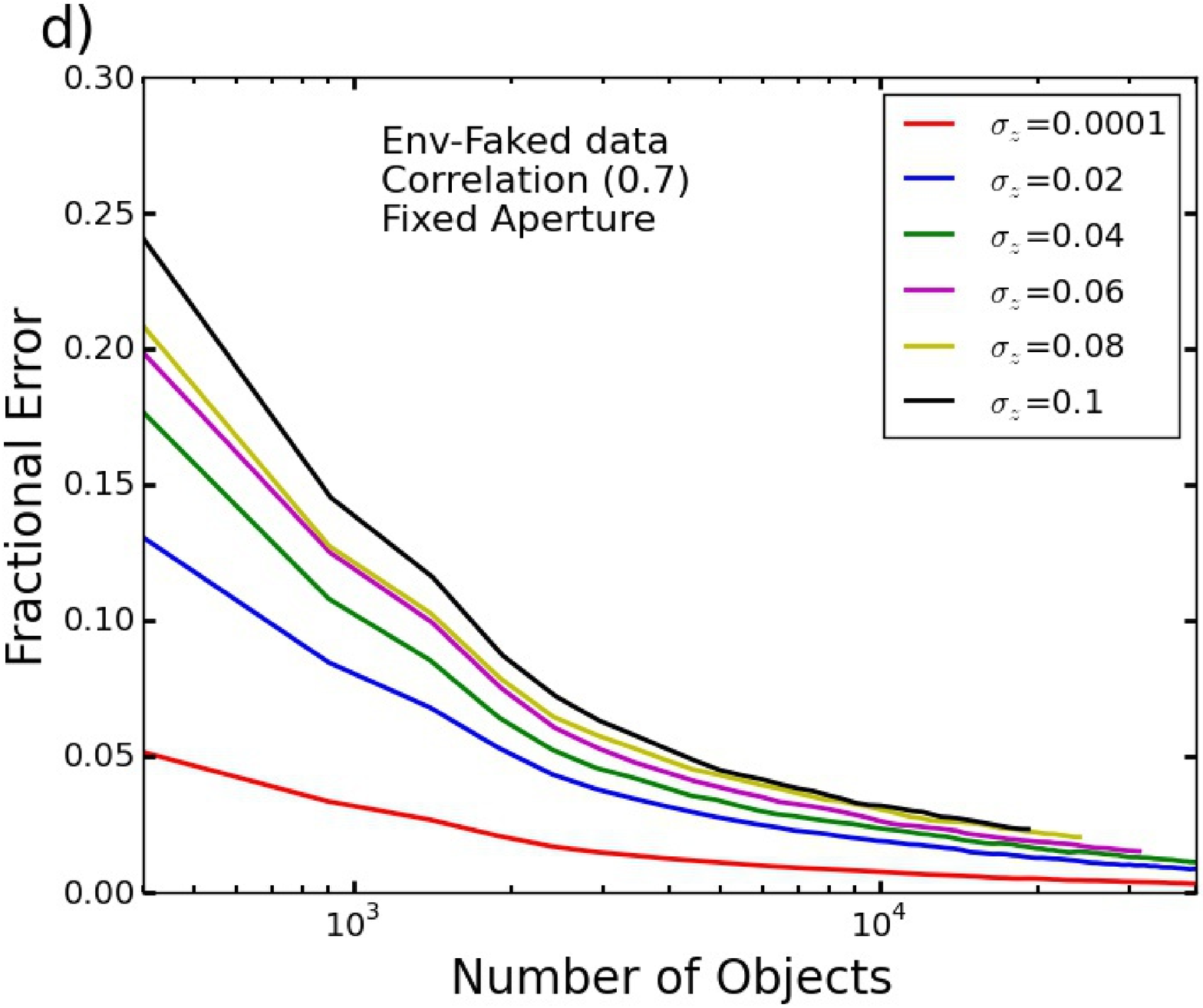}
\end{minipage}
\caption{Shows the fractional error of the Spearman rank correlation coefficient between the environments and the faked data where the correlation of the faked data to the spectroscopic benchmark environments was constructed to be 0.3 (a and b) and 0.7 (c and d) for the $N$\textsuperscript{th} nearest neighbour method (a and c) and the fixed aperture method (b and d) as a function of the number of objects in the sample for six redshift uncertainties: $\sigma_z=0.0001,0.02,0.04,0.06,0.08$ and $0.1$. }  
\label{fig:fractional_error_number}
\end{figure*}

\begin{figure*}
\begin{minipage}[t]{0.49\linewidth}
   \centering 
  \includegraphics[width=0.95\linewidth]{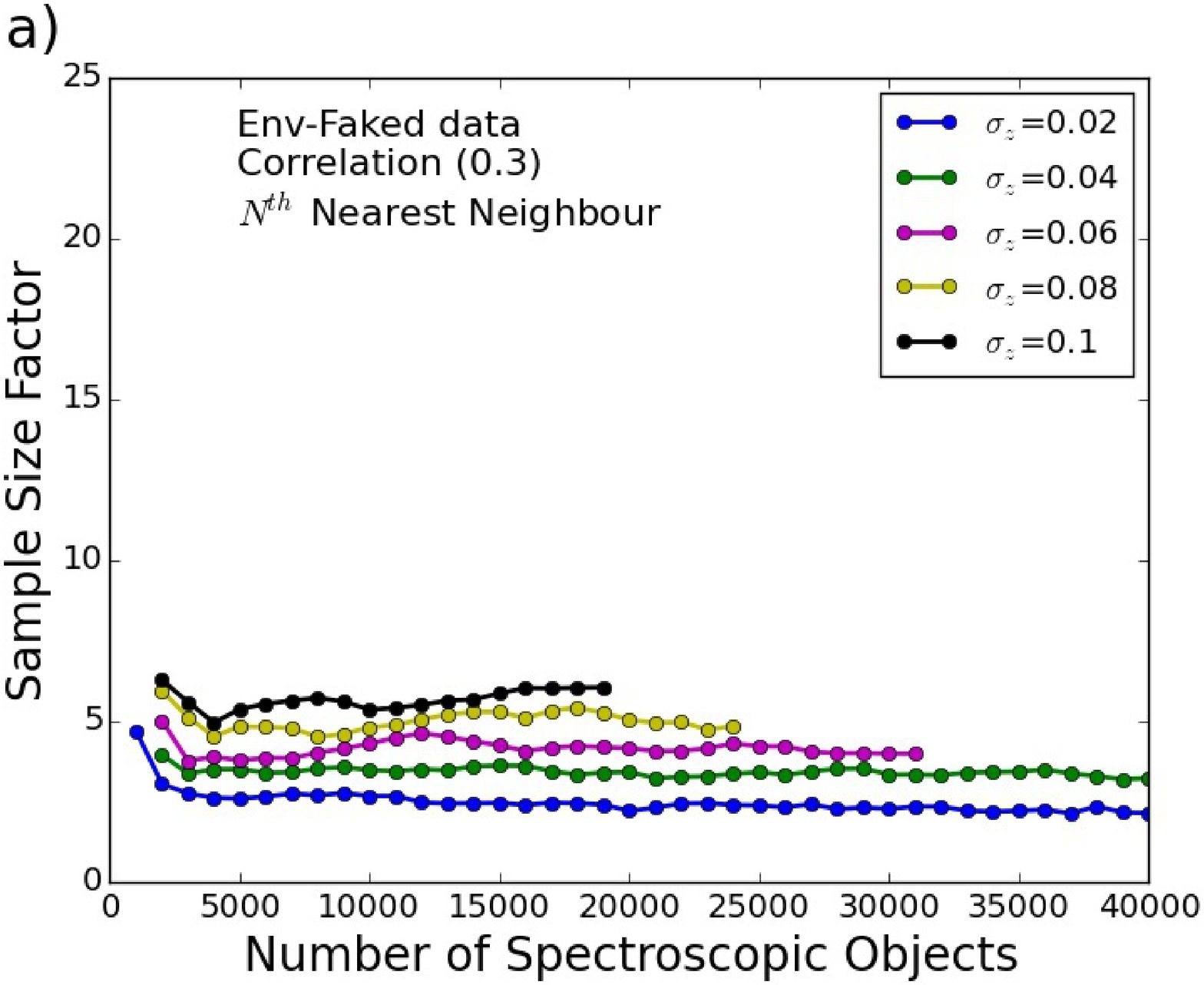}
\end{minipage}
\begin{minipage}[t]{0.49\linewidth}
   \centering 
   \includegraphics[width=0.95\linewidth]{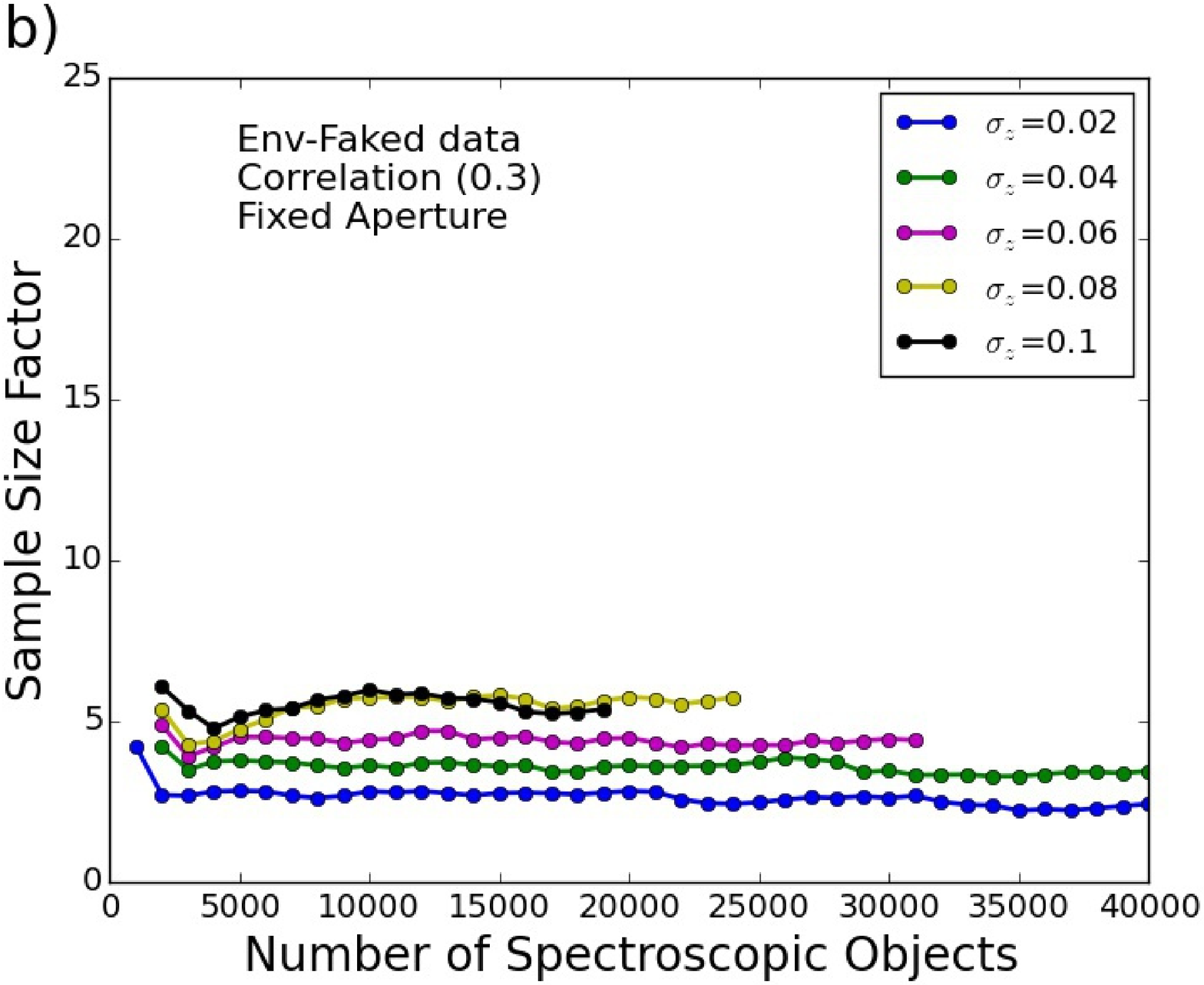}
\end{minipage}
\begin{minipage}[t]{0.49\linewidth}
   \centering 
  \includegraphics[width=0.95\linewidth]{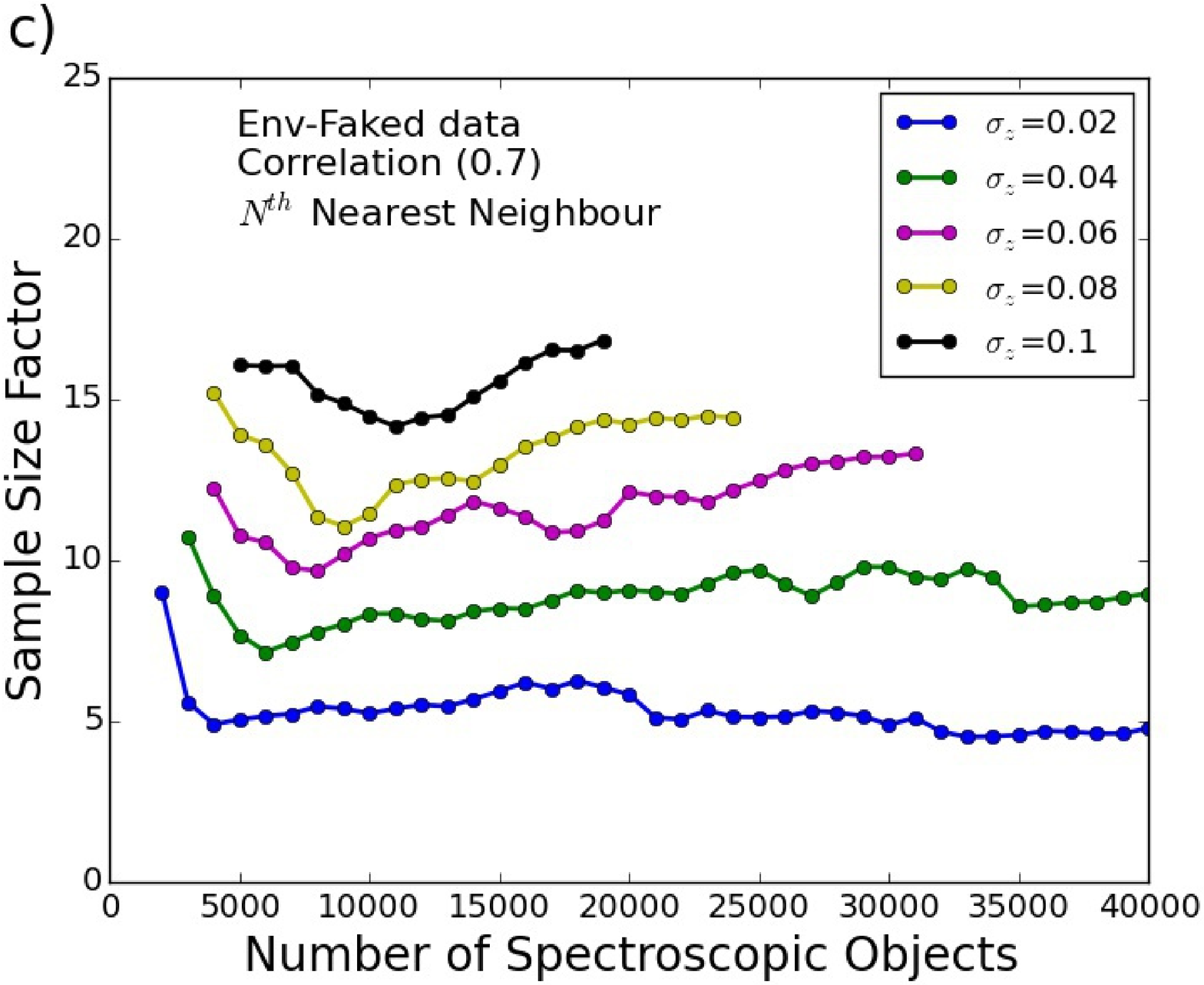}
\end{minipage}
\begin{minipage}[t]{0.49\linewidth}
   \centering 
   \includegraphics[width=0.95\linewidth]{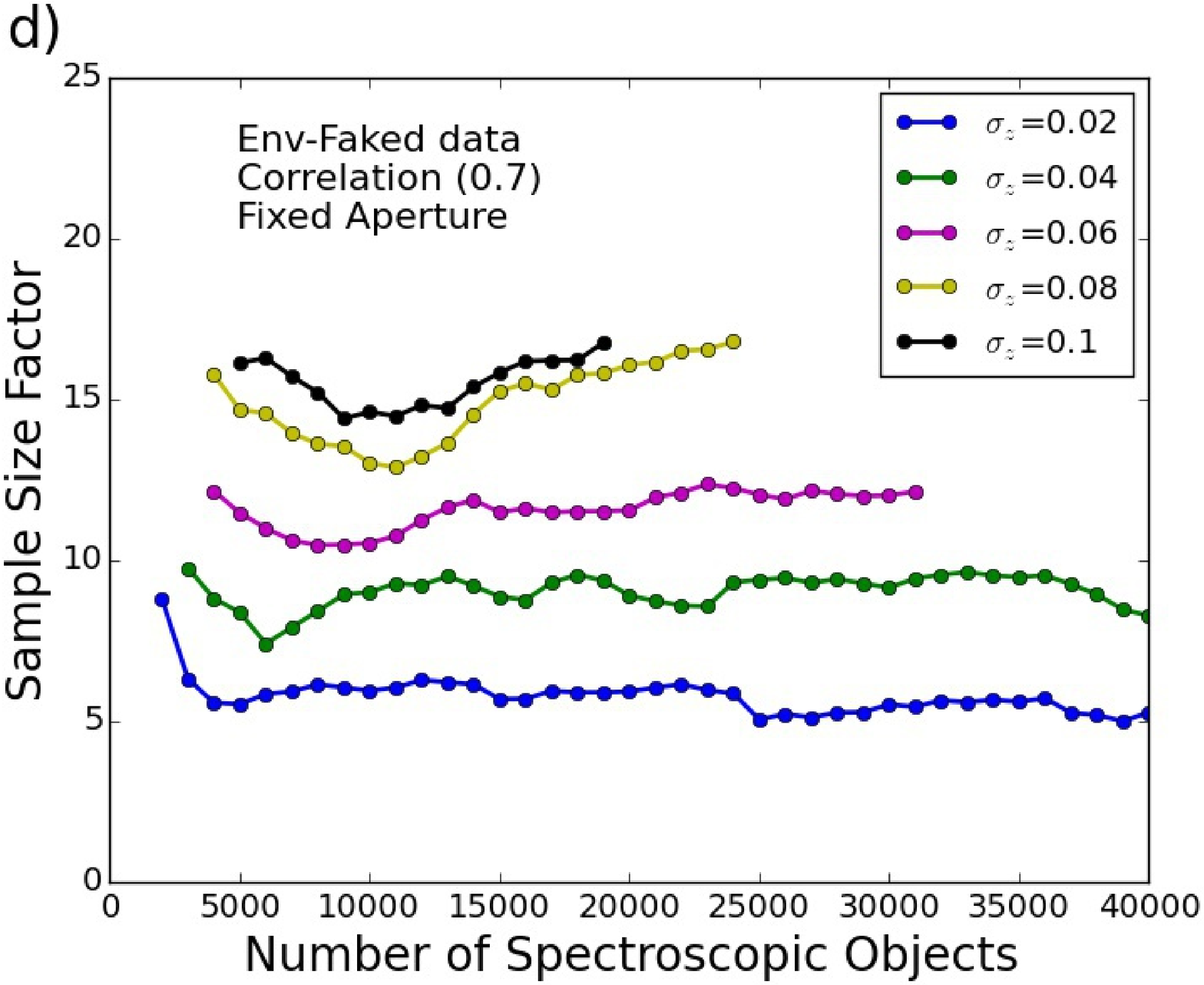}
\end{minipage}
\caption{Shows the number of times larger than the spectroscopic sample the photometric sample must be to obtain equivalent fractional errors for the Spearman rank correlation coefficient between environment and faked data where the correlation of the faked data to the spectroscopic benchmark environments was constructed to be 0.3 (a and b) and 0.7 (c and d) for the $N$\textsuperscript{th} nearest neighbour method (a and c) and the fixed aperture method (b and d) as a function of the number of spectroscopic objects for five redshift uncertainties: $\sigma_z=0.02,0.04,0.06,0.08$ and $0.1$. }  
\label{fig:sample_size_factor}
\end{figure*}

\subsection{Photometric sample size for environment correlations} \label{subsec:sample_size}

One of the main goals of this paper is to quantify the impact of uncertain redshifts on measurements of environment trends. The questions we consider here are: (i) How many times larger than the spectroscopic sample must the photometric sample be to measure environment correlations to the same accuracy? (ii) How does this factor depend on the trend correlation with spectroscopic measurements? (iii) How does this factor vary with redshift uncertainty?

To calibrate our measurements to enable comparisons between the results obtained with different methods, aperture parameter values and redshift uncertainties we constructed two sets of fake data which we label: F-Low and F-High. The fake data represents some galaxy property that increases monotonically with environment. More specifically the fake properties were constructed to be linearly related to both the $N$\textsuperscript{th} nearest neighbour spectroscopic benchmark environments ($N=4$ and $dv=1000\;$km/s) and the fixed aperture spectroscopic benchmark environments ($r=1.8\;$Mpc and $dv=1000\;$km/s) with a fixed level of Gaussian scatter. F-Low was constructed to have a Spearman rank correlation coefficient of $0.3$ whereas F-High was constructed to have a Spearman rank correlation coefficient of $0.7$ with both sets of spectroscopic benchmark environments. Two important simplification to note are that the value of the fake property does not change with the value of galaxy's redshift and no mass dependence is built into the fake properties. Figure \ref{fig:property_env_trend} shows plots of the fake properties as a function of the spectroscopic benchmark environments for the $N$\textsuperscript{th} nearest neighbour (a and c) and fixed aperture (b and d) methods.

We calculated the Spearman rank correlation coefficient between the environments and the fake properties as a function of redshift uncertainty and the aperture parameters. We made figures similar to Figures \ref{fig:env_nnn_benchmark_correlation} and \ref{fig:env_conical_benchmark_correlation} and then found the aperture parameters that result in the largest correlations for each redshift uncertainty. The parameters for each redshift uncertainty are shown in Table \ref{table:aperture_parameters_realizations}. The second and third columns of the table show the parameters for the $N$\textsuperscript{th} nearest neighbour method. The fourth and fifth columns of the table show the parameters for the fixed aperture method. 

Next we generated $224$ different realizations of the redshift catalogues. To obtain different realizations of the spectroscopic catalogue we assigned a nominal redshift uncertainty of $0.0001$ \citep[e.g.][]{Cooper2005} to the spectroscopic redshifts. For each realization we followed the same procedure as before but used the parameters specified in Table \ref{table:aperture_parameters_realizations}. We calculated the Spearman rank correlation coefficient between the environments and the fake galaxy properties as a function of the number of objects in the samples.

\begin{table} 
\begin{minipage}{90mm}
\caption{Aperture parameters to optimize the correlation between 
the faked data and environment for the $N^{th}$ nearest neighbour 
method and the fixed aperture method for a range of redshift uncertainties.}
\label{table:aperture_parameters_realizations}
 \begin{tabular}{@{}lcccc}
    \hline
   $\sigma_{z}$ & $N$  & $dv$ (km/s) & $r$ (Mpc) & $dv$ (km/s)  \\ 
   \hline
    $0.0001$ & 4 & 1000 & 1.8 & 1000 \\ 
    $0.02$ & 5 & 11000 & 1.6 & 8000  \\ 
    $0.04$ & 7 & 13000 & 1.7 & 13000 \\ 
    $0.06$ & 6 & 12000 & 2.0 & 14000 \\ 
    $0.08$ & 6 & 16000 & 2.0 & 17000 \\ 
    $0.1$ & 9 & 20000 & 2.0 & 20000  \\ 
    \hline
    \end{tabular}
\end{minipage}
The second and third columns are the optimal parameter values for the $N^{th}$ nearest neighbour method and the fourth and fifth colums are the optimal parameter values for the fixed aperture method.
\end{table}

Figure \ref{fig:nn_correlation_hist} shows histograms of the Spearman rank correlation coefficients between environment and F-Low for $224$ realizations of the spectroscopic catalogue (red) and the simulated photometric catalogue (blue) with $\sigma_z = 0.02$. In this example we calculated the Spearman rank correlation coefficient using $10,000$ randomly selected objects from each realization. The plot shows that a larger redshift uncertainty leads to a smaller Spearman rank correlation coefficient. The median correlations are $0.30$ and $0.70$ for the spectroscopic realizations and $0.19$ and $0.45$ for the photometric realizations (with $\sigma_{z}=0.02$) for F-Low and F-High respectively. In addition the correlation distribution is slightly wider for the photometric realizations. The standard deviation of the Spearman rank correlation coefficients of the spectroscopic realizations are $0.0085$ and $0.005$ and the standard deviation of the photometric realizations are $0.0093$ and $0.0085$ for F-Low and F-High respectively. We compute the fractional error for each set of realizations by taking the ratio of standard deviation of the distribution and the median of the distribution. The fractional errors for the spectroscopic realizations are $0.029$ and $0.0072$ and for the photometric realizations the fractional errors are $0.048$ and $0.019$ for F-Low and F-High respectively.

Figure \ref{fig:fractional_error_number} shows four plots of the fractional error as a function of the number of objects in the sample for six different redshift uncertainties: $\sigma_z=$ $0.0001$, $0.02$, $0.04$, $0.06$, $0.08$ and $0.1$. The top row of plots show the fractional error for the correlation between environment and F-Low. The bottom row of plots show the fractional error for the correlation between environment and F-High. The left hand plots are for the $N$\textsuperscript{th} nearest neighbour method and the right hand plots are for the fixed aperture method. 

There are several points to note about the plots. Firstly the plots show that the fractional error decreases as the number of objects in the sample increases. Secondly for a particular environment correlation (strong or weak) and and for a particular method, larger redshift uncertainties have larger fractional errors. Thirdly for a particular correlation (strong or weak) the fractional error for the spectroscopic redshift uncertainty as a function of the number of objects in the sample is approximately the same for the $N$\textsuperscript{th} nearest neighbour and fixed aperture methods. Fourthly by comparing plots a) and b) for the weak correlation to plots c) and d) for the strong correlation it is clear that a larger spectroscopic trend correlation leads to a smaller fractional error for the same redshift uncertainty and this is true for both the $N$\textsuperscript{th} nearest neighbour method and the fixed aperture method. 

We used the fractional error curves to obtain the number of spectroscopic objects required to achieve a particular fractional error. In this way we obtained the ratio of the number of photometric objects to the number of spectroscopic objects - the sample size factor. i.e how many times larger the photometric sample must be relative to the spectroscopic sample to achieve the same fraction error.

Figure \ref{fig:sample_size_factor} shows four plots of the sample size factor as a function of the number of spectroscopic objects for five different redshift uncertainties: $\sigma_z=$ $0.02$, $0.04$, $0.06$, $0.08$ and $0.1$. The arrangement of the plots is the same as for Figure \ref{fig:fractional_error_number}. The top row of plots are for the correlation between environment and F-Low. The bottom row of plots are for the correlation between environment and F-High. The left hand plots are for the $N$\textsuperscript{th} nearest neighbour method and the right hand plots are for the fixed aperture method. The sample size factor is fairly insensitive to the number of objects in the sample as the curves are nearly horizontal. The sample size factor increases with redshift uncertainty. 

At a redshift uncertainty of $0.1$ approximately $6-16$ times more photometric objects than spectroscopic objects are required to measure environment correlations with equivalent fractional errors for spectroscopic correlations of $0.3-0.7$ respectively. Galaxy properties that have a strong spectroscopic environment dependence are easier to detect in absolute terms with photometry than properties that correlate weakly with the spectroscopic environment measurements. However we calculate the sample size factor required to obtain a photometric measurement of the correlation with the \textit{same} fractional error as was obtained with spectroscopy. We find that galaxy properties that have a strong spectroscopic environment dependence actually require a larger sample size factor to obtain the same fractional error than galaxy properties that depend weakly on spectroscopic environment. This is because spectroscopic measurements of strong correlations have a small error whereas the uncertain redshifts still contribute a large error to photometric measurements. Even more photometric objects are therefore required to bring the photometric error down to a competitive level. 

We note that the sample size factors for the $N$\textsuperscript{th} nearest neighbour method and the fixed aperture method are very similar and so there is no preferred method. Excluding considerations of the dependence of galaxy properties on mass we therefore find that relatively small sample size factors are required to make equivalent detections of the correlations between galaxy properties and environment.

\subsection{Brief comparison with literature} \label{subsec:lit_compare}

We complete this section with a few paragraphs to compare this work with the literature. \cite{Cooper2005} presented the most extensive range of tests to date regarding measures of galaxy environment and the impact of photometric redshifts. There are a number of important differences between the work by \cite{Cooper2005} and this work. Using mock galaxy catalogues to mimic galaxy surveys Cooper et al. investigated a higher and larger redshift range: $0.7<z<1.4$ and considered a 1 squared degree field that contained {\raise.17ex\hbox{$\scriptstyle\sim$}}22,000 galaxies. We have studied a low redshift range: $0.02<z<0.085$ using data extracted from the SDSS DR7 which covers nearly 7000 squared degrees and includes {\raise.17ex\hbox{$\scriptstyle\sim$}}150,000 galaxies. \cite{Cooper2005} concluded that photometric redshift measurements with errors $>0.02$ severely limit studies of the local galaxy environment. Nevertheless they found a Spearman rank correlation coefficient of {\raise.17ex\hbox{$\scriptstyle\sim$}}0.3 between real space and photometric measures of galaxy environment for $\sigma_z=0.05$ \citep[see Table 3 in][]{Cooper2005}. This is non-zero and so suggests that there actually is still a measurable environment signal, albeit a weak one. We point out that \cite{Cooper2005} employed velocity cuts $<2000\;$km/s in their analysis and argued that deeper apertures would not measure the \textit{local} environment. We build on the work of \cite{Cooper2005} by explicitly testing deeper apertures and use velocity cuts of $1000-20,000\;$km/s. We show that the correlation between environments based on photometric redshifts and a set of spectroscopic benchmark environments depends on the aperture parameters. To obtain the optimal correlation we find that the aperture depth must increase with redshift uncertainty. We argue that using deeper apertures is a measurement integrated over the large scale structure (along the line of sight) however it is also a proxy for the local galaxy environment in photometric surveys. Referring to Figure \ref{fig:optimal_aperture_parms} in this work we estimate the optimal Spearman rank correlation coefficient at $\sigma_{z}=0.05$ to be {\raise.17ex\hbox{$\scriptstyle\sim$}}0.5. This is larger than what was reported in \cite{Cooper2005}. However the difference can be explained by referring to plots g) and h) in Figures 8 or 9. Adopting a non-optimal velocity cut (i.e. {\raise.17ex\hbox{$\scriptstyle\sim$}}$2000\;$km/s) as was chosen in \cite{Cooper2005} results in a degraded Spearman rank correlation coefficient.

We also compare our work with a more recent study by \cite{Fossati2015}. This study is based on semi-analytic models of galaxy formation and investigates a redshift snapshot at $z${\raise.17ex\hbox{$\scriptstyle\sim$}}$1$. Figure 4 in \cite{Fossati2015} shows the galaxy density measured within cylindrical apertures with radii of $0.75\;$Mpc for high-res measurements (i.e. simulating high quality spectroscopic redshift measurements) and for photometric redshift measurements (obtained by convolving the high-res measurements with a Gaussian distribution leading to a redshift accuracy of {\raise.17ex\hbox{$\scriptstyle\sim$}}$0.015$ at $z${\raise.17ex\hbox{$\scriptstyle\sim$}}$1.0$). In the study by \cite{Fossati2015} velocity cuts of $1500\;$km/s and $7000\;$km/s were used for the high-res measurements and photometric redshift measurements respectively. Although there are differences in redshift and aperture size we compare their Figure 4 with our Figure \ref{fig:spec_vs_photo_envs}. In a qualitative sense the plots are similar. In Figure 4 by \cite{Fossati2015} the photometric measurements of the high densities recover the high-res density measurements as the blue contours follow the dashed 1-to-1 line. We also recover the high density environments better than the low density environments however our photometric environment measurements fall short of the 1-to-1 line. This is because we compute a volume density normalized using the mean density and \cite{Fossati2015} presents a surface density. Both studies report a reduced dynamic range for the photometric measurements compared to the spectroscopic measurements.

\cite{Fossati2015} also investigate trends with galaxy environment. They present a method to ``clean'' the population of low mass centrals and classify the central and satellite galaxies. Figure 10 in \cite{Fossati2015} shows the red fraction for the central galaxies as a function of the surface density and stellar mass for the high-res and photometric redshift measurements. The trend obtained with the high-res measurements is reproduced using the photometric measurements; although this conclusion comes with caveats \citep[see][for more details]{Fossati2015}. In this paper we also report that we are able to reproduce trends, albeit some degradation, using photometric redshifts but instead of using simulations we use observed data from the SDSS. We present the red fraction as a function of mass and environment \citep{Peng2010} using photometric redshifts (see the second row of Figure \ref{fig:surface_plots})

Lastly we add to earlier works by attempting to give a ``ball-park'' estimate of how much larger than spectroscopic samples photometric redshift samples must be to measure environment correlations.

\section{Conclusions} \label{sec:conclusions}

The next generation of photometric surveys such as the DES and Euclid will deliver enormous datasets. The cost of surveying enormous cosmological volumes is redshift precision. The aim of this paper is to examine the impact of uncertain redshifts on the measurement of galaxy environment. 

We use a low redshift ($0.02<z<0.085$) sample of galaxies from the seventh data release of the SDSS and investigate two methods to measure galaxy environment: (i) the $N$\textsuperscript{th} nearest neighbour method (ii) and the fixed aperture method. We study a range of aperture parameters. 

For the $N$\textsuperscript{th} nearest neighbour we study $N=1-10$. This corresponds to median comoving projected scales of $0.3-3.4\;$Mpc. To ensure that the projected size of the apertures are well matched between the two methods we use fixed apertures with radii of $0.1-3.0\;$Mpc centred on the galaxy targets. For both methods we employ a range of velocity cuts: $1000-20,000\;$km/s. A velocity cut of $20,000\;$km/s extends through the entire redshift range. The possible comoving depths of the apertures are $27.9-1271.7\;$Mpc.

We establish a set of spectroscopic benchmark environments for the $N$\textsuperscript{th} nearest neighbour method and the fixed aperture method. The aperture parameters for these benchmark measurements are $N=4$ and $dv=1000\;$km/s for the $N$\textsuperscript{th} nearest neighbour method and $r=1.8\;$Mpc and $dv=1000\;$km/s for the fixed aperture method. 

We calculate environment measurements for a spectroscopic and photometric redshift sample and find the redshift uncertainty of the SDSS photometric redshift sample is $0.0185$. We find that the $N$\textsuperscript{th} nearest neighbour method gives a smoother environment distribution compared to the fixed aperture method, especially for sparse environments. The $N$\textsuperscript{th} nearest neighbour method typically also has a larger dynamic range than the fixed aperture method.

We compare the spectroscopic environment distributions to the photometric environment distributions and find for both methods that the photometric range is smaller than the spectroscopic range for fixed aperture parameter values. We attribute the reduction in dynamic range to the uncertain photometric redshifts. Uncertain redshift measurements tend to scatter galaxies away from dense regions and into less dense regions.

We study the impact of the aperture parameter values by calculating the Spearman rank correlation coefficient between environment measurements and the spectroscopic benchmark environments. 

We construct a set of simulated photometric redshift catalogues by displacing the spectroscopic redshifts and correcting the r-band absolute magnitudes. The catalogues cover a range of redshift uncertainties: $\sigma_{z}=0.0025-0.1$. We use the simulated photometric redshift catalogues to study the impact of uncertain redshifts on the Spearman rank correlation coefficient between the photometric environments and the spectroscopic benchmark environments. 

We find that the Spearman rank correlation coefficient between the photometric environments and the spectroscopic benchmark measurements generally decreases as the redshift uncertainty increases. The correlation depends also on the choice of aperture parameter values. For low redshift uncertainties ($\sigma_z < 0.02$) there exists a relative `sweet spot'. For larger redshift uncertainties providing the aperture parameter are large enough (e.g. $N>3$, $r>1.2\;$Mpc, $dv>10,000\;$km/s and smaller than the maximum values we study) the correlation is relatively insensitive to the aperture parameters. 

In addition we find the parameter values that result in the strongest correlation between the environments and the spectroscopic benchmark measurements at each redshift uncertainty. The parameters that control the projected size of the apertures i.e. $N$ for the $N$\textsuperscript{th} nearest neighbour method and $r$ for the fixed aperture method are fairly insensitive to the redshift uncertainty. The optimal values for these parameters are essentially the same as the benchmark values. The optimal value for the velocity cut however increases with redshift uncertainty. This is to ensure that galaxies that are increasingly scattered along the line of sight by the uncertain redshift measurements are capture within the apertures. 

Studying the Spearman rank correlation coefficient as a function of the redshift uncertainty using the optimal parameter values at each redshift uncertainty we find that the correlation decays rapidly at low uncertainties ($\sigma_{z} < 0.02$) and then more gradually at larger redshift uncertainties ($\sigma_{z} > 0.02$) 

Adopting the optimal parameter values at a redshift uncertainty of $0.1$ we find that Spearman rank correlation coefficient between the photometric environments and the benchmark environments is {\raise.17ex\hbox{$\scriptstyle\sim$}}$0.4$ for both the $N$\textsuperscript{th} nearest neighbour method and the fixed aperture method. 
 
As a case study we examine the bivariate dependence of the red fraction of galaxies on log(mass) and environment. Using the spectroscopic redshift measurements and the $N$\textsuperscript{th} nearest neighbour method ($N=4$ and $dv=1000\;$km/s) we reproduce the result presented by \cite{Peng2010}. In addition we present red fraction surfaces using the fixed aperture method ($r=1.8\;$Mpc and $dv=1000\;$km/s). We also study the red fraction bivariate dependence with the SDSS photometric redshifts and are able to obtain similar, albeit slightly deteriorated surfaces. The information to split the red fraction by mass and environment is still present in the SDSS photometric redshift measurements. Using the simulated photometric redshift catalogues and the optimal parameter values we find that the deterioration of the red fraction surfaces increases with redshift uncertainty. Nevertheless for the $N$\textsuperscript{th} nearest neighbour method even with large redshift uncertainties ($\sigma_z=0.06$) and including an additional scatter of $0.3$ dex in the mass estimates the environmental dependence of the red fraction does not break down entirely. 
  
Using a Monte-Carlo type approach we calculate how much larger, than a spectroscopic sample, a photometric sample needs to be to obtain equivalent fractional errors for trends between galaxy properties and galaxy environment. To do this we use sets of faked data that correlate with the spectroscopic benchmark environments but not with mass. We define the fractional error as the ratio of the median Spearman rank correlation coefficient between the faked data and environment and the standard deviation of the correlation coefficient distribution from a set of $224$ realizations.

We find that photometric redshift samples with the expected DES redshift uncertainty of $0.1$ must be approximately $6-16$ times larger than spectroscopic samples to measure environment correlations with equivalent fractional errors assuming that the correlation between the galaxy properties and the spectroscopic environment measurements was $0.3-0.7$ respectively. The performance of the $N$\textsuperscript{th} nearest neighbour and the fixed aperture method are similar when the optimal parameter values are adopted.  

This study was limited to a low redshift sample and the results may not hold at higher redshifts. Nevertheless the largest spectroscopic pencil beam surveys, such as DEEP2 and COSMOS only cover a few squared degrees on the sky. DES is expected to measure $5000$ square degrees and Euclid will be a full sky survey. Hence such sample size factors are easily achievable in the next generation of large scale photometric surveys potentially enabling competitive measurements of the correlations between galaxy properties and their environments.

\section*{Acknowledgements}

The Science, Technology and Facilities Council is acknowledged for support through the ’Survey Cosmology and Astrophysics’ rolling grant, ST/I001204/1. Numerical computations were done on the Sciama High Performance Compute (HPC) cluster which is supported by the ICG, SEPnet and the University of Portsmouth.

Funding for the SDSS and SDSS-II has been provided by the Alfred P. Sloan Foundation, the Participating Institutions, the National Science Foundation, the U.S. Department of Energy, the National Aeronautics and Space Administration, the Japanese Monbukagakusho, the Max Planck Society, and the Higher Education Funding Council for England. The SDSS Web Site is http://www.sdss.org/.

The SDSS is managed by the Astrophysical Research Consortium for the Participating Institutions. The Participating Institutions are the American Museum of Natural History, Astrophysical Institute Potsdam, University of Basel, University of Cambridge, Case Western Reserve University, University of Chicago, Drexel University, Fermilab, the Institute for Advanced Study, the Japan Participation Group, Johns Hopkins University, the Joint Institute for Nuclear Astrophysics, the Kavli Institute for Particle Astrophysics and Cosmology, the Korean Scientist Group, the Chinese Academy of Sciences (LAMOST), Los Alamos National Laboratory, the Max-Planck-Institute for Astronomy (MPIA), the Max-Planck-Institute for Astrophysics (MPA), New Mexico State University, Ohio State University, University of Pittsburgh, University of Portsmouth, Princeton University, the United States Naval Observatory, and the University of Washington.

We thank Claudia Maraston, Robert Crittenden and Diego Capozzi for useful discussions. We also thank the anonymous referee for thoroughly reviewing the work and for the suggestions that lead to significant improvements.

\bibliography{paperbib}{}

\end{document}